\documentclass[aps,prb,twocolumn,floatfix,citeautoscript,longbibliography]{revtex4-1}

\usepackage{amsmath}
\usepackage{amssymb}
\usepackage{amsfonts}
\usepackage{graphicx,graphics}
\usepackage{bm}

\newcommand{\ket}[1]{\lvert #1\rangle}
\newcommand{\bra}[1]{\langle#1 \rvert}
\newcommand{\abs}[1]{\lvert #1 \rvert}
\newcommand{\expect}[1]{\langle #1\rangle}
\newcommand{\braket}[2]{\langle #1 \rvert #2\rangle}

\newcommand{\br}{\mathbf{r}}
\newcommand{\bk}{\mathbf{k}}
\newcommand{\bq}{\mathbf{q}}
\newcommand{\bR}{\mathbf{R}}

\newcommand{\bT}{\mathbf{T}}

\begin{document}

\title{Atomistic $T$-matrix theory of disordered two-dimensional materials:
  Bound states, spectral properties, quasiparticle scattering, and transport}

\author{Kristen Kaasbjerg}
\email{kkaasbjerg@gmail.com}
\affiliation{Center for Nanostructured Graphene (CNG), Department of Physics, 
  Technical University of Denmark, DK-2800 Kongens Lyngby, Denmark}
\date{\today}
\begin{abstract}
  In this work, we present an atomistic first-principles framework for modeling
  the low-temperature electronic and transport properties of disordered
  two-dimensional (2D) materials with randomly distributed point defects
  (impurities). The method is based on the $T$-matrix formalism in combination
  with realistic density-functional theory (DFT) descriptions of the defects and
  their scattering matrix elements. From the $T$-matrix approximations to the
  disorder-averaged Green's function (GF) and the collision integral in the
  Boltzmann transport equation, the method allows calculations of, e.g., the
  density of states (DOS) including contributions from bound defect states, the
  quasiparticle spectrum and the spectral linewidth (scattering rate), and the
  conductivity/mobility of disordered 2D materials. We demonstrate the method by
  examining these quantities in monolayers of the archetypal 2D materials
  graphene and transition metal dichalcogenides (TMDs) contaminated with vacancy
  defects and substitutional impurity atoms. By comparing the Born and
  $T$-matrix approximations, we also demonstrate a strong breakdown of the Born
  approximation for defects in 2D materials manifested in a pronounced
  renormalization of, e.g., the scattering rate by the higher-order $T$-matrix
  method. As the $T$-matrix approximation is essentially exact for dilute
  disorder, i.e., low defect concentrations ($c_\text{dis} \ll 1$) or density
  ($n_\text{dis}\ll A_\text{cell}^{-1}$ where $A_\text{cell}$ is the unit cell
  area), our first-principles method provides an excellent framework for
  modeling the properties of disordered 2D materials with defect concentrations
  relevant for devices.
\end{abstract}

\maketitle

\section{Introduction}

Over the past decade, there has been an explosive development in theoretical
predictions~\cite{Marzari:Two,Thygesen:The,Thygesen:Discovering,Marzari:Abundance}
and experimetal fabrication of new two-dimensional (2D) materials hosting
exciting electronic properties. This holds great promise for novel applications
in electronics, optoelectronics, and other emerging (spin-, valley-, strain-,
twist-)tronics disciplines. However, atomic disorder which degrades the material
properties is still a major hindrance, and fabrication platforms that can
deliver high-quality materials with low disorder concentrations are
needed. Recently, there have been advances in some of the most widely studied 2D
materials such as, e.g., monolayers of graphene and transition-metal
dichalcogenides (TMDs), where devices based on high-quality materials
encapsulated in ultra-clean van der Waals (vdW) heterostructures have shown
promising electrical and optical properties~\cite{Hone:Disorder}. Such
developments are essential for the realization of quantum devices based on 2D
materials~\cite{Kim:Electrical,Ensslin:Gate}.

The initial characterization of atomic disorder due to point defects in 2D
materials often proceeds by means of scanning tunneling microscopy/spectroscopy
(STM/STS) which provides valuable insight into the defect type as well as the
structural and electronic properties of the defects. In, for example,
graphene~\cite{Strocio:Scattering,Pasupathy:Visualizing,Arkady:Structural} and
TMDs~\cite{Andrei:Bandgap,Pasupathy:Atomic,Pasupathy:Approaching,Bargioni:Large,Bargioni:Identifying,Bargioni:How}
this has been useful for the identification of the most common types of defects
as well as probing for bound defect states which lead to strong modifications of
the electronic properties of the pristine material. In addition, measurements of
quasiparticle interference in various 2D
materials~\cite{Strocio:Scattering,Veuillen:Quasi,Veuillen:Role,Wu:Evidence,Xie:Observation,LeRoy:Local,Khajetoorians:Probing,Busse:Energy,Busse:Suppression},
i.e., spatial ripples in the local density of states (LDOS) in the vicinity of a
defect, is a direct fingerprint of the defect-induced scattering processes which
may govern the electron dynamics and limit the electrical and optical
performance of materials at low temperatures. Theoretical methods which can
supplement such experiments in predicting the impact of defects on the electron
dynamics are of high value for the understanding of electrical and optical
properties of new materials.

In this work, we introduce an atomistic first-principles method for modeling the
electronic properties of disordered 2D materials. Our method is based on
realistic density-functional theory (DFT) calculations of the defect scattering
potential and matrix elements, in combination with the $T$-matrix
formalism~\cite{Rammer,Flensberg} for the description of the interaction with
the random disorder potential. From the disorder-averaged Green's function (GF),
accurate descriptions of experimentally relevant quantities such as, e.g., the
density of states, in-gap bound and resonant quasibound defect states, spectral
properties, and the disorder-induced quasiparticle scattering rate/lifetime can
be obtained. Furthermore, using the $T$-matrix scattering amplitude in the
calculation of the momentum relaxation time in the Boltzmann transport equation
allows for theoretical predictions of the disorder-limited low-temperature
conductivity/mobility as well as its dependence on the Fermi energy (carrier
density). This is thus complementary to our previous
first-principles $T$-matrix study of the LDOS and quasiparticle interference in
2D materials~\cite{Jauho:Symmetry}.

In comparison with analytic and tight-binding based $T$-matrix studies of
defects in, e.g.,
graphene~\cite{Neto:Disorder,Neto:Modeling,Sarma:Density,Neto:Electronic,Katsnelson:Resonant,Ducastelle:Electronic,Fabian:Resonant,Ast:Band}
and black phosphorus~\cite{Chang:Impurity}, our first-principles method permits
for parameter-free modeling of realistic defects in disordered materials. It
furthermore goes beyond other first-principles studies of defects and their
transport-limiting effects based on the Born
approximation~\cite{Pantelides:First,Aberg:Charge,Mertig:Impact,Windl:A,Bernardi:Efficient},
which we here demonstrate breaks down for point defects in 2D materials. The
first-principles $T$-matrix method introduced in this work is therefore
of high relevance for the further development of first-principles transport
methodologies with high predictive
power~\cite{PhysRevB.94.201201,Brandbyge:First,Gibertini:Mobility,Giustino:Towards,Bernardi:Efficient,Giustino:First,Bernardi:Ab}.

The details of our method which is implemented in the GPAW electronic-structure
code~\cite{GPAW,GPAW1,GPAW2} are described in Secs.~\ref{sec:method}
and~\ref{sec:Tmatrix}. In Secs.~\ref{sec:TMD} and~\ref{sec:Graphene}, we
demonstrate the power of our method on a series of timely problems in disordered
monolayers of TMDs and graphene. For the TMDs MoS$_2$ and WSe$_2$ with vacancies
and oxygen substitutionals, we analyze (i) bound and quasibound defect states,
respectively, in the gap and in the bands, (ii) the linewidth in the
quasiparticle spectrum, the energy dependence of the scattering rate in the
$K,K'$ valleys, and the complete suppression of intervalley scattering by the
spin-orbit splitting and a symmetry-induced selection
rule~\cite{Jauho:Symmetry}, and (iii) a prediction of unconventional transport
characteristics in $p$-type WSe$_2$ with a mobility that \emph{decreases} with
increasing Fermi energy. In graphene we focus on vacancies and nitrogen
substitutionals and examine (i) the position of quasibound defect states on the
Dirac cone, (ii) their signature in the quasiparticle spectrum and the presence
(absence) of a band-gap opening for sublattice asymmetric (symmetric) defect
configurations, and (iii) the pronounced electron-hole asymmetry in the
transport characteristics induced by strong resonant scattering.

An important finding of this work is that the Born approximation breaks down for
point defects in both TMDs and graphene, and hence most likely also in other 2D
materials. While it is well-known that the description of quasibound states and
resonant scattering in graphene is beyond the Born approximation, our finding
that it severely overestimates the disorder-induced scattering rate in the 2D
TMDs by up to several orders of magnitude is remarkable, and only emphasizes the
high relevance of a first-principles $T$-matrix approach for the modeling of
disordered 2D materials.

\section{Atomistic defect potentials}
\label{sec:method}

We start by introducing an atomistic first-principles method for the calculation
of the single-defect (or impurity) potential $\hat{V}_i$ and its matrix elements
which are the basic building block in the diagrammatic $T$-matrix formalism for
disordered systems outlined in Sec.~\ref{sec:Tmatrix}. The method is analogous
to the method for calculating the electron-phonon
interaction~\cite{Kaasbjerg:MoS2,Kaasbjerg:Unraveling,Kaasbjerg:MoS2Acoustic},
and is based on DFT within the projector augmented-wave (PAW)
method~\cite{Blochl:PAW}, an LCAO supercell representation of the defect
potential, and is implemented in the GPAW electronic-structure
code~\cite{GPAW,GPAW1,GPAW2}.

In this work, we restrict the considerations to nonmagnetic spin-diagonal
defects in which case the defect potential for a defect of type $i$ takes the
form
\begin{equation}
  \label{eq:Vhat}
  \hat{V}_i = V_i(\hat{\br}) \otimes \hat{s}_0 ,
\end{equation}
where $V_i(\br)$ is the scalar spin-independent defect potential and $\hat{s}_0$
is the identity operator in spin space. We thus neglect defect-induced changes
in the spin-orbit interaction, which are, in general, small relative to the
spin-independent potential. The spin dependence is, of course, important for
spin relaxation and spin-orbit scattering, but this is outside the scope of the
present work.

Here, the spin-diagonal defect potential is defined as the change in the
microscopic crystal potential induced by the defect, and is obtained from DFT as
the difference in the crystal potential between the lattice with a defect and
the pristine lattice, i.e.
\begin{equation}
  \label{eq:Vi}
  V_i(\hat{\br}) = V_\text{def}^i(\hat{\br}) - V_\text{pris}(\hat{\br}) .
\end{equation}
The two potentials have contributions from the atomic cores (ions), which define
the overall potential landscape in the lattice, as well as from the valence
electrons which describe interactions between the valence electrons at a
mean-field level (see Sec.~\ref{sec:paw} below). The defect potential in
Eq.~\eqref{eq:Vi} thus carries information about (i) the defect-induced lattice
imperfection (e.g., vacancy, substitutional, or impurity atom), and (ii) the
electronic relaxation in the vicinity of the defect. Both are important for a
quantitative description of the defect potential.

In practice, the defect potential is expressed in a basis of Bloch states
$\ket{n\bk s}$ of the pristine lattice, where $n$ is the band index, $\bk\in 1$st
Brillouin zone (BZ) is the electronic wave vector, and $s$ is the spin
index. For brevity, we combine in the following the band and spin indices in a
composite ``band'' index. The matrix elements of the defect potential becomes
\begin{align}
  \label{eq:V}
  V_{i,\bk\bk'}^{mn} &  = \bra{m\bk} \hat{V}_i \ket{n\bk'} 
                       \nonumber \\
                     & = \sum_{s_z} \bra{m\bk; s_z} V_i(\hat{\br}) \ket{n\bk';s_z} ,
\end{align}
where as a consequence of the spin-orbit mixing of up and down spin
($s_z=\pm 1$) in the Bloch states, the matrix elements, in general, have
contributions from both spin components $\ket{\cdot; s_z}$ in spite of the fact
that the defect potential itself is spin-diagonal.

The following two subsections summarize our DFT-based supercell method for the
calculation of the defect matrix elements. The two main technical aspects of the
method concern (i) the representation of the defect potential in an LCAO basis,
and (ii) the calculation of the defect potential in the PAW
formalism~\cite{Blochl:PAW}.

\subsection{LCAO supercell representation} 
\label{sec:lcao}

The numerical evaluation of the defect matrix element in Eq.~\eqref{eq:V} is
based on an LCAO expansion of the Bloch functions of the pristine lattice,
$\ket{\psi_{n\bk}} = \sum_{\mu s_z} c_{n \bk}^{\mu s_z} \ket{\phi_{\mu \bk}}$,
where $\mu=(\alpha, i)$ is a composite atomic ($\alpha$) and orbital index ($i$)
and
\begin{equation}
  \label{eq:bloch_sum}
  \ket{\phi_{\mu \bk}} 
  = \frac{1}{\sqrt{N}} \sum_l e^{i \bk \cdot \mathbf{R}_l}
                \ket{\phi_{\mu l}} ,
\end{equation}
are Bloch expansions of the spin-independent LCAO basis orbitals
$\ket{\phi_{\mu l}}$, where $N$ is the number of unit cells in the lattice and
$\mathbf{R}_l=l_1 \mathbf{a}_1 + l_2 \mathbf{a}_2$, $l_i\in \mathbb{Z}$, is the
lattice vector to the $l$'th unit cell with $\mathbf{a}_i$ denoting the primitive
lattice vectors.

\begin{figure}[!t]
  \includegraphics[width=0.6\linewidth]{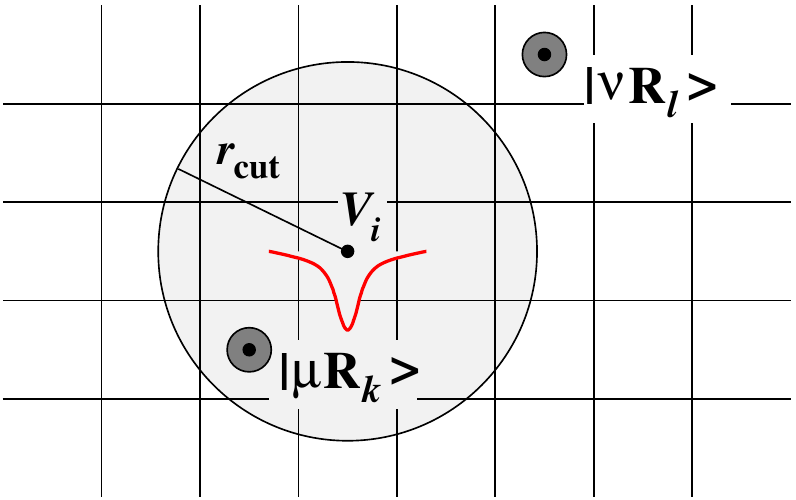}
  \hfill
  \caption{Schematic illustration of the LCAO supercell representation of the
    defect potential $V_i$ given by the matrix elements in the last line of
    Eq.~\eqref{eq:V_lcao}. The square lattice indicates the unit cells of the
    lattice. The defect potential is cut off in real space in order to ensure an
    isotropic range. The real-space cutoff $r_\text{cut}$ is measured from the
    position of the defect site.}
\label{fig:supercell_matrix}
\end{figure}

Inserting in the expression for the matrix element in Eq.~\eqref{eq:V}, we find 
\begin{align}
  \label{eq:V_lcao}
  V_{i,\bk\bk'}^{mn} &  
  = \sum_{s_z} \sum_{\mu\nu} (c_{m \bk}^{\mu s_z})^* c_{n \bk'}^{\nu s_z} 
                         \bra{\phi_{\mu \bk}} V_i(\hat{\br}) \ket{\phi_{\nu \bk'}}  
         \nonumber \\  
     & = \frac{1}{N} \sum_{s_z} 
             \sum_{\mu\nu} (c_{m \bk s}^{\mu s_z})^* c_{n \bk' s'}^{\nu s_z} 
         \nonumber \\  
     & \quad \times \sum_{kl} e^{i ( \bk'\cdot\mathbf{R}_l - \bk\cdot\mathbf{R}_k) }
          \bra{\phi_{\mu k}} V_i(\hat{\br}) \ket{\phi_{\nu l}} ,
\end{align}
where the factor of $1/N$ stems from the normalization of the Bloch sum in
Eq.~\eqref{eq:bloch_sum} to the lattice, the last factor in the second equality
is the LCAO representation of the defect potential $V_i(\br)$ illustrated in
Fig.~\ref{fig:supercell_matrix}, and the $k,l$ sums run over the cells in the
lattice.

In practice, the defect potential is calculated in a finite $N_1\times N_2$
supercell constructed by repeating the primitive unit cell $N_i$ times in the
direction of the $i$'th primitive lattice vector, and with the defect site
located at the center. Due to periodic boundary conditions in the in-plane
directions, the supercell must be chosen large enough that defect sites in
neighboring supercells do not interact. In the direction perpendicular to the
material plane, the cell boundaries are imposed with Dirichlet boundary
conditions, which ensures a common reference for the two potentials on the
right-hand side of Eq.~\eqref{eq:Vi} and avoids spurious interactions between
repetitions of the defect in the perpendicular direction.

In order to impose an isotropic range of the defect potential, matrix elements
involving LCAO basis functions located beyond a cutoff distance $r_\text{cut}$
from the defect site $\mathbf{R}_0$ are zeroed, i.e.,
\begin{equation}
  \label{eq:cut}
  \bra{\phi_{\mu k}} V_i(\hat{\br}) \ket{\phi_{\nu l}}
  = 0 \quad \text{if} \quad 
  \abs{\mathbf{R}_{\mu k,\nu l} - \mathbf{R}_0} > r_\text{cut}  ,
\end{equation}
where $\mathbf{R}_{\mu k}=\mathbf{R}_k + \mathbf{R}_\alpha$ is the center of the
LCAO orbital $\ket{\phi_{\mu k}}$ at the atomic site $\alpha$ in unit cell $k$.

Once the LCAO representation of the defect potential in the supercell has been
obtained, the defect matrix elements can be evaluated efficiently at arbitrary
$\bk,\bk'=\bk+\bq$ vectors using Eq.~\eqref{eq:V_lcao}.

It should be noted that the LCAO procedure for the calculation of the defect
matrix elements outlined here, bears close resemblance to methods based on
Wannier functions~\cite{Louie:e-ph,Bernardi:Ab}. However, the use of a fixed
LCAO basis has the advantage that the additional step for the generation of the
Wannier functions, which is not always trivial, is avoided.

\subsection{PAW method}
\label{sec:paw}

% \begin{itemize}
% \item PAW reminder from PHYSICAL REVIEW B 93, 205147 (2016)
% \item PAW charge density: PHYSICAL REVIEW B 89, 045116 (2014)
% \end{itemize}

In the PAW formulation to DFT~\cite{Blochl:PAW}, the basic idea is to transform
the \emph{all-electron} Hamiltonian, whose eigenstates $\ket{\psi_{n\bk}}$
oscillate strongly in the vicinity atomic cores, into an auxiliary Hamiltonian
with smooth pseudo eigenstates $\ket{\tilde{\psi}_{n\bk}}$, thereby eliminating
the numerical complications associated with an accurate description of rapidly
varying functions.

The physically relevant all-electron wave functions and the auxiliary
pseudo-wave functions are connected via the transformation $\hat{\mathcal{T}}$
defined as
\begin{align}
  \label{eq:psi_paw}
  \ket{\psi_{n\bk}} & = \ket{\tilde{\psi}_{n\bk}} 
        + \sum_{a,i} \left[ \ket{\phi_i^a} - \ket{\tilde{\phi}_i^a}
               \right] \braket{\tilde{p}_i^a}{\tilde{\psi}_{n\bk}}
      \nonumber \\
      & \equiv \hat{\mathcal{T}} \ket{\tilde{\psi}_{n\bk}} ,
\end{align}
where the terms in the sum over atomic sites $a$, respectively, add and subtract
expansions of the all-electron and pseudo-wave functions inside so-called
augmentation spheres $\Omega_a$ centered on the atoms. Here, $\ket{\phi_i^a}$
are the correct all-electron wave functions inside the augmentation spheres, and
the pseudo partial waves $\ket{\tilde{\phi}_i^a}$ and projector functions
$\ket{\tilde{p}_i^a}$ are constructed to obey the completeness relation
$\sum_i \ket{\tilde{\phi}_i^a}\bra{\tilde{p}_i^a}=1$. This ensures the
orthogonality of the all-electron wave functions,
\begin{equation}
  \label{eq:overlap}
  \braket{\psi_{m\bk}}{\psi_{n\bk'}} = 
  \bra{\tilde{\psi}_{m\bk}} \hat{\mathcal{T}}^\dagger \hat{\mathcal{T}} 
  \ket{\tilde{\psi}_{n\bk'}} = \delta_{mn} \delta_{\bk\bk'} .
\end{equation}
via the operator $\hat{\mathcal{T}}^\dagger \hat{\mathcal{T}}$.

Likewise, the all-electron matrix elements of the defect potential $\hat{V}_i$
can be expressed as a matrix element of a transformed operator with respect to
the smooth wave functions $\ket{\tilde{\psi}_{n\bk'}}$, i.e.,
\begin{align}
  \label{eq:O}
  V_{i,\bk\bk'}^{mn} & = \bra{\psi_{m\bk}} \hat{V}_i \ket{\psi_{n\bk'}} 
  %    \nonumber \\
   = \bra{\tilde{\psi}_{m\bk}} \hat{\mathcal{T}}^\dagger \hat{V}_i \hat{\mathcal{T}} 
      \ket{\tilde{\psi}_{n\bk'}}
      \nonumber \\
  & \equiv \bra{\tilde{\psi}_{m\bk}} \hat{\tilde{V}}_i
      \ket{\tilde{\psi}_{n\bk'}} ,
\end{align}
where the transformed operator $\hat{\tilde{V}}_i$ in the last line is given by
\begin{align}
  \label{eq:Otrans}
  \hat{\tilde{V}}_i & = \hat{\mathcal{T}}^\dagger \hat{V}_i \hat{\mathcal{T}} 
  \nonumber \\
  & = \bigg( 1 + \sum_{a,i_1} \ket{\tilde{p}_{i_1}^a} 
    \left[ \bra{\phi_{i_1}^a} - \bra{\tilde{\phi}_{i_1}^a}
               \right]  \bigg)
      \hat{V}_i \nonumber \\
  & \quad\times \bigg( 1 + \sum_{a,i_2} \left[ \ket{\phi_{i_2}^a} - \ket{\tilde{\phi}_{i_2}^a}
    \right] \bra{\tilde{p}_{i_2}^a} \bigg) 
  \nonumber \\
  & \approx \hat{V}_i 
    + \sum_\alpha \sum_{i_1 i_2} \; \ket{\tilde{p}_{i_1}^a} 
        \Delta V_{i_1 i_2}^a \bra{\tilde{p}_{i_2}^a} , 
\end{align}
and the atomic coefficients are defined as
\begin{equation}
  \label{eq:DeltaO}
  \Delta V_{i_1 i_2}^a = 
  \bra{\phi_{i_1}^a} \hat{V}_i \ket{\phi_{i_2}^a} 
  - \bra{\tilde{\phi}_{i_1}^a} \hat{V}_i \ket{\tilde{\phi}_{i_2}^a} .
\end{equation}
The last line in Eq.~\eqref{eq:Otrans} holds for local operators and
furthermore assumes that the bases are complete and that the atomic augmentation
spheres $\Omega_a$ do not overlap. However, in practical PAW calculations,
finite bases and small overlaps between different augmentation spheres can be
tolerated without substantial loss of accuracy.

The transformed operator, which incorporates the full details of the potential
due to the all-electron density (frozen core + valence electrons), can be
expressed as
\begin{equation}
  \label{eq:V_paw}
  \hat{\tilde{V}} = v_\text{eff}(\hat{\br})
    + \sum_\alpha \sum_{i_1 i_2} \; \ket{\tilde{p}_{i_1}^a} 
                                \Delta V_{i_1 i_2}^a
                                \bra{\tilde{p}_{i_2}^a} ,
\end{equation}
where $v_\text{eff}= v_\mathrm{H}+v_\mathrm{xc}$ is the effective potential
given by the sum of the electrostatic Hartree potential $v_\mathrm{H}$
(including the potential due to the atomic cores) and the exchange-correlation
potential $v_\mathrm{xc}$, and the last term is the all-electron corrections
given by the atomic coefficients $\Delta V_{i_1 i_2}^\alpha$ defined in
Eq.~\eqref{eq:DeltaO}. Full PAW expressions for $v_\text{eff}$ and
$\Delta V_{i_1 i_2}^\alpha$ can be found in, e.g., Ref.~\onlinecite{GPAW2}.

In the defect potential in Eq.~\eqref{eq:Vi}, the two contributions to the
potential in Eq.~\eqref{eq:V_paw} describe, respectively, perturbations in the
crystal potential and atomic-core states on the defect. The latter term can be
regarded as the PAW analog of a L{\"o}wdin downfolding of atomic defect states
onto the Bloch functions~\cite{Fabian:Resonant}.

\subsection{Examples}
\label{sec:examples}

In this section, we show examples of matrix elements for the defects in 2D TMDs
and graphene studied in Secs.~\ref{sec:TMD} and~\ref{sec:Graphene} below.
\begin{figure*}[!t]
  \centering
  \includegraphics[width=0.75\linewidth]{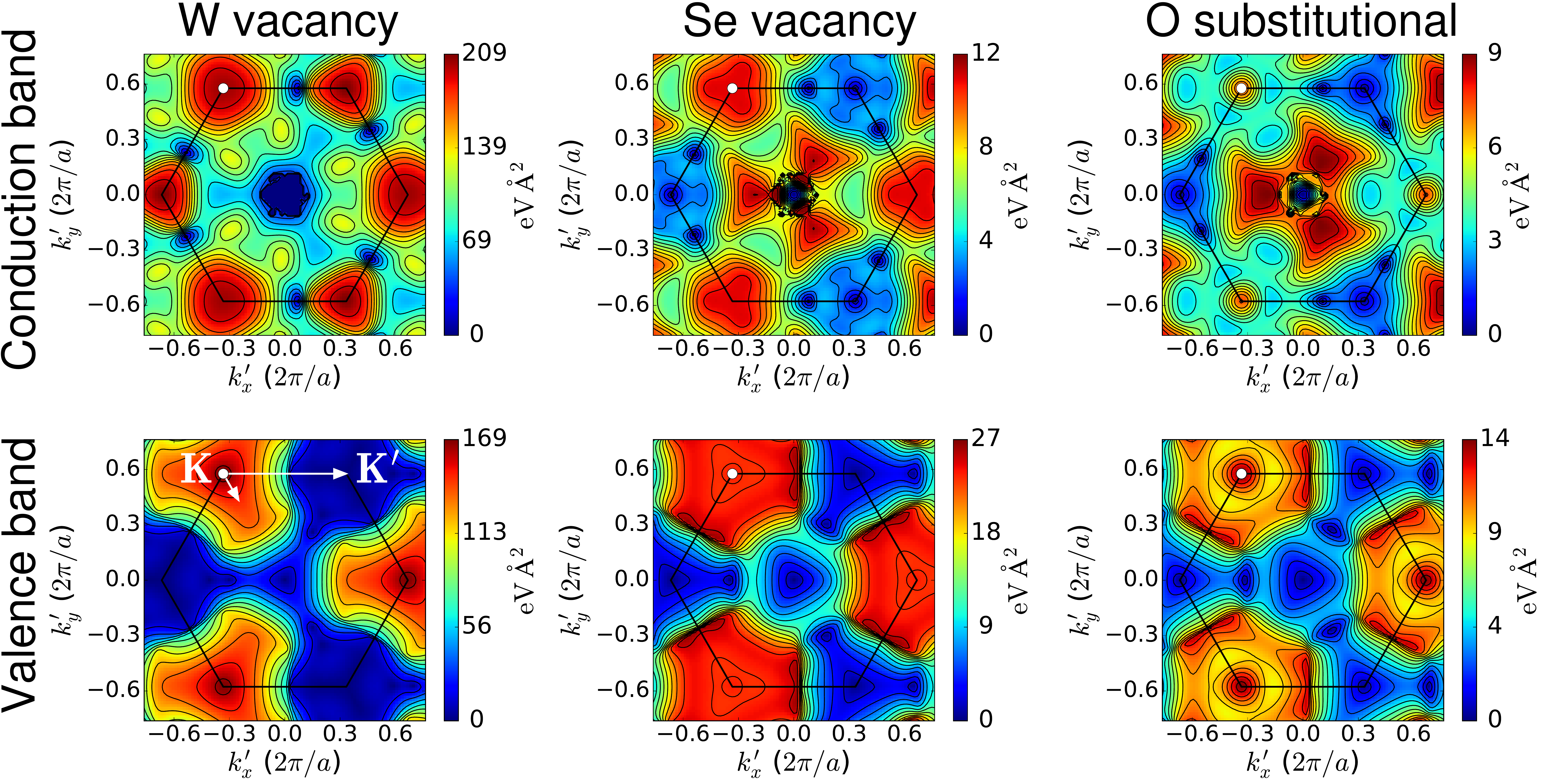}
  \caption{Defect matrix elements for vacancies ($\mathrm{V}_i$,
    $i=\mathrm{W},\mathrm{Se}$) and substitutional oxygen
    ($\mathrm{O}_\mathrm{Se}$) in 2D WSe$_2$. The plots show the absolute value
    of the spin-diagonal intraband matrix elements $V_{i, \bk\bk'}^{nn}$ in the
    valence (bottom) and conduction (top) bands, with the initial state fixed to
    $\bk=\mathbf{K}$ (marked with dots $\circ$) and as a function of $\bk'$. The
    small (large) arrow in the lower left plot corresponds to intravalley
    (intervalley) couplings. DFT parameters: $11\times 11$ supercell with
    10~{\AA} of vacuum to the cell boundaries in the vertical direction.}
\label{fig:M_BZ}
\end{figure*}

In order to relate the DFT-calculated matrix elements (which have units of
energy) to the impurity strength $V_0$ of the $\delta$-function potential in
continuum descriptions of defects, $V_i(\br) = V_0 \delta(\br - \bR_0)$, it is
instructive to rewrite the matrix element as
\begin{equation}
  \label{eq:Vbar}
  V_{i,\bk\bk'}^{mn} \equiv \frac{1}{N} \bar{V}_{i,\bk\bk'}^{mn}
  \equiv \frac{1}{A} \widetilde{V}_{i,\bk\bk'}^{mn} ,
\end{equation}
where the definition of $\bar{V}_i$ in the first step follows trivially from
Eq.~\eqref{eq:V_lcao}, and in the second step we have used that
$A = N A_\text{cell}$, where $A$ and $A_\text{cell}$ are, respectively, the
sample and unit-cell area. In the last equation,
$\widetilde{V}_i = A_\text{cell} \bar{V}_i$ has units of
$\mathrm{eV}\,\mathrm{\AA}^2$ like the impurity strength $V_0$ above. In the
following, the first symbol in Eq.~\eqref{eq:Vbar} is used interchangeably for
the different matrix elements.

\subsubsection{Defects in 2D TMDs}
\label{sec:examples_tmds}

The semiconducting TMD monolayers are some of the most well-studied 2D materials
in terms of electrical, optical, and structural properties. This includes
numerous STM/STS studies of their atomic defects, showing that the most common
types of defects are
monovacancies~\cite{Idrobo:Intrinsic,Suenaga:Threefold,Zhang:Exploring,Krasheninnikov:Two,Robertson:Sulfur,Kim:Stability,Krasheninnikov:Native,Wu:Defect,Muller:Electron,Uemura:Magnetism,Silverman:Micro,Bargioni:Large},
oxygen
substitutionals~\cite{Quek:Origin,Bargioni:Identifying,Bargioni:How,Lischner:Resonant},
i.e., an oxygen atom substituting a chalcogenide atom, and antisite
defects~\cite{Uemura:Magnetism,Pasupathy:Approaching}. The variability in the
predominant defect type stems from the different fabrication
techniques~\cite{Hone:Disorder}, where so far CVD/CVT yield rather low material
quality in comparison to recent flux-grown
materials~\cite{Pasupathy:Approaching,Strauf:Deterministic} with defect
densities as low as $10^{10}$--$10^{11}$~cm$^{-2}$.

In this work, we focus on atomic monovacancies and oxygen substitutionals. In
our DFT calculations of the defect supercell, we find in agreement with previous
works~\cite{Krasheninnikov:Two,Robertson:Sulfur,Kim:Stability,Krasheninnikov:Native}
that structural relaxation around the defect site is minor and is therefore
disregarded here.

Figure~\ref{fig:M_BZ} summarizes the defect matrix elements in 2D WSe$_2$ for W
and Se monovacancies (V$_\mathrm{W,Se}$) as well as oxygen substitutionals
(O$_\mathrm{Se}$). The plots show the absolute value of the spin- and
band-diagonal matrix element in the valence (bottom row) and conduction (top
row) bands, with the initial state of the matrix element $V_{i,\bk\bk'}^{nn}$
fixed to $\bk=\mathbf{K}$ which is the position of the band edges in most of the
semiconducting monolayer TMDs~\cite{Wilson:Visualizing}. The $K,K'$ intravalley
and intervalley matrix elements are indicated with, respectively, a small and a
large arrow in the lower left plot.

Overall, the matrix elements exhibit a nontrivial wave-vector dependence as a
function of $\bq=\bk'-\bk$. Only in the vicinity of the high-symmetry $K,K'$
points are the matrix elements characterized by regions with trigonal symmetry
where a relatively constant value is attained. The magnitude of the $K,K'$
intravalley and intervalley matrix elements for the W vacancy are about an order of
magnitude larger than the matrix elements for the Se vacancy and O
substitutional. This can be understood from the fact that the Bloch states are
dominated by the transition-metal $d$ orbitals~\cite{Yao:SpinValley}, and
therefore have a larger overlap with defects on the transition-metal site
compared to defects on the chalcogenide sites. On the contrary, the matrix
elements for the Se vacancy and O substitutional resemble each other, indicating
that the two types of defects will have similar impact on the electronic
properties of WSe$_2$.

As expected for atomic point defects, the W vacancy gives rise to both strong
intravalley and intervalley matrix elements which are comparable in magnitude. On
the other hand, the matrix elements for the Se vacancy and O substitutional, as
well as the valence-band matrix element for the W vacancy, show a highly
unconventional feature; their intervalley matrix elements are strongly
suppressed and vanishes identically between the two high-symmetry $K,K'$
points. This is in spite of the fact that we here consider the spin-conserving
matrix element where the spin is the same for the two Bloch functions in the
matrix element in Eq.~\eqref{eq:V}. That is, this feature is unrelated to the SO
splitting of the bands~\cite{Schwing:GiantSO,Rossier:Large}. In a recent
work~\cite{Jauho:Symmetry}, we have shown that this originates from the $C_3$
symmetry of the defect sites together with the valley-dependent orbital
character of the Bloch functions~\cite{Jauho:Symmetry}, which give rise to a
symmetry-induced selection rule that makes the $K\leftrightarrow K'$ valence and
conduction band intervalley matrix elements vanish identically, except for
defects on the transition-metal site where it only vanishes in the valence band.

Another important selection rule is the one imposed by time-reversal symmetry
on the intervalley matrix element between states of opposite spin at the $K$ and
$K'$ points, 
\begin{equation}
  \label{eq:VTRS}
  \bra{nK s} \hat{V}_i \ket{nK' \bar{s}} = 0 ,
\end{equation}
where $\bar{s} \neq s$. Note that this holds even in the presence of spin-orbit
coupling in the defect potential which does not break time-reversal symmetry.

\subsubsection{Graphene}
\label{sec:examples_graphene}

Graphene is a host of a large variety of defects ranging from vacancies and
lattice reconstructions like, e.g., Stone-Wales defects, to adatoms and
substitutional atoms involving alkali-metal, halogen, and other nonmetallic
atoms, or molecules~\cite{Arkady:Structural,Hone:Disorder}. In this work, we
restrict the considerations to single carbon
vacancies~\cite{Williams:Defect,Rodriguez:Missing,Fuhrer:Tunable,Andrei:Realization}
and nitrogen
substitutionals~\cite{Pasupathy:Visualizing,Vyalikh:Nitrogen,Henrard:Localized}.
\begin{figure}[!b]
  \centering
  \includegraphics[width=1.05\linewidth]{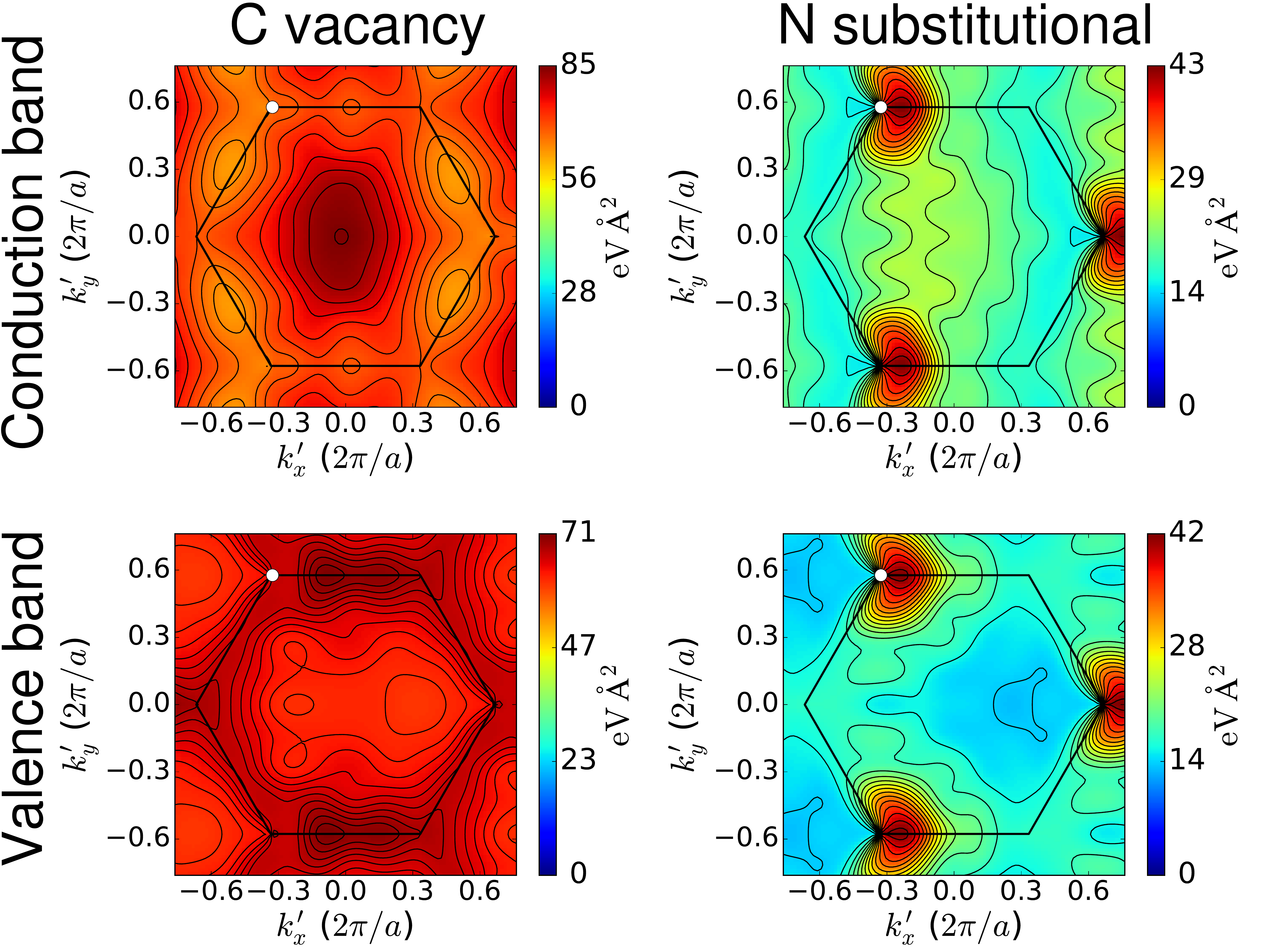}
  \caption{Defect matrix elements for vacancies ($\mathrm{V}_{A,B}$) and
    substitutional nitrogen ($\mathrm{N}_{A,B}$) in graphene. The plots show the
    absolute value of the spin-diagonal intraband matrix elements
    $V_{i, \bk\bk'}^{nn}$ in the valence (bottom) and conduction (top) bands,
    with the initial state fixed to $\bk=\mathbf{K}+ \delta\hat{\mathbf{x}}$,
    $\delta\ll \pi/a$, (marked with dots $\circ$), and as a function of
    $\bk'$. DFT parameters: $11\times 11$ supercell with 10~{\AA} of vacuum to
    the cell boundaries in the vertical direction.}
\label{fig:M1_BZ}
\end{figure}

In Fig.~\ref{fig:M1_BZ} we show the valence and conduction-band matrix elements
for a carbon vacancy (left) and a nitrogen substitutional (right). In contrast
to the matrix elements for the TMDs in Fig.~\ref{fig:M_BZ}, there are no
selection rules on the matrix elements for vacancies and substitutionals in
graphene. Consequently, their matrix elements show less variation as a function
of $\bk'$ and the intravalley and intervalley matrix elements are both significant.

It is instructive to analyze the matrix elements in Fig.~\ref{fig:M1_BZ} in
terms of the standard tight-binding (TB) model for
vacancies~\cite{Neto:Disorder,Wiesendanger:Local,Neto:Modeling,Ducastelle:Electronic,Ferreira:Impurity,Silva:Symmetry,Ast:Band}
and substitutional atoms~\cite{Roche:Charge,Henrard:Long,Charlier:Electronic},
where the defect is often described by a change $V_0$ in the onsite energy of
the defect site. In the $A,B$ sublattice (pseudospin) basis, the defect
potential can therefore be expressed as
\begin{equation}
  \label{eq:Vab_i}
  \hat{V}_i = \frac{V_0}{2}
  \left( \hat{\sigma}_0 \pm \hat{\sigma}_z \right) ,
\end{equation}
where $\hat{\sigma}_i$, $i=x,y,z$, are Pauli matrices ($i=0$ denotes the
identity matrix) in the pseudospin basis, and $\pm$ is for defects on the $A$
and $B$ sublattice, respectively.

For wave vectors in the vicinity of the $K,K'$ points, the graphene TB
Hamiltonian can be approximated by the Dirac model,
$\hat{H}_{\tau\bk} = \hbar v_F \bm{\sigma}_\tau \cdot \bk$, where $\tau=\pm1$ is
the $K,K'$ valley index, and
$\bm{\sigma}_\tau=(\tau \hat{\sigma}_x,\hat{\sigma}_y)$, with eigenstates
$\chi_{n\tau \bk} = \tfrac{1}{\sqrt{2}} (1, n\tau e^{i \tau \theta_\bk} )^T$ and
eigenenergies $\varepsilon_{n\bk} = n \hbar v_F k$, where $n=\pm 1$ is the band
index for the conduction ($c$) and valence ($v$) bands, respectively.

Without loss of generality, we now for simplicity consider a defect on an $A$
site. Performing a unitary transformation to the eigenstate basis, the matrix
elements of the defect potential in Eq.~\eqref{eq:Vab_i} become
\begin{equation}
  \label{eq:Vmn_i}
  V_{i,\bk\bk'}^{mn}
  = \frac{1}{2} V_0  ,
\end{equation}
which are independent on the wave vectors $\bk,\bk'$ and band indices $m,n$, and
thus correspond to identical intravalley and intervalley as well as intraband and
interband couplings.

For the vacancy defect in Fig.~\ref{fig:M1_BZ}, Eq.~\eqref{eq:Vmn_i} is seen to
be a reasonable approximation in the vicinity of the $K,K'$ points. From the
value of the intra- and intervalley matrix elements,
$\sim 70$~$\mathrm{eV}\,\mathrm{\AA}^2$, we find via Eq.~\eqref{eq:Vbar} that
the energy shift at the vacancy site is $V_0 \approx +27$~eV
($A_\text{cell}=5.24\,\text{\AA}^2$). The positive sign can be attributed to the
missing attractive core potential as well as unpaired $\sigma$ electrons left at
the vacancy site which yield an overall repulsive defect potential in
Eq.~\eqref{eq:V_paw}.

For the N substitutional in Fig.~\ref{fig:M1_BZ}, the different values of the
intra- and intervalley matrix elements as well as the anisotropy of the
intravalley matrix element indicate that Eq.~\eqref{eq:Vmn_i} is a less good
approximation. This is due to the fact that substitutional nitrogen donates a
fraction of an electron to the graphene lattice and thereby ends up as a
positively charged impurity characterized by a strong intravalley matrix
element~\cite{Sarma:Carrier}. The defect potential due to chemical
substitutionals therefore presents both short range (since there is a
substantial onsite chemical energy shift) as well as some long-range
features~\cite{Roche:Charge}. In Sec.~\ref{sec:Graphene} below, we find that an
average value of $V_0 \approx -10$~eV ($\sim 26$~$\mathrm{eV}\,\mathrm{\AA}^2$),
where the minus sign is due to the partially positively charged N
substitutional, yields good agreement with our full DFT-based results.

\section{$T$-matrix formalism}
\label{sec:Tmatrix}

In this section, we introduce the $T$-matrix formalism~\cite{Rammer,Flensberg}
for the description of (i) a single isolated defect at $\bR_0$ with defect
potential $V_i(\br - \bR_0)$, and (ii) disordered systems with a random
configuration of defects and total disorder potential
$V_\text{dis}(\br) = \sum_{i,\bR_i} V_i(\br - \bR_i)$ where $\bR_i$ denotes the
positions of defects of type $i$. The latter case is the main focus of this work.

\subsection{Single defects}

The situation of a single isolated defect in an otherwise perfect infinite
lattice is relevant for, e.g., STM/STS studies which probe the LDOS in the
vicinity of the defect site which can be obtained as
\begin{equation}
  \label{eq:ldos}
  \rho(\br,\varepsilon) = -\frac{1}{\pi} \mathrm{Im} \,
       G(\br,\br; \varepsilon)  ,
\end{equation}
where $G(\br,\br; \varepsilon) = \bra{\br} \hat{G}(\varepsilon) \ket{\br}$ is
the real-space representation of the Green's function (GF; all Green's functions
are assumed to be retarded in this work).

For the single-defect problem, the \emph{exact} GF can be expressed in terms of
the $T$ matrix which describes scattering off the defect to infinite order in
the defect potential,
$\hat{T}_i(\varepsilon) = \hat{V}_i + \hat{V}_i \hat{G}^0(\varepsilon)
\hat{T}_i(\varepsilon)$,
where $\hat{G}^0(\varepsilon) = 1/[\varepsilon - \hat{H}_0]$ is the GF of the
pristine lattice. In the basis of the Bloch states $\{\psi_{n\bk}\}$ of the
pristine lattice, the GF becomes
\begin{equation}
  \label{eq:GF_T}
  \hat{G}_{\bk \bk'}(\varepsilon) 
      = \delta_{\bk, \bk'} \hat{G}_{\bk}^0(\varepsilon)
      + \hat{G}_{\bk}^0(\varepsilon)
        \hat{T}_{\bk \bk'}(\varepsilon)
        \hat{G}_{\bk'}^0(\varepsilon) ,
\end{equation}
where $G_{n\bk}^0(\varepsilon) = (\varepsilon - \varepsilon_{n\bk} +
i\eta)^{-1}$, and 
\begin{equation}
  \label{eq:Tmatrix}
  \hat{T}_{i,\bk\bk'}(\varepsilon) = \hat{V}_{i,\bk\bk'}
  + \sum_{\bk''} \hat{V}_{i,\bk\bk''} \hat{G}_{\bk''}^0(\varepsilon)
     \hat{T}_{i,\bk''\bk'}(\varepsilon) ,
\end{equation}
Here, $\hat{V}_{i,\bk\bk'}$ are the matrix elements of the defect potential in
Eq.~\eqref{eq:V} and the sum over $\bk''\in 1$st~BZ is over virtual intermediate
states. In contrast to $\hat{V}_{i,\bk\bk'}$, the $T$ matrix is, in general, not
Hermitian.

Given the GF in Eq.~\eqref{eq:GF_T}, the real-space LDOS in Eq.~\eqref{eq:ldos}
which contains information about the electronic properties of the defect can be
obtained via a Fourier transform~\cite{Jauho:Symmetry}. For example,
defect-induced bound states manifest themselves in a high LDOS intensity at
energies corresponding to the bound-state energy. They arise when the $T$ matrix
introduces new poles in the GF via the correction
$\delta \hat{G} = \hat{G}^0 \hat{T}\hat{G}^0$ in the last term of
Eq.~\eqref{eq:GF_T}. From the full matrix form of the $T$-matrix
\begin{equation}
  \label{eq:T_matrix}
  \bT(\varepsilon) = \left[ \mathbf{1} - 
       \mathbf{V} \mathbf{G}^0(\varepsilon) \right]^{-1} \mathbf{V} ,
\end{equation}
where the boldface symbols denote matrices in the band ($n$) and wave-vector
($\bk$) indices, the poles of the $T$ matrix are seen to appear at energies
where the determinant of the matrix in the square brackets vanishes, i.e.,
\begin{equation}
  \label{eq:det}
  \det\left[ \mathbf{1} - 
       \mathbf{V} \mathbf{G}^0(\varepsilon) \right] = 0  .
\end{equation}
Therefore, the positions of the bound states depend sensitively on the defect
matrix elements and the band structure, and since also high-energy bands can be
involved in the formation of bound states, their exact position can, in general,
not be inferred from low-energy models. For in-gap bound states residing in the
band gap of a semiconductor, the bound states form discrete energy levels and
will be strongly localized to the defect site due to a weak interaction with the
delocalized Bloch states. On the other hand, quasibound resonant state in the
bands acquire a finite width and tend to be more delocalized.

An in-depth study of the LDOS and the associated quasiparticle interference is
beyond the scope of this work and has been deferred to other
works~\cite{Jauho:Symmetry,Kaasbjerg:Unified}.

\subsection{Disordered systems}

In disordered systems with a random configuration of defects, experimental
observables are often self-averaging and must be obtained on the basis of the
disorder-averaged GF. In contrast to the single-defect problem discussed above,
the problem for the disorder-averaged GF \emph{cannot} be solved exactly.

The disorder-averaged GF is given by the Dyson equation
\begin{equation}
  \label{eq:dyson}
  \hat{G}_{\bk}(\varepsilon) = \hat{G}_{\bk}^0(\varepsilon) + 
      \hat{G}_{\bk}^0(\varepsilon) \hat{\Sigma}_{\bk}(\varepsilon) \hat{G}_{\bk}(\varepsilon) ,
\end{equation}
where the disorder self-energy $\hat{\Sigma}_\bk$ accounts for the interaction
with the disorder potential, and introduces spectral shifts, broadening, and
potentially bound defect states. As the disorder average restores translational
symmetry, the GF is diagonal in $\bk$. However, the matrix structure in the band
and spin indices is retained, and Eq.~\eqref{eq:dyson} must be solved by matrix
inversion.

\begin{figure}[!t]
  \centering
  \vspace{5mm}
  \includegraphics[scale=0.99]{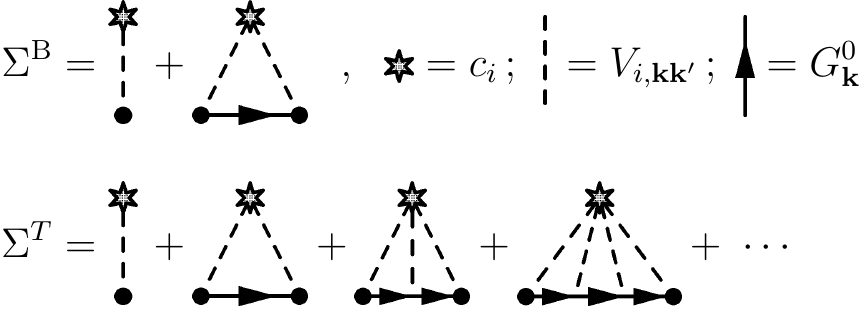}
  \caption{Feynman diagrams for the Born $\Sigma^\text{B}$ (top) and $T$-matrix
    $\Sigma^T$ (bottom) approximations for the disorder self-energy. The
    $T$-matrix self-energy sums up all diagrams involving multiple scattering
    off the same defect and is therefore exact to lowest order in the defect
    concentration $c_i=N_i/N$ where $N_i$ and $N$ are, respectively, the number
    of defects and unit cells in the lattice.}
\label{fig:feynman}
\end{figure}

In this work, the self-energy is described at the level of the Born and
$T$-matrix (full Born) approximations~\cite{Rammer,Flensberg}, which apply to
dilute concentrations of defects. The two self-energies are illustrated with
Feynman diagrams in Fig.~\ref{fig:feynman}, where the individual diagrams
describe repeated scattering off single defects to different orders in the
scattering potential.

The $T$-matrix self-energy $\hat{\Sigma}^T$ in the bottom equation of
Fig.~\ref{fig:feynman} takes into account multiple scattering off defects to
all orders in the defect potential, and is therefore \emph{exact} to lowest
order in the disorder concentration $c_i=N_i/N$ (or density $n_i=N_i /A$) where
$N_i$ is the number of defects of type $i$. The self-energy is given
by~\cite{Rammer,Flensberg}
\begin{align}
  \label{eq:fullborn}
  \hat{\Sigma}_{i,\bk}^T(\varepsilon) = N_i   
  \bigg[
       \hat{V}_{i,\bk\bk} +  \sum_{\bk'} \hat{V}_{i,\bk\bk'} 
     \hat{G}_{\bk'}^0(\varepsilon) \hat{T}_{i,\bk'\bk}(\varepsilon)
  \bigg] ,
\end{align}
where $\hat{T}$ is the $T$ matrix in Eq.~\eqref{eq:Tmatrix}, and can be
expressed in terms of the $\bk$-diagonal of the $T$ matrix as
\begin{equation}
  \label{eq:sigma_T}
  \hat{\Sigma}_{i,\bk}^T(\varepsilon) 
  = N_i \hat{T}_{i,\bk\bk}(\varepsilon)
  \equiv c_i \hat{\bar{T}}_{i,\bk\bk}(\varepsilon) 
  \equiv n_i \hat{\widetilde{T}}_{i,\bk\bk}  .
\end{equation}
Here, the definitions of the symbols in the two last equalities are analogous to
the ones for the defect matrix elements in Eq.~\eqref{eq:Vbar}, and express the
self-energy in terms of the disorder concentration $c_i$ or density $n_i$,
respectively. In the latter case, the $T$ matrix
$\widetilde{T}=A_\text{cell}\bar{T}$ has units of $\mathrm{eV}\,\mathrm{\AA}^2$.

In the Born approximation in the top equation of Fig.~\ref{fig:feynman}, the
self-energy is truncated after the second-order term in Eq.~\eqref{eq:fullborn},
i.e.,
\begin{align}
  \label{eq:born}
  \hat{\Sigma}_{i,\bk}^\text{B}(\varepsilon) = N_i 
  \bigg[
      \hat{V}_{i,\bk\bk} + \sum_{\bk'} \hat{V}_{i,\bk\bk'} 
      \hat{G}_{\bk'}^0(\varepsilon) \hat{V}_{i,\bk'\bk}
  \bigg] .
\end{align}
Here, the first lowest-order term given by the $\bk$-diagonal matrix elements of
the defect potential is purely real and gives rise to a shift of the unperturbed
band energies $\varepsilon_{n\bk}$. The second-order contribution in the last
term gives the leading-order contribution to the scattering rate, or linewidth
broadening, corresponding to the Fermi's golden rule expression
\begin{equation}
  \label{eq:tau_born}
  \tau_{n\bk}^{-1} = \frac{2\pi}{\hbar} N_i \sum_{m\bk'}
        \abs{V_{i,\bk\bk'}^{nm}}^2 
        \delta(\varepsilon_{n\bk} - \varepsilon_{m\bk'}) .
\end{equation}
In Sec.~\ref{sec:qp} below, we discuss the $T$-matrix generalization of this
expression via the Optical theorem.

In the above, we have only considered defects of a single type $i$. For disorder
consisting of different types of defects, the disorder average involves an
average over the defect types in addition to the usual average over their random
positions. In the Born and $T$-matrix approximations which neglect coherent
scattering off different defect sites, this amounts to averaging over the
self-energies of the different defect types, i.e.
\begin{equation}
  \label{eq:sigma_avg}
  \expect{\hat{\Sigma}_\bk}_\text{dis} = \sum_i \hat{\Sigma}_{i,\bk} =
  N_\text{dis} \sum_i x_i \hat{T}_{i,\bk\bk} ,
\end{equation}
where $N_\text{dis}=\sum_i N_i$ is the total number of defects, and
$x_i=N_i/N_\text{dis}$ is the fraction of defects of type $i$. Note that this
procedure does \emph{not} apply to self-energies containing, e.g., diagrams with
crossed impurity lines~\cite{Rammer,Flensberg}. In the following, we omit the
sum over defect types for brevity.

The difference between the Born and $T$-matrix approximations is that the
Born approximation is only valid for \emph{weak} defects, while the
infinite-order $T$-matrix approximation applies to defects of \emph{arbitrary}
strength. As a consequence, the $T$ matrix generally renormalizes the Born
results due to multiple scattering processes if the defect is not weak. The
formation of bound defect states is another good example where the Born
approximation fails to capture the correct physical picture described by the
$T$-matrix approximation.

Both the Born and $T$-matrix self-energies are first order in the disorder
concentration $c_i = N_i /N$ (defect sites per unit cell), and hence valid for
dilute defect concentrations, $c_i \ll 1$ (or $n_i\ll A_\text{cell}^{-1}$). To
demonstrate the wide range of disorder densities where this is fulfilled in 2D
materials, we consider graphene ($A_\text{cell}=5.24\,\text{\AA}^2$) where a
disorder concentration of $c_i=1\,\%$ is equivalent to a density of
$n_i \sim 2\times 10^{13}\,\mathrm{cm}^{-2}$. This corresponds to a rather poor
material quality, why the $T$-matrix self-energy is an excellent approximation
for most experimentally relevant disorder concentrations.

At the level of the two-particle GF for the conductivity there are, however,
effects not captured by the $T$-matrix approximation even at low disorder
concentrations. One well-known example is the weak-localization correction to
the conductivity which arises due to interference between scattering processes
at different defect sites~\cite{Lee,Lee:rmp}. Nevertheless, the $T$-matrix
approach presented here provides a good compromise between wide applicability
and practicality for applications with realistic defects and band structures.

\subsubsection{Quasiparticle spectrum and scattering}
\label{sec:qp}

As a consequence of disorder scattering, the pristine band structure is
renormalized and broadened, yielding quasiparticle (QP) states with a finite
lifetime which can be probed in, e.g., ARPES. The measured spectral function
$A_\bk(\varepsilon)=\sum_n A_{n\bk}(\varepsilon)$ is given by the diagonal
elements of the GF as
\begin{equation}
  \label{eq:spectral}
  A_{n\bk}(\varepsilon) 
  = - 2 \mathrm{Im} \, G_{\bk\bk}^{nn}(\varepsilon) ,
\end{equation}
where $A_{n\bk}$ obeys the sum rule
$\int \! \tfrac{d\varepsilon}{2\pi} \, A_{n\bk} (\varepsilon) = 1$.

While our numerical calculations of the spectral function and DOS presented
below are based on the full matrix form of the GF in Eq.~\eqref{eq:dyson}, it is
instructive to assume a diagonal form of the self-energy and GF,
\begin{equation}
  \label{eq:G_diag}
  G_{n\bk}(\varepsilon) = 
      \frac{1}{\varepsilon -\varepsilon_{n\bk} - \Sigma_{n\bk}(\varepsilon)},
\end{equation}
in order to analyze the effects of disorder scattering in closer detail.

In the diagonal approximation for the GF in Eq.~\eqref{eq:G_diag}, the spectral
function can in the vicinity of the QP energies $\tilde{\varepsilon}_{n\bk}$
given by the solution to the QP equation 
\begin{equation}
  \label{eq:qpeq}
  \varepsilon - \varepsilon_{n\bk} - \mathrm{Re}\Sigma_{n\bk} (\varepsilon) = 0 ,
\end{equation}
be approximated as
\begin{equation}
  \label{eq:spectral_qp}
    A_{n\bk}(\varepsilon) \approx Z_{n\bk}
      \frac{\gamma_{n\bk}}
           {(\varepsilon - \tilde{\varepsilon}_{n\bk})^2 
             + (\gamma_{n\bk}/2)^2} ,
\end{equation}
where the QP weight is given by
$Z_{n\bk}=[1- \partial_\varepsilon \mathrm{Re}\,
\Sigma_{n\bk}(\tilde{\varepsilon}_{n\bk})]^{-1}$,
and the linewidth broadening is given by the imaginary part of the self-energy,
\begin{align}
  \label{eq:gamma_nk}
  \gamma_{n\bk} & = -2 Z_{n\bk} \mathrm{Im}\, \Sigma_{n\bk}(\tilde{\varepsilon}_{n\bk})
                  \nonumber \\
  & = - 2 Z_{n\bk} N_i \mathrm{Im}\, T_{i,\bk\bk}^{nn}(\tilde{\varepsilon}_{n\bk}) ,
\end{align}
evaluated at the on-shell QP energy $\varepsilon=\tilde{\varepsilon}_{n \bk}$,
and where the last equality holds for the $T$-matrix self-energy.

Via the optical theorem~\cite{Rammer,Flensberg}, the diagonal elements of the
imaginary part of the $T$ matrix can be expressed as
\begin{align}
  \label{eq:optical}
  -2 \mathrm{Im} T_{i,\bk\bk}^{nn}(\varepsilon) 
  & = -2 \mathrm{Im} \sum_{m\bk'}
       \frac{\abs{T_{i,\bk\bk'}^{nm}(\varepsilon)}^2}
             {\varepsilon - \varepsilon_{m \bk'} + i \eta} \nonumber \\
    & = 2\pi \sum_{m\bk'}
        \abs{T_{i,\bk\bk'}^{nm}(\varepsilon)}^2 
        \delta(\varepsilon-\varepsilon_{m\bk'}) ,
\end{align}
and the lifetime broadening (or scattering rate) in Eq.~\eqref{eq:gamma_nk} can
be brought on a form which resembles the Born expression in
Eq.~\eqref{eq:tau_born}. This allows to identify the elements
$T_{i,\bk\bk'}^{nm}$ of the $T$ matrix as the renormalized Born scattering
amplitude given by the bare matrix element $V_{i,\bk\bk'}^{nm}$. Furthermore,
the optical theorem in Eq.~\eqref{eq:optical} can be used to separate out the
contributions to the lifetime broadening from, e.g., intravalley and intervalley
scattering by splitting the $\bk'$ sum into sums over intravalley and
intervalley processes,
$\sum_{\bk'} \rightarrow \sum_{\bk'\in\text{intra}} +
\sum_{\bk'\in\text{inter}}$.
This may be desirable in order to extract, e.g., the disorder-limited valley
lifetime.

In context of the discussion of scattering above, it is important to note that
selection rules in the defect matrix elements $V_{i,\bk\bk'}^{nm}$ imposed by a
symmetry $\Theta$ common to the lattice and defect potential,
$\hat{V}_i = \Theta \hat{V}_i \Theta^{-1}$, are transferred to the elements of
the $T$ matrix. This follows straight forwardly from the fact that the $T$
matrix transforms as the defect potential,
$\hat{T}_i = \Theta \hat{T}_i \Theta^{-1}$, under such symmetry
transformations. Thus, scattering processes which are forbidden by symmetry due
to vanishing matrix elements in the Born approximation, are also forbidden in
$T$-matrix approximations, in spite of the fact that scattering processes in the
latter case proceed via virtual intermediate states.

\subsubsection{Bound defect states}

When bound states appear in the single-defect problem, it is interesting to ask
how they manifest themselves in the spectral function and the DOS (here defined
per unit cell) of the disordered system,
\begin{equation}
  \label{eq:dos}
  \rho(\varepsilon) = -\frac{1}{N \pi} \mathrm{Im} 
      \left[ \mathrm{Tr}\, \hat{G}(\varepsilon) \right] ,
\end{equation}
where $N$ is the number of unit cells in the lattice. Naively, one would expect
peaks at the bound-state energies of the isolated defect. To shed light on the
the bound-state DOS of a dilute disordered system, we consider a situation where
the single-defect GF in Eq.~\eqref{eq:GF_T} has an in-gap bound state stemming
from a pole of the $T$ matrix with energy $E_\text{b}$.

To facilitate a simple analysis, we resort again to the diagonal form of the GF
in Eq.~\eqref{eq:G_diag}. In this case, the self-energy is given by the diagonal
elements of the $T$ matrix, which in the vicinity of the pole can be
approximated as
\begin{equation}
  \label{eq:Tpole}
  T_{n\bk} (\varepsilon) = \frac{1}{N}\frac{a_{n\bk}}
      {\varepsilon - E_\text{b} + i \eta} ,
  \quad \varepsilon \approx E_\text{b} ,
\end{equation}
where $a_{n\bk}$ (with unit $\mathrm{eV}^2$) is the strength of the pole for a
given band $n$ and $\bk$ point, and the positive infinitesimal $\eta=0^+$
ensures the correct analytic behavior of the $T$ matrix.

\begin{figure}[!t]
  \centering
  \includegraphics[width=0.65\linewidth]{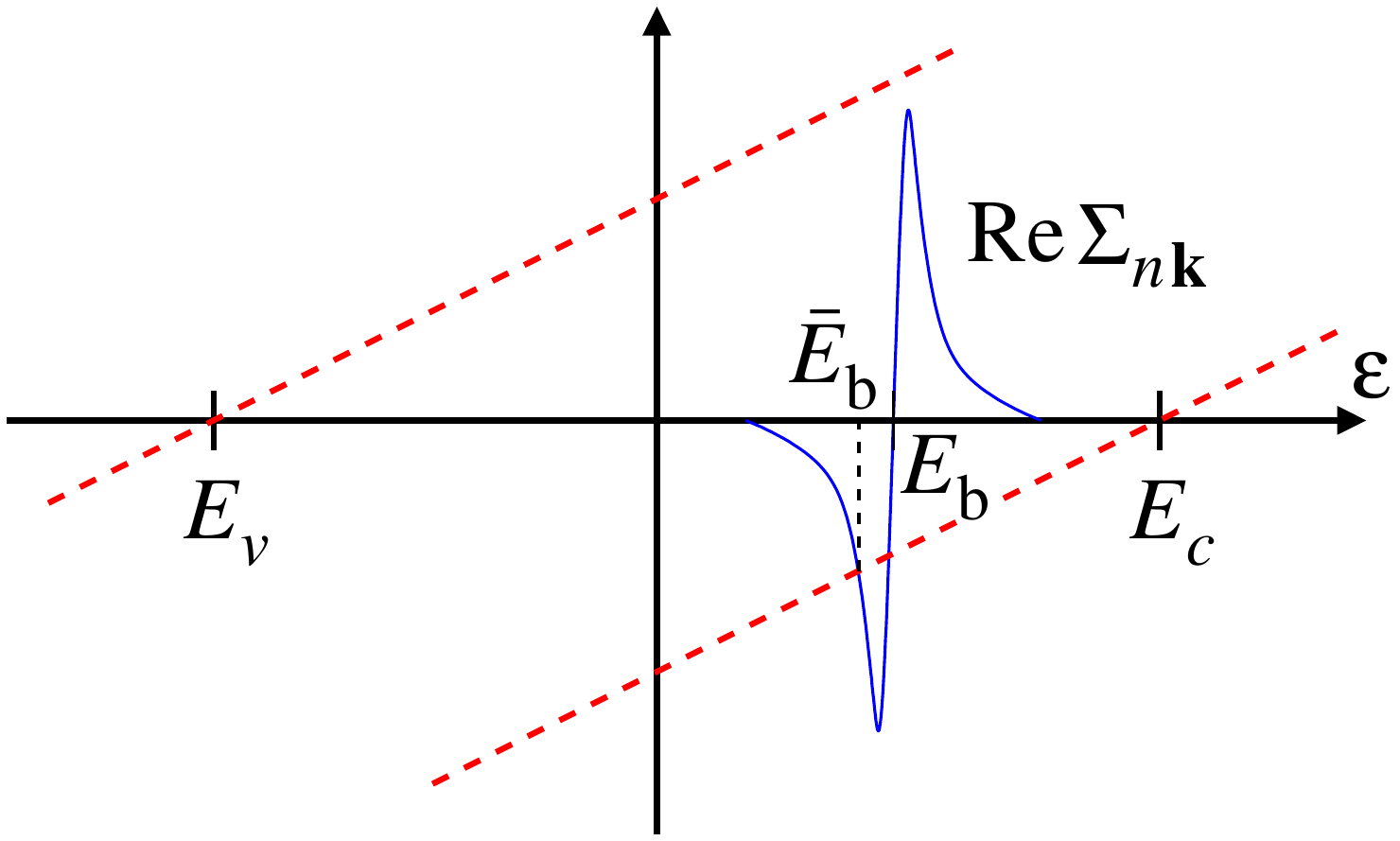}
  \caption{Graphical illustration of the bound-state solution
    $\varepsilon=\bar{E}_\text{b}$ to the QP equation in Eq.~\eqref{eq:qpeq} in
    the presence of an in-gap pole in the $T$-matrix self-energy
    $\Sigma_{n\bk}^T = N_i T_{n\bk}$. The bound-state position
    $\bar{E}_\text{b}$ is determined by the intersection [with
    $\partial_\varepsilon \mathrm{Re}\Sigma_{n\bk}^T<0$ in order for
    $Z_{n\bk}^\text{b}$ in Eq.~\eqref{eq:G_e0} to be $>0$] between
    $\mathrm{Re}\Sigma_{n\bk}^T$ (solid blue line) and
    $\varepsilon - \varepsilon_{n\bk}$ (dashed red lines) here sketched for
    states at the band edge, i.e. $\varepsilon_{n\bk}=E_{v,c}$. }
\label{fig:graphical}
\end{figure}

We now demonstrate how the pole of the $T$ matrix gives rise to a new
bound-state pole in the disorder-averaged GF, whose energy we denote
$\bar{E}_\text{b}$ in order to distinguish it from the $T$-matrix pole at
$\varepsilon=E_\text{b}$. The bound state, being a well-defined quasiparticle of
the disordered system, emerges as a new solution to the QP equation in
Eq.~\eqref{eq:qpeq} with energy $\varepsilon = \bar{E}_\text{b}$ as sketched
graphically in Fig.~\ref{fig:graphical} for states at the band edges,
i.e., $\varepsilon_{n\bk}=E_{v,c}$. Here, the generic shape of the real part of
the $T$ matrix (self-energy) originates from its pole form in
Eq.~\eqref{eq:Tpole}.

Expanding the self-energy around $\varepsilon=\bar{E}_\text{b}$,
$\Sigma(\varepsilon) \approx \mathrm{Re}\Sigma(\bar{E}_\text{b}) + (\varepsilon
-
\bar{E}_\text{b}) \partial_\varepsilon\mathrm{Re}\Sigma\vert_{\varepsilon=\bar{E}_\text{b}}$,
the in-gap GF takes the form
\begin{equation}
  \label{eq:G_e0}
  G_{n\bk} (\varepsilon) \approx
  \frac{Z_{n\bk}^\text{b}}{\varepsilon - \bar{E}_\text{b} + i \eta} ,
  \quad \varepsilon \approx \bar{E}_\text{b} ,
\end{equation}
where
$Z_{n\bk}^\text{b} = (1 - \partial_\varepsilon
\mathrm{Re}\Sigma_{n\bk}\vert_{\varepsilon=\bar{E}_\text{b}})^{-1}$
is the contribution to the spectral weight of the bound states from the state
$n\bk$. Note that bound states may have contributions from several bands, and by
virtue of the sum-rule for the spectral function in Eq.~\eqref{eq:spectral},
$Z_{n\bk}^\text{b} \ll 1$ in order for the pristine bands to remain well-defined.

In the presence of bound states, the spectral function in
Eq.~\eqref{eq:spectral} thus becomes a sum of two contributions
$A_{n\bk}=A_{n\bk}^\text{qp} + A_{n\bk}^\text{b}$ given, respectively, by (i)
Eq.~\eqref{eq:spectral_qp} describing the QPs associated with the pristine
bands, and (ii) the imaginary part of Eq.~\eqref{eq:G_e0} for the bound state.
Note that the bound-state solution to the QP equation may, in fact, depend on
both $n$ and $\bk$, such that dispersive defect bands, i.e.
$\bar{E}_\text{b} \rightarrow \bar{\varepsilon}_{\mathrm{b},n\bk}$, may arise
even though hybridization between the defects is not accounted for in the
$T$-matrix approximation. As we demonstrated in a recent work, this situation
arises in, e.g., alkali-metal-decorated graphene~\cite{Kaasbjerg:Spectral}.

The bound-state position $\bar{E}_\text{b}$ naturally depends on the disorder
concentration via Eq.~\eqref{eq:qpeq}. With the self-energy given by the pole
form of the $T$ matrix in Eq.~\eqref{eq:Tpole}, it follows straight forwardly
that $\bar{E}_\text{b}\rightarrow E_\text{b}$ in the limit of
$c_i \rightarrow 0$. This is not surprising as the $c_i \rightarrow 0$ limit is
equivalent to setting $N_i = 1$ and letting $N \rightarrow \infty$, i.e., the
single-defect limit. In addition, the quasiparticle weight vanishes as
$Z_{n\bk} \approx c_i a_{n\bk}/(\varepsilon - \varepsilon_{n\bk})^2$, implying
that the in-gap form of the imaginary part of the GF becomes
\begin{equation}
  \label{eq:G_1}
  -\mathrm{Im} G_{n\bk} (\varepsilon) \approx 
   \frac{\pi}{N}
  \frac{a_{n\bk}}{(\varepsilon - \varepsilon_{n\bk})^2 }
  \delta (\varepsilon - E_\text{b})
  , \quad % \text{for} \quad
  c_i \rightarrow 0 ,
\end{equation}
which is identical to the in-gap form of the imaginary part of the single-defect
GF (its diagonal elements) in Eq.~\eqref{eq:GF_T}. Thus, the bound-state DOS of
a dilute system approaches the single-defect DOS for $c_i\rightarrow 0$ as
anticipated, and its weight vanishes as $1/N$ compared to the DOS of the
pristine lattice. At higher disorder concentrations (but still $ \ll 1$), the
graphical solution of Eq.~\eqref{eq:qpeq} in Fig.~\ref{fig:graphical} indicates
that the GF pole $\bar{E}_\text{b}$ drifts away from the $T$-matrix pole at
$E_\text{b}$.

When the disorder concentration becomes so high that electronic states on
neighboring defect sites start to hybridize and form impurity bands, the
single-site $T$-matrix approximation considered here breaks down and more
advanced methods are required~\cite{RevModPhys.46.465,Roche:Linear}.

\subsubsection{Transport}

At low temperatures where electron-phonon scattering is frozen out, the
longitudinal conductivity is often limited by the intrinsic disorder of the
material. Within the framework of Boltzmann transport theory, the
disorder-limited longitudinal conductivity $\sigma$ can be obtained from the
current density,
\begin{equation}
  \label{eq:conductivity}
  \mathbf{j} = q \sum_{n\bk} \mathbf{v}_{n\bk} \delta f_{n\bk} 
    \equiv \sigma \mathbf{E},
\end{equation}
where $q$ is the charge of the carriers, $\mathbf{v}_{n\bk}=1/\hbar \nabla_\bk
\varepsilon_{n\bk}$ is the band velocity, and $\delta f_{n\bk}= f_{n\bk} -
f_{n\bk}^0$ is the deviation of the distribution function away from the
equilibrium Fermi-Dirac distribution, $f_{n\bk}^0\equiv
f^0(\varepsilon_{n\bk})$, to first order in the applied field $\mathbf{E}$.

The deviation function is given by the linearized Boltzmann equation which for
elastic disorder scattering in a multiband system takes the form
\begin{equation}
  \label{eq:BE}
  q \mathbf{v}_{n\bk} \cdot \mathbf{E}
  \left.\frac{\partial f^0}{\partial \varepsilon} \right\vert_{\varepsilon=\varepsilon_{n\bk}}
      = -\sum_{n'\bk'} P_{n\bk,n'\bk'}
      \left[ \delta f_{n\bk} - \delta f_{n'\bk'} \right] ,
\end{equation}
where
\begin{equation}
  \label{eq:Pnknk}
  P_{n\bk,n'\bk'} = \frac{2\pi}{\hbar} N_i
      \abs{T_{i,\bk\bk'}^{nn'}(\varepsilon_{n\bk})}^2
      \delta(\varepsilon_{n\bk} - \varepsilon_{n'\bk'}) ,
\end{equation}
is the transition rate in the $T$-matrix approximation, which follows from the
optical theorem in Eq.~\eqref{eq:optical}.

In Appendix~\ref{sec:BE} we outline a least-square method for the solution of the
Boltzmann equation~\eqref{eq:BE} on general $\bk$-point grids and with
first-principles inputs for the band structure, band velocities, and elastic
scattering rate. The method does not rely on any assumptions about the
functional form of the deviation function $\delta f_{n\bk}$ or a relaxation-time
approximation. Other approaches for the solution of the BE based on
first-principles input for inelastic electron-phonon scattering have been
discussed in the literature~\cite{PhysRevB.94.201201,Brandbyge:First,Gibertini:Mobility}. Our method
in Appendix~\ref{sec:BE} was recently applied in calculations of the transport
properties of Li-doped graphene within a TB description of the graphene bands
and the Li-induced carrier scattering~\cite{Folk:Weak}.

In the calculations presented in this work, we restrict the discussion to
transport involving a single band (spin-degenerate or spin-orbit split), and
furthermore assume that the band structure is isotropic with a constant
effective mass $m^*$ (or Fermi velocity $v_F$ in the case of graphene), which is
a good approximation for the transport-relevant energy range close to the band
edges. In this case, the conductivity can be expressed in terms of the
relaxation time given by the $T$-matrix scattering amplitude as
\begin{align}
  \label{eq:tau_tr_iso}
  \tau_{n\bk}^{-1}  & = \frac{2\pi}{\hbar} N_i
      \sum_{\bk'} \abs{T_{i,\bk\bk'}^{nn}(\varepsilon_{n\bk})}^2
  \nonumber \\ & \quad \times
      \left[
        1 - \cos\theta_{\bk\bk'} 
      \right] \delta(\varepsilon_{n\bk'} - \varepsilon_{n\bk}) ,
\end{align}
where $\theta_{\bk\bk'}=\theta_{\bk}- \theta_{\bk'}$ is the scattering angle.
The only difference between the QP scattering rate in Eqs.~\eqref{eq:gamma_nk}
and~\eqref{eq:optical} and the inverse transport relaxation time is the factor
$1-\cos\theta_{\bk\bk'}$ in the square brackets, which accounts for the fact
that the transport is insensitive to \emph{small-angle} scattering while the QP
lifetime is equally sensitive to \emph{all} scattering processes. For isotropic
scattering, the two scattering times become identical as the angular integral of
the $\cos\theta_{\bk\bk'}$ term vanishes.

With the above-mentioned assumptions and considering the low-temperature limit
($k_BT\ll E_F$), the conductivity in Eq.~\eqref{eq:conductivity} simplifies to
the well-known Drude form given by
\begin{equation}
  \label{eq:drude}
  \sigma = \frac{n e^2 \tau(E_F)}{m^*}
  \quad \text{and} \quad 
  \sigma = \frac{e^2 v_F^2  }{2} \rho(E_F) \tau(E_F) ,
\end{equation}
in, respectively, a 2D semiconductor and graphene, where $n$ is the carrier
density and $\rho$ is the density of states. In combination with
first-principles calculations of the $T$-matrix transport relaxation time in
Eq.~\eqref{eq:tau_tr_iso}, these expressions provide a simple and accurate
framework for calculating the low-temperature conductivity in disordered 2D
materials.

\subsection{Numerical and calculational details}
\label{sec:numerics}

The calculation of the $T$ matrix in Eq.~\eqref{eq:Tmatrix} is the most
demanding step in the evaluation of the above-mentioned quantities. Rather than
solving the equation by direct matrix inversion as in~\eqref{eq:T_matrix}, it is
numerically more stable to recast it as a system of coupled linear equation (one
set of coupled equations for each column in $\mathbf{T}$ and $\mathbf{V}$),
\begin{equation}
  \label{eq:Tmatrix_matrix}
  \left[\mathbf{1} - \mathbf{V} \mathbf{G}^0(\varepsilon) \right]
  \mathbf{T}(\varepsilon) =  \mathbf{V},
\end{equation}
and solve it with a standard linear solver. This requires one factorization
followed by a matrix-vector multiplications and scales as $O(M^3)$ where $M$
denotes the matrix dimension.
\begin{figure}[!b]
  \centering
  \includegraphics[width=0.49\linewidth]{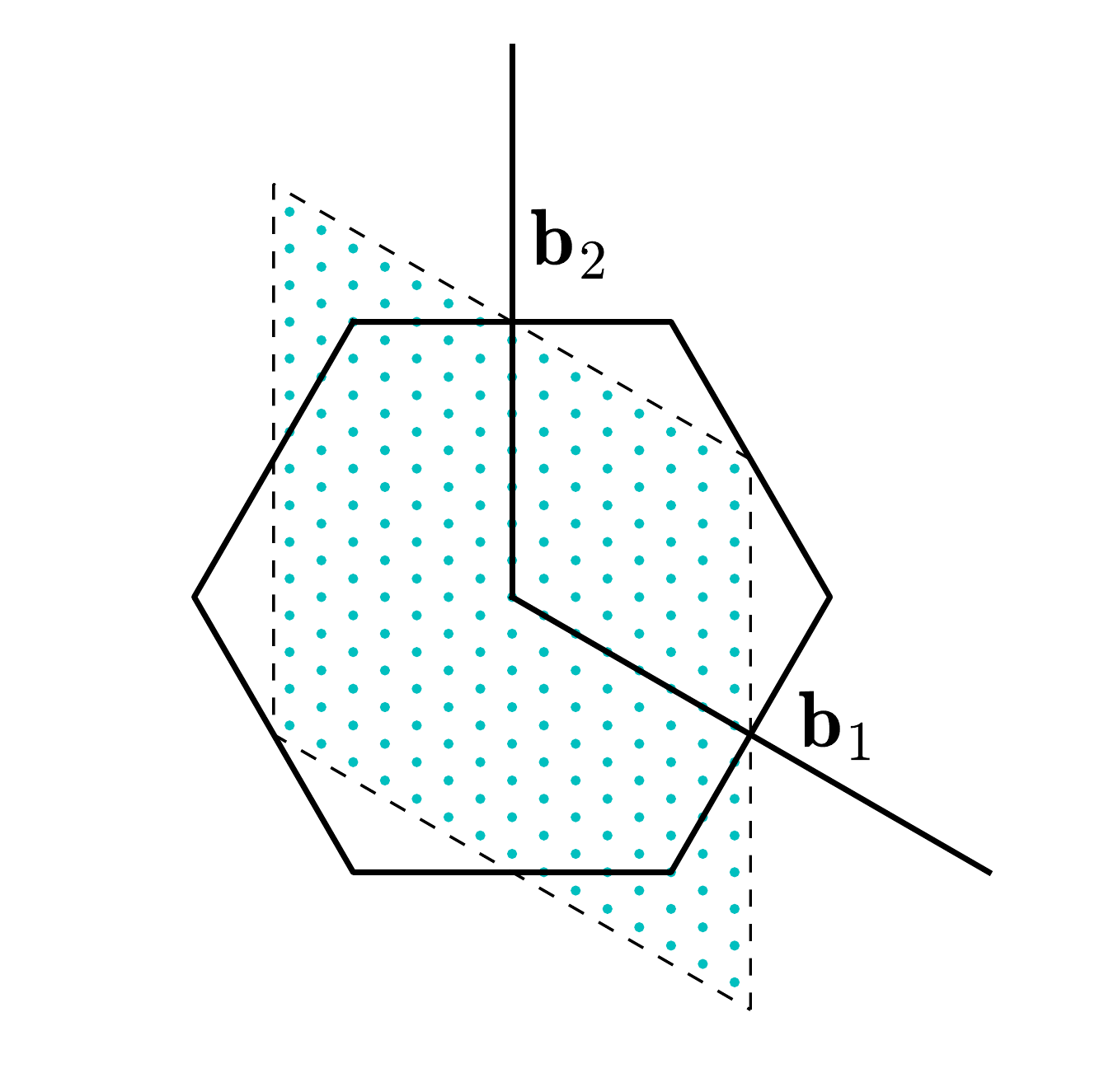}
  \includegraphics[width=0.49\linewidth]{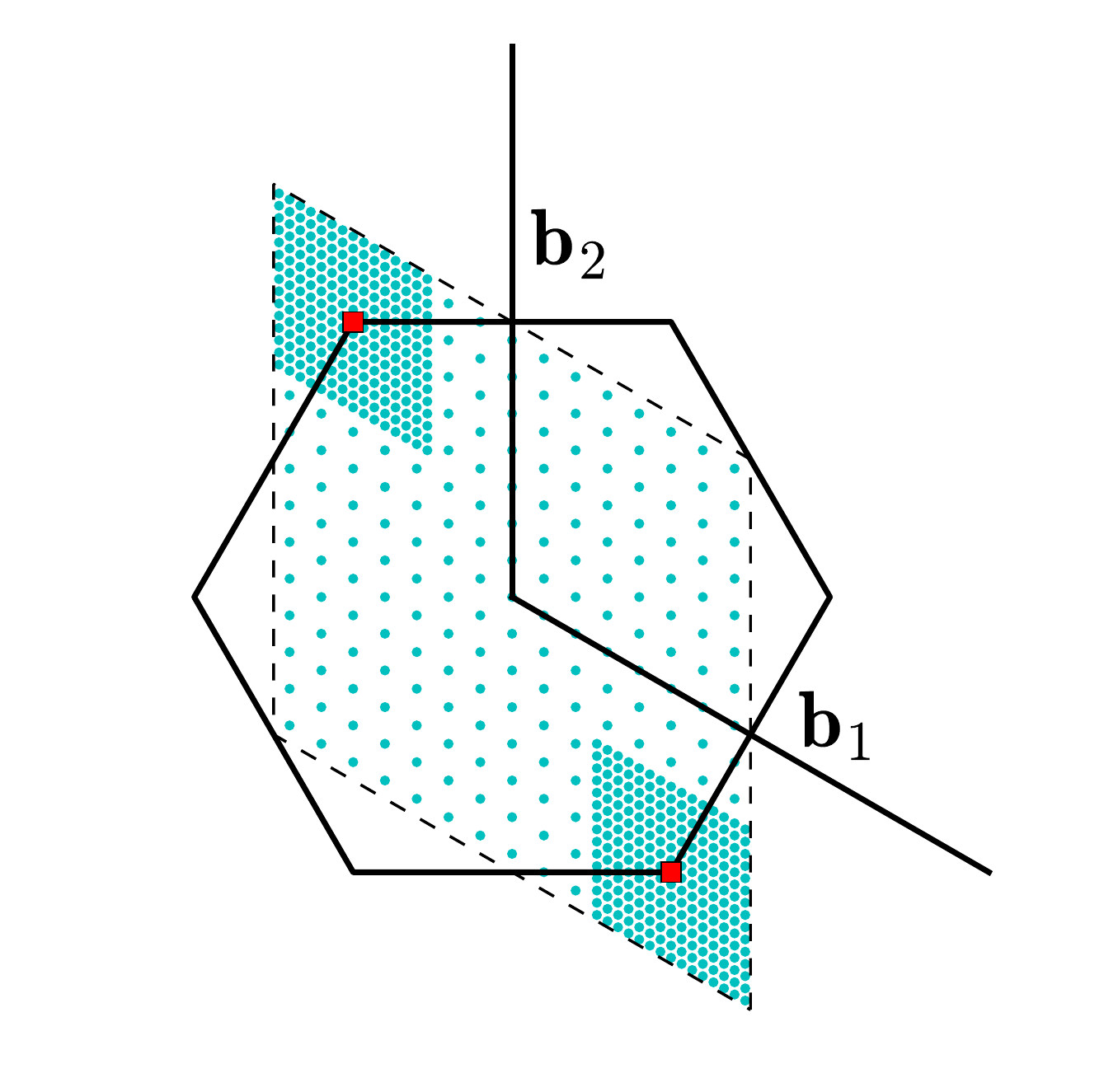}
  \caption{Brillouin zone grids for a hexagonal lattice with, respectively,
    (left) uniform, and (right) nonuniform $\bk$-point sampling. In the
    nonuniform grids a denser sampling is used in a small region around the
    high-symmetry points (red squares) of particular interest. In this case,
    only the dense $\bk$-point sampling is specified in the text, and it is
    marked with an asterisk as ${N_{k_1}\times N_{k_2}}^*$ in order to indicate
    that it is a nonuniform grid.}
\label{fig:bzgrid}
\end{figure}
\begin{figure*}[!t]
  \centering
  \includegraphics[width=0.325\linewidth]{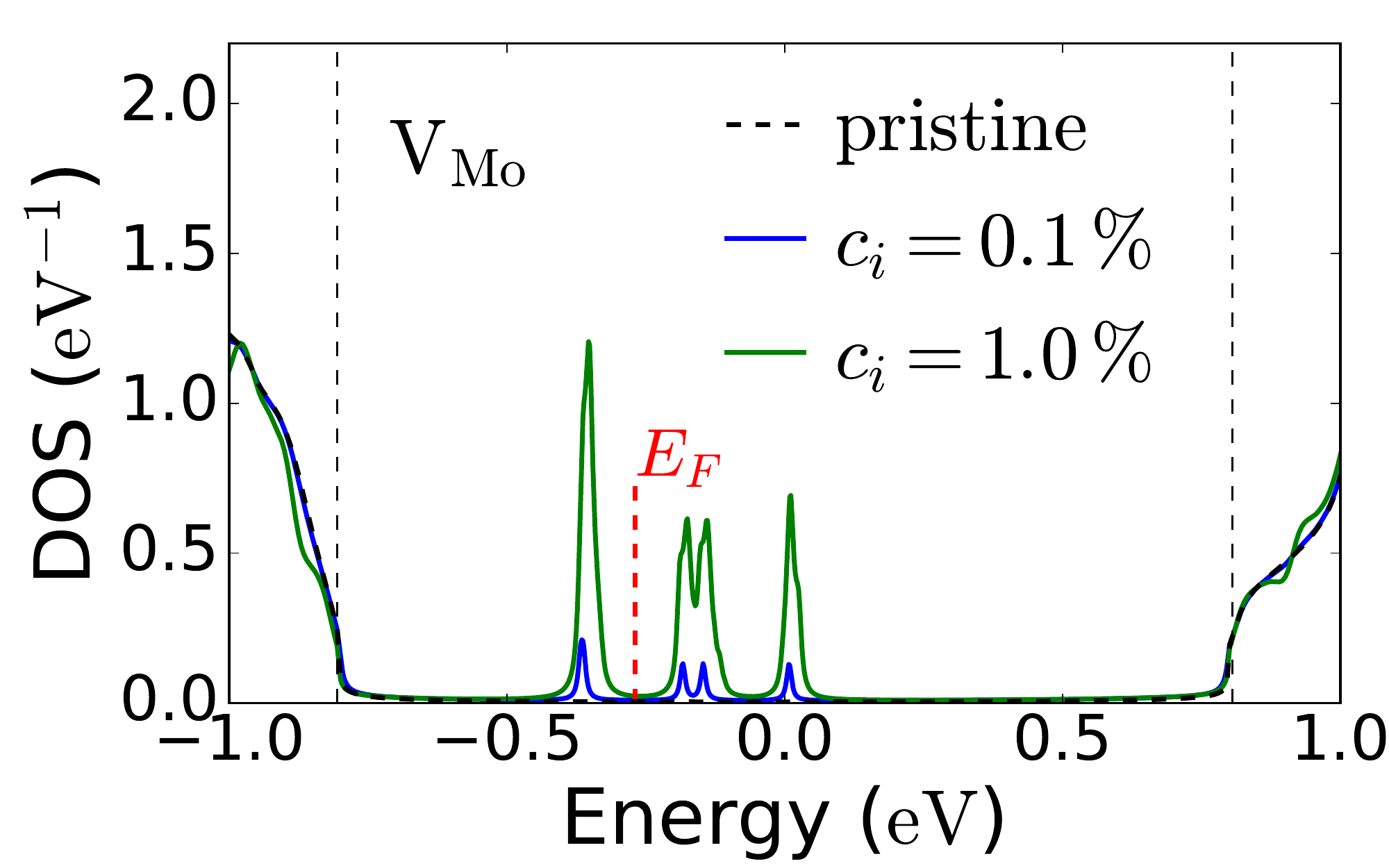}
  \includegraphics[width=0.325\linewidth]{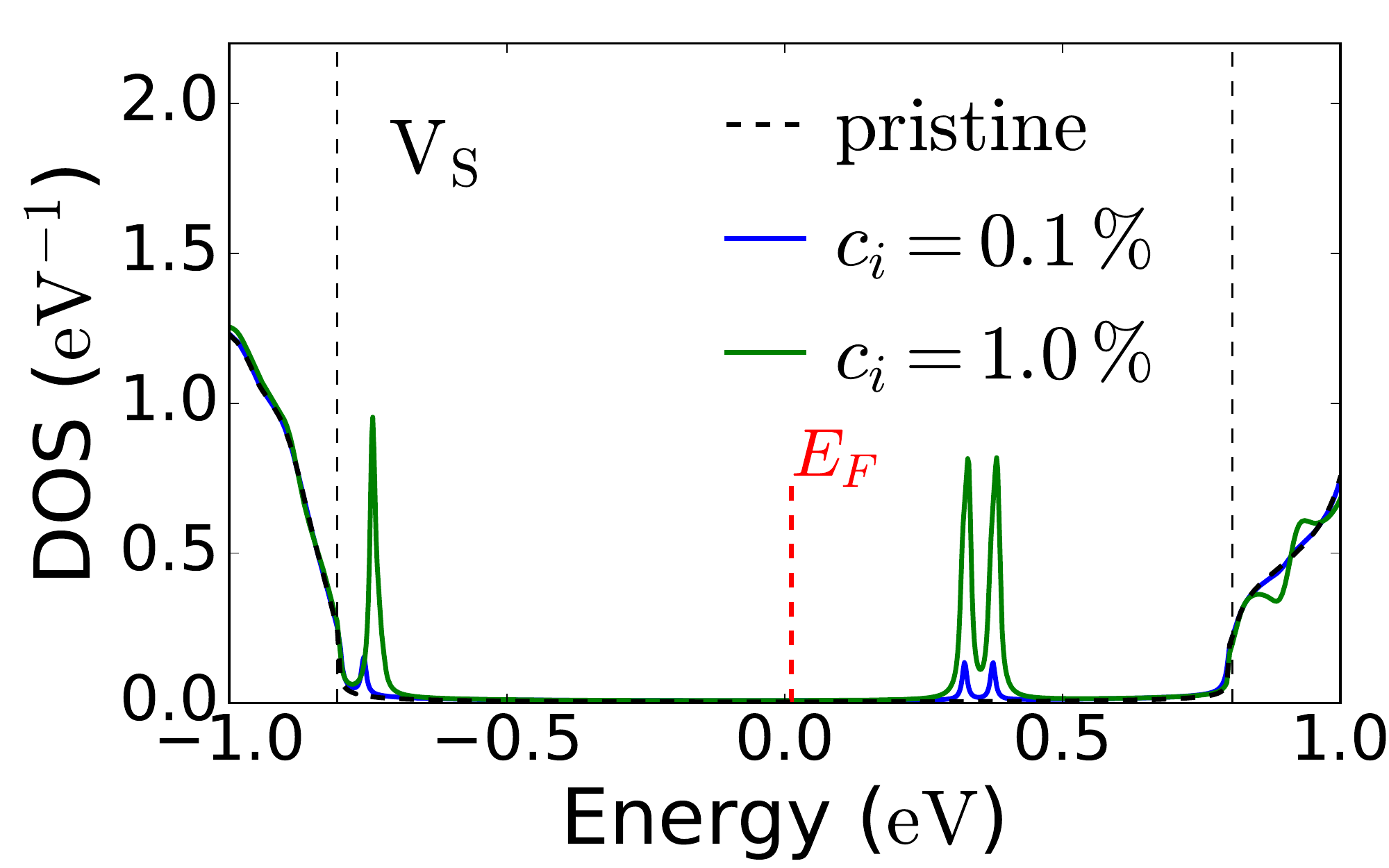}
  \\
  \hfill
  \includegraphics[width=0.325\linewidth]{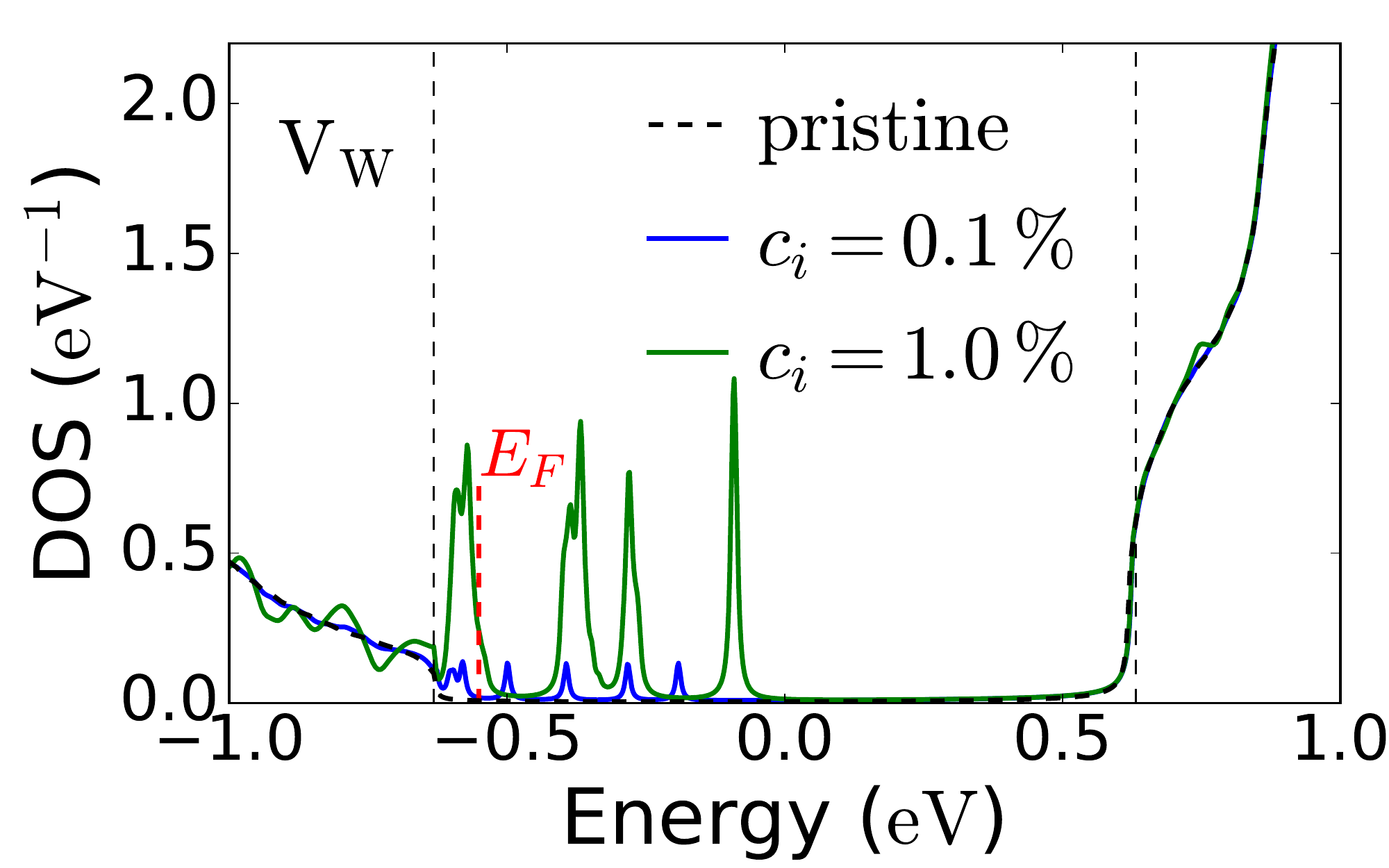}
  \includegraphics[width=0.325\linewidth]{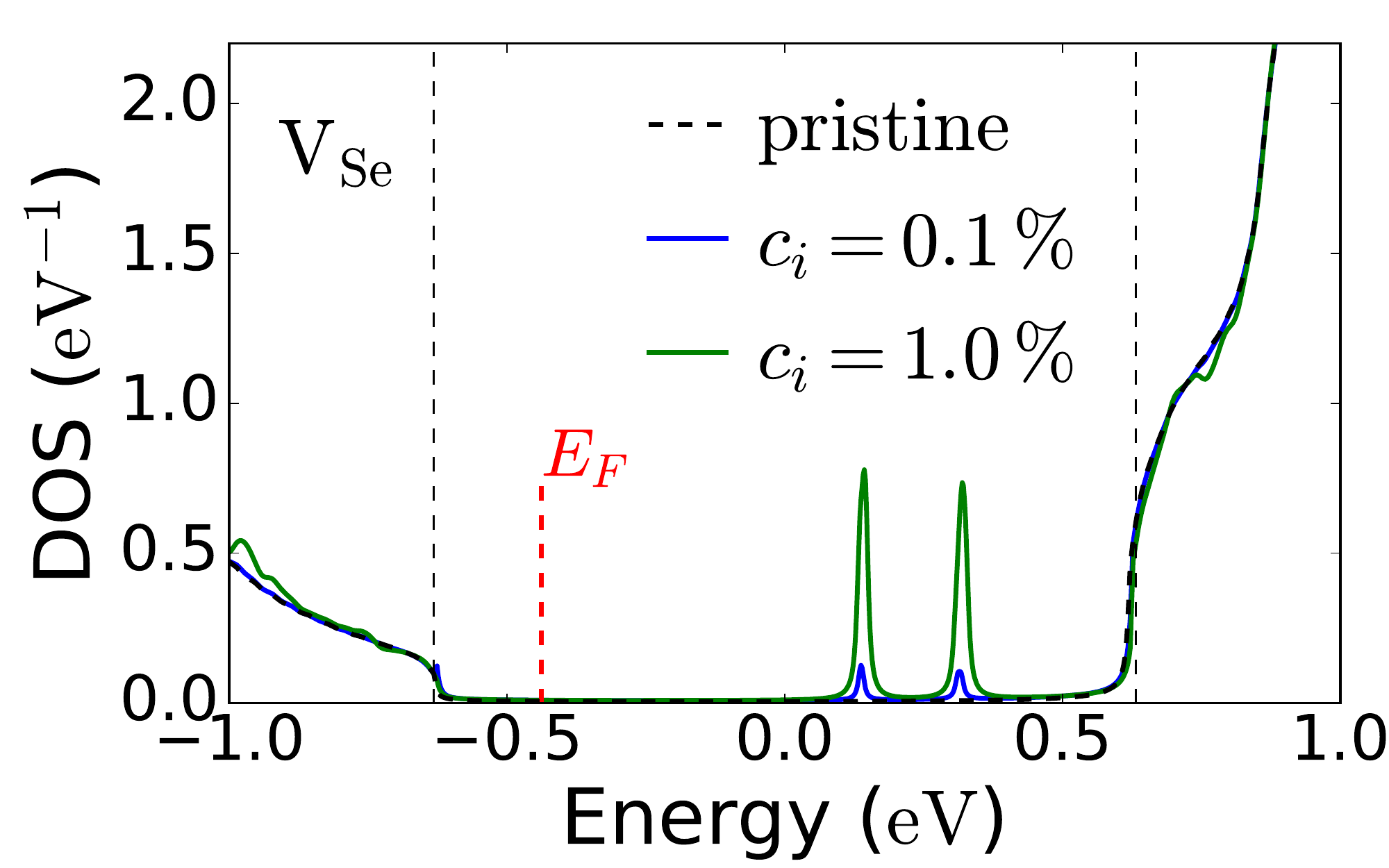}
  \includegraphics[width=0.325\linewidth]{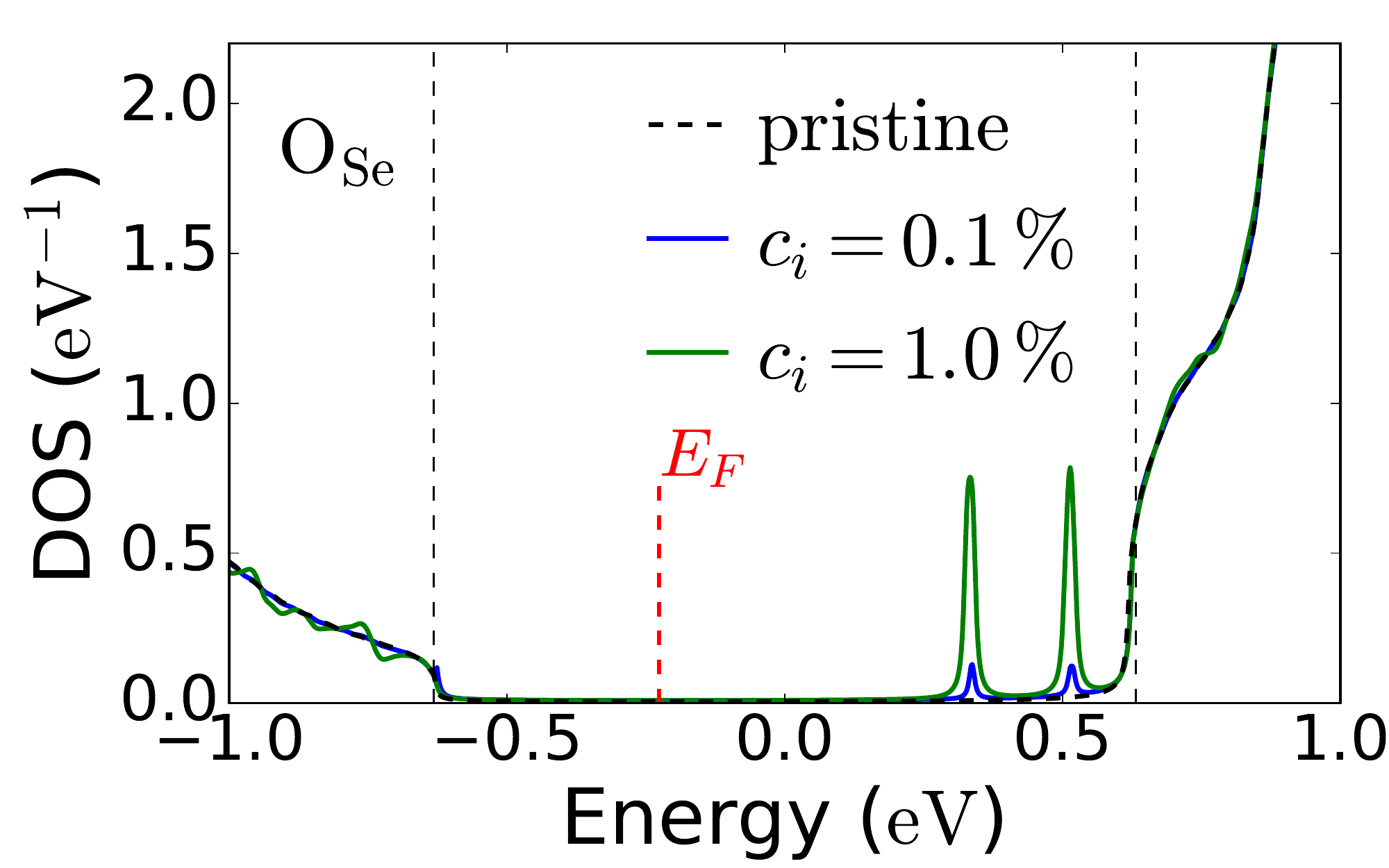}
  \caption{Density of states of disordered 2D TMDs for different types of
    defects and defect concentrations: (top) MoS$_2$, and (bottom) WSe$_2$. The
    energy is measured with respect to the center of the band gap, the vertical
    dashed lines indicate the position of the valence and conduction band edges,
    and the small (red) dashed lines indicate the Fermi energy
    $E_F$. Parameters: $21\times 21$ $\bk$ points ($135 \times 135$ $\bk$ for
    the pristine DOS), 60 bands, and $\eta=5$~meV (20~meV in the bands).}
\label{fig:tmd_dos}
\end{figure*}

The calculation of the $T$ matrix must be checked for convergence with respect
to (i) the BZ sampling with $N_\bk$ $\bk$ points on either
$N_{k_1}\times N_{k_2}$ uniform BZ grids or nonuniform grids with a higher
density of grid points in the vicinity of important high-symmetry points (see
Fig.~\ref{fig:bzgrid}), and (ii) the number of bands $N_b$ included in the
calculation of the $T$ matrix. In general, we have found that the convergence of
the position of bound defect states in the gap of 2D semiconductors requires a
large number of bands ($N_b>50$) starting from the bottom of the spectrum and up
to high energies, while only a moderate $\bk$ sampling ($\sim 21\times 21$) is
required. On the other hand, the linewidth broadening in Eq.~\eqref{eq:gamma_nk}
[and Eq.~\eqref{eq:optical}] requires only a few bands ($N_b \sim 2$--$4$) (as
long as there are no bound states in the bands), but a dense $\bk$-point grid
($135 \times 135$) in order to sample the constant-energy surface on which the
quasiparticle scattering takes place. In the case of graphene, quasibound
resonant states are inherent to the Dirac cone dispersion, and hence both the
the resonant state and the scattering rate can be calculated with a dense
$\bk$-point sampling including only a few bands.

With the above-mentioned number of bands and $\bk$-point samplings, the
dimension $M$ of the matrices in Eqs.~\eqref{eq:Tmatrix_matrix} becomes
$M=N_b \times N_\bk \sim 20000$--$50000$. With the matrix elements represented
as 128-bit complex floating-point numbers, the memory requirement for each of
the dense complex matrices in Eq.~\eqref{eq:Tmatrix_matrix} becomes
$M^2\times 128 / 8\,\mathrm{bytes}\approx 10$--$30$~GBs. To tackle the large
matrix dimensions in the solution of the matrix equation in
Eq.~\eqref{eq:Tmatrix_matrix}, we exploit the automatic openMP multithreading of
the LAPACK linear solvers.

In the calculation of the DOS in Eq.~\eqref{eq:dos}, a very fine $\bk$-point
sampling is needed to converge the DOS of the bands near the band edges. We
therefore (i) first calculate the difference
$\delta \rho = \rho_\text{dis} - \rho_0$ between the DOS of the disordered and
pristine materials on a coarse $\bk$-point grid (in order to include enough
bands to capture bound states), and (ii) subsequently add $\delta \rho$ to the
DOS of the pristine material obtained on a fine $\bk$-point grid. In this way,
we avoid spiky artifacts in the DOS of the bands due to insufficient $\bk$-point
sampling, while at the same time capturing potential defect states in the bands.

For the results presented in the following sections, the band structures and
defect matrix elements have been obtained with the GPAW electronic-structure
code~\cite{GPAW,GPAW1,GPAW2}, using DFT-LDA within the projector augmented-wave
(PAW) method, a DZP LCAO basis, and including spin-orbit
interaction~\cite{olsen:designing}. The parameters used in the individual
calculations are listed in the figure captions.

\section{Disordered 2D TMDs}
\label{sec:TMD}

The experimental consensus on the prevalent types of defects in the 2D TMDs (see
Sec.~\ref{sec:examples_tmds}) has led to numerous theoretical DFT studies of
their structural and electronic
properties~\cite{Kim:Stability,Nieminen:Charged,Neto:Donor,Krasheninnikov:Native,Sanyal:Systematic,Tay:Theoretical,Thygesen:Defect,Jauho:Symmetry,Leuenberger:Electronic,Yazyev:Point}. On
the other hand, first-principles calculations of the impact on electron dynamics
in disordered 2D TMDs remain few~\cite{Neaton:Defect,Kaasbjerg:Transport}. In
the following subsections, we analyze in detail the electronic (DOS and
quasiparticle spectrum) and transport (conductivity and mobiliy) properties.

\begin{figure*}[!t]
  \centering
  \includegraphics[width=0.4\linewidth]{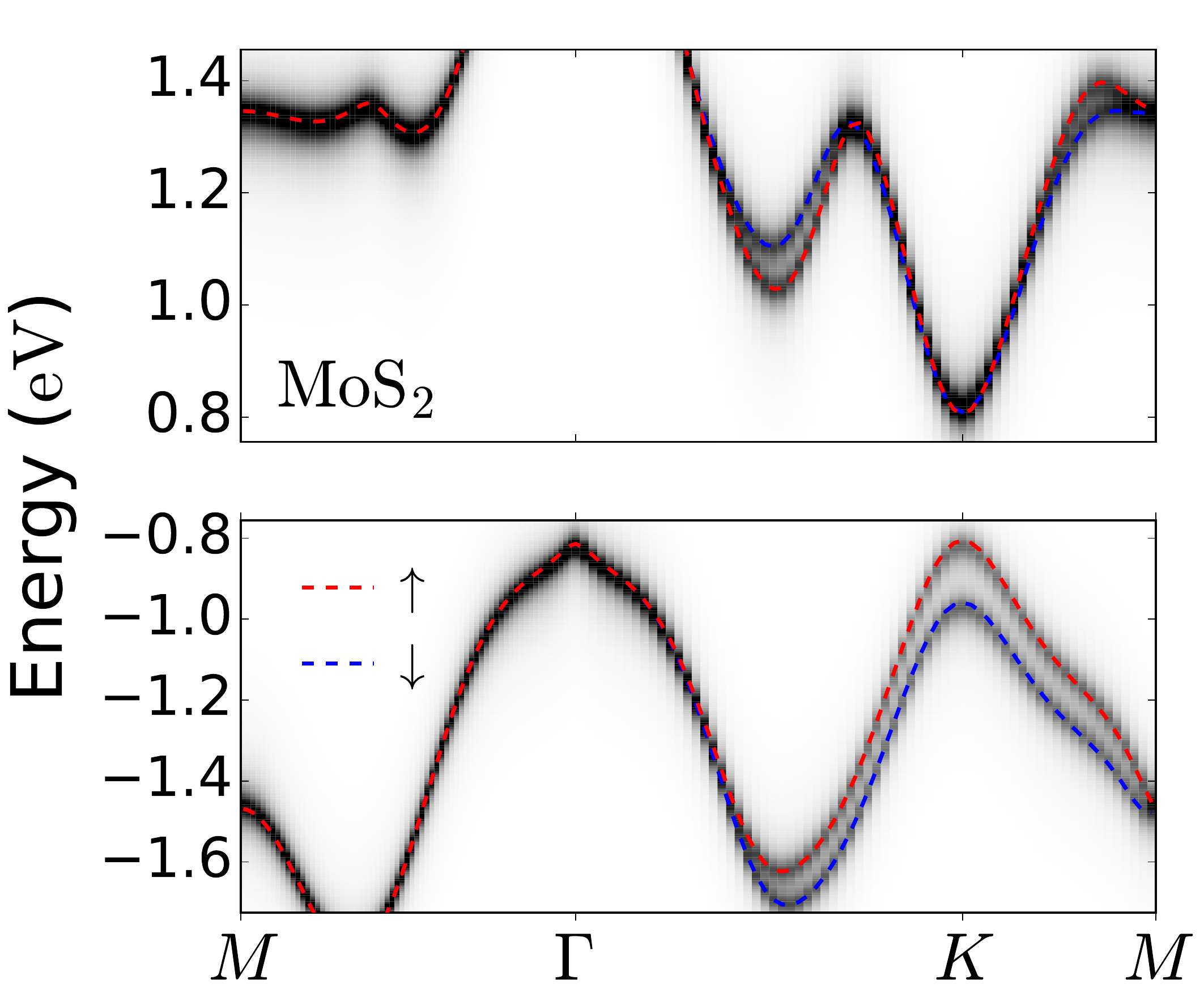}
  \hspace{1cm}
  \includegraphics[width=0.4\linewidth]{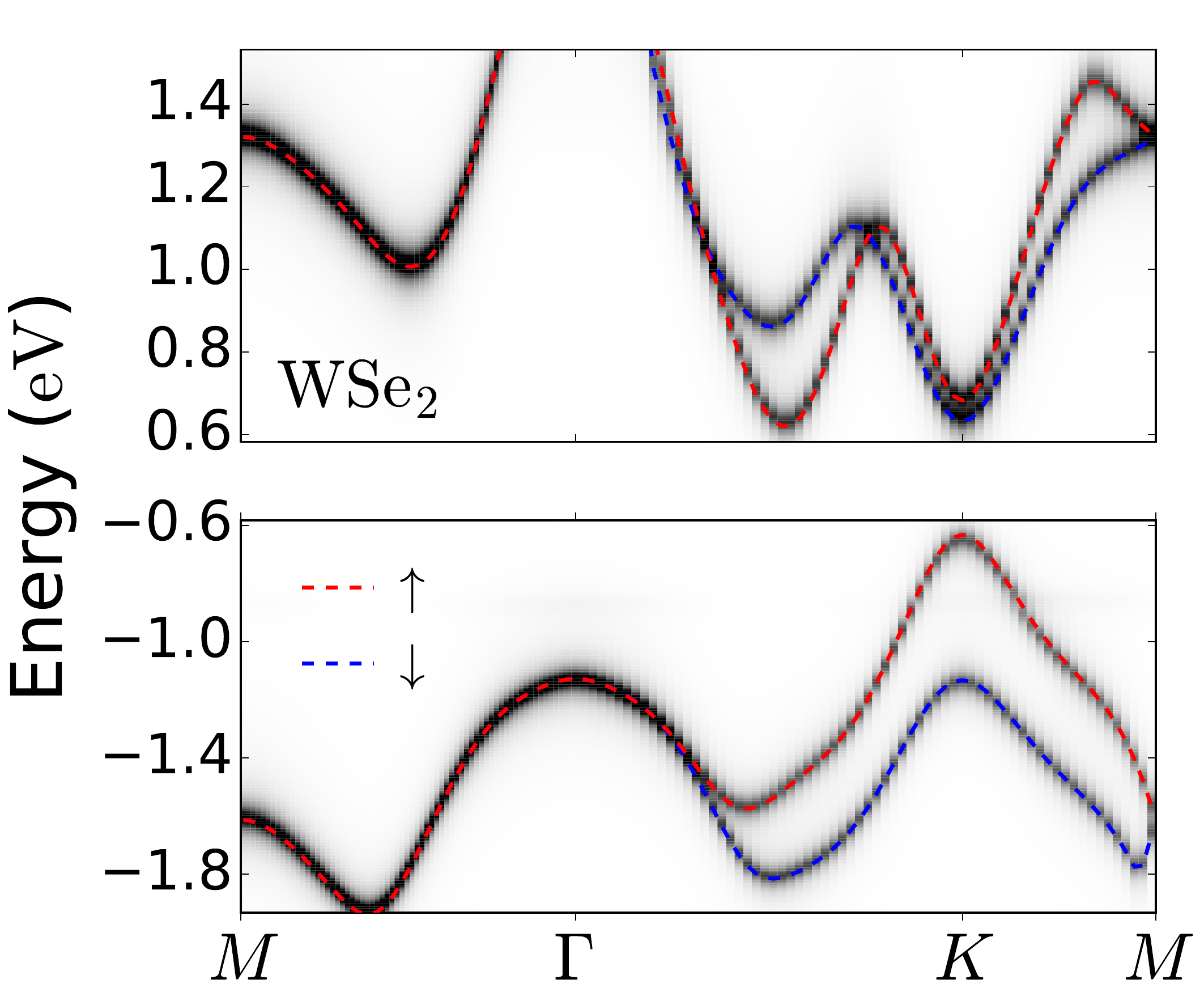}
  \caption{Spectral function of disordered 2D TMDs showing the two first
    spin-orbit split valence and conduction bands for: (left) MoS$_2$ with Mo
    vacancies, and (right) WSe$_2$ with Se vacancies. The dashed lines show the
    bands of the pristine materials. Parameters: $c_i=1\%$, $135\times 135$
    $\bk$ points, 4 bands, and $\eta=20$~meV.}
\label{fig:tmd_spectral}
\end{figure*}

\subsection{DOS and in-gap bound states}

We start by discussing the impact of defects on the DOS, and in particular the
defect-induced in-gap states observed in various STM/STS
experiments~\cite{Andrei:Bandgap,Pasupathy:Atomic,Pasupathy:Approaching,Bargioni:Large,Bargioni:Identifying,Bargioni:How}.
Figure~\ref{fig:tmd_dos} shows the DOS for disordered MoS$_2$ (top) and WSe$_2$
(bottom) with different types of defects. The dashed vertical lines mark the
position of the valence and conduction band edges (black) as well as the Fermi
energy ($E_F$; red dashed line). All the defects, in particular transition-metal
vacancies, introduce a series of defect-localized in-gap states at positions in
good agreement with previously reported DFT supercell
calculations~\cite{Kim:Stability,Huis:Strong,Leuenberger:Electronic,Neaton:Defect}.

Our results in Fig.~\ref{fig:tmd_dos} show that some of the orbitally-degenerate
bound states are subject to a notable spin-orbit induced spin splitting. This is
also in agreement with previous supercell calculations of defect-induced in-gap
states taking into account spin-orbit
interaction~\cite{Huis:Strong,Neaton:Defect,Bargioni:Large}. In MoS$_2$, the
splitting is of the order of $\sim 50$~meV for the two
$\mathrm{V}_{\mathrm{Mo}}$ states above the Fermi energy and the two
$\mathrm{V}_{\mathrm{S}}$ top states, whereas a significantly larger splitting
of $\sim 270$~meV is observed for the unoccupied $\mathrm{V}_{\mathrm{Se}}$ and
$\mathrm{O}_{\mathrm{Se}}$ states in WSe$_2$. It should be noted that this
effect is captured in spite of the fact that we here consider spin-independent
defect potentials, i.e. the spin-orbit interaction enters only through the
unperturbed band structure.

It is interesting to note that most of the considered defects introduce both
occupied and unoccupied in-gap states occur, except for the
$\mathrm{V}_{\mathrm{Se}}$ and $\mathrm{O}_{\mathrm{Se}}$ defects in WSe$_2$
which only introduce unoccupied in-gap states. For all the other defects, the
shallow occupied states above the valence-band edge act as hole traps in the
$p$-doped (gated) materials. The converse holds for the unoccupied in-gap states
which act as deep electron traps in the $n$-doped materials. As we have recently
demonstrated, the charging of the defect sites resulting from such carrier
trapping has detrimental impact on the transport properties of gated 2D
TMDs~\cite{Kaasbjerg:Transport}.

In addition to in-gap bound state, defects may also introduce quasibound states
inside the bands as predicted in, e.g., MoSe$_2$ and
WS$_2$~\cite{Neaton:Defect,Bargioni:Large}. As witnessed by the $c_i=1\,\%$
curves in Fig.~\ref{fig:tmd_dos} which show pronounced deviations from the
pristine DOS inside the bands, this seems to be the case also in MoS$_2$ and
WSe$_2$. However, we find that some of these features are artifacts from the
procedure we have used to calculate the DOS in the bands of the disordered system
(see Sec.~\ref{sec:numerics} above). Only the features in the valence band for
Mo vacancies in MoS$_2$ ($\sim 350$~meV below the band edge), W ($\sim 150$~meV
below the band edge) and Se vacancies ($\sim 350$~meV below the band edge) in
WSe$_2$ correspond to true quasibound states. As their positions are reasonably
far away from the band edges, resonant scattering off the quasibound states can
be neglected in calculations of the disorder-limited transport
properties~\cite{Kaasbjerg:Transport}. By contrast, both bound and quasibound
defect states have been demonstrated to alter the optical properties of 2D TMDs
by binding the excitons in the defect
states~\cite{Wu:Defect,Silverman:Micro,Neaton:Defect,Quek:Origin}.

\subsection{Spectral function and quasiparticle scattering}

In Fig.~\ref{fig:tmd_spectral}, we show the valence and conduction band spectral
functions (grayscale intensity plots) for disordered MoS$_2$ and WSe$_2$ with a
$c_i=1\,\%$ concentration of, respectively, Mo and Se vacancies together with
the unperturbed band structure of the pristine materials (dashed
lines). Although our DFT calculations indicate that the direct and indirect band
gaps in MoS$_2$ and WSe$_2$ are almost identical, and that the band gap in some
cases is indirect~\cite{Shih:Probing}, recent microARPES experiments have given
conclusive evidence that the band gap in monolayers of the semiconducting TMDs
$MX_2$ with $M=\mathrm{Mo},\mathrm{W}$ and $X=\mathrm{S},\mathrm{Se}$ is
direct~\cite{Wilson:Visualizing}.

Overall, the spectral functions overlap almost perfectly with the unperturbed
band structures, indicating that disorder-induced renormalization of the bands
is small at $c_i=1\,\%$ in the $T$-matrix approximation. This is in stark
contrast to the Born approximation (not shown), where the first term in
Eq.~\eqref{eq:born} given by the bare defect matrix element gives rise to a
giant shift of the bands. In the $T$-matrix approximation, this shift is
strongly renormalized by the matrix inverse on the right-hand side of
Eq.~\eqref{eq:T_matrix}.
\begin{figure}[!b]
  \centering
  \includegraphics[width=0.49\linewidth]{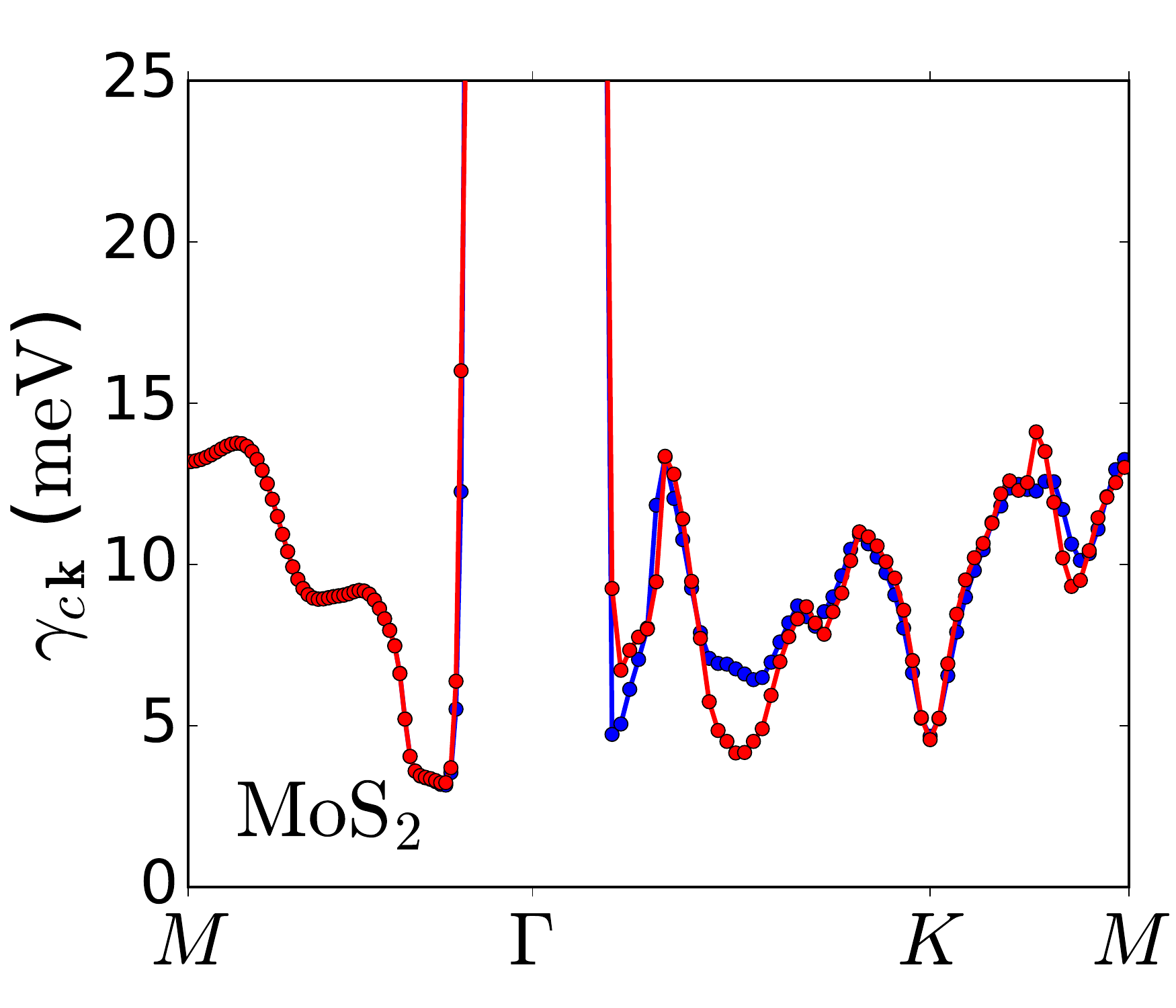}
  \includegraphics[width=0.49\linewidth]{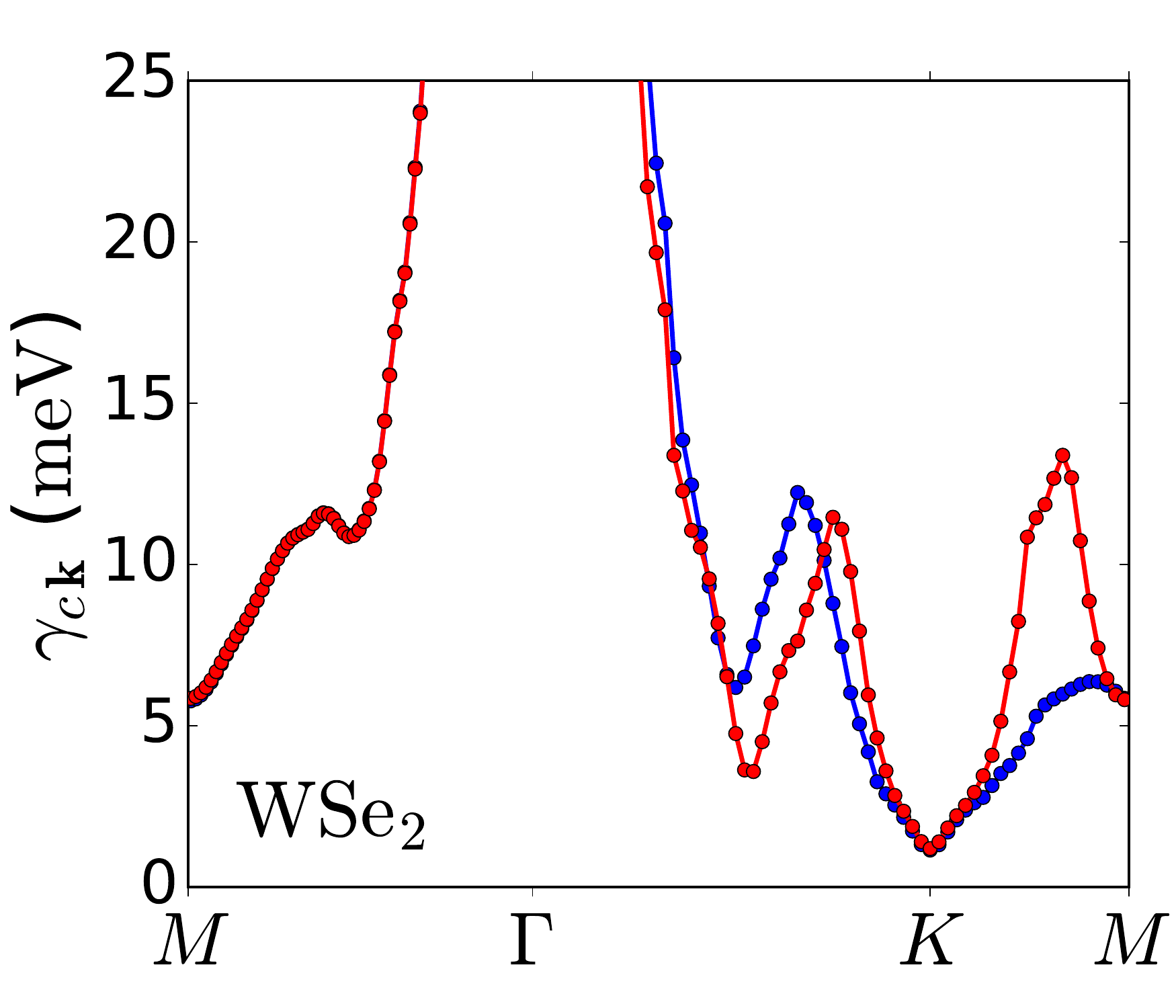}
  \includegraphics[width=0.49\linewidth]{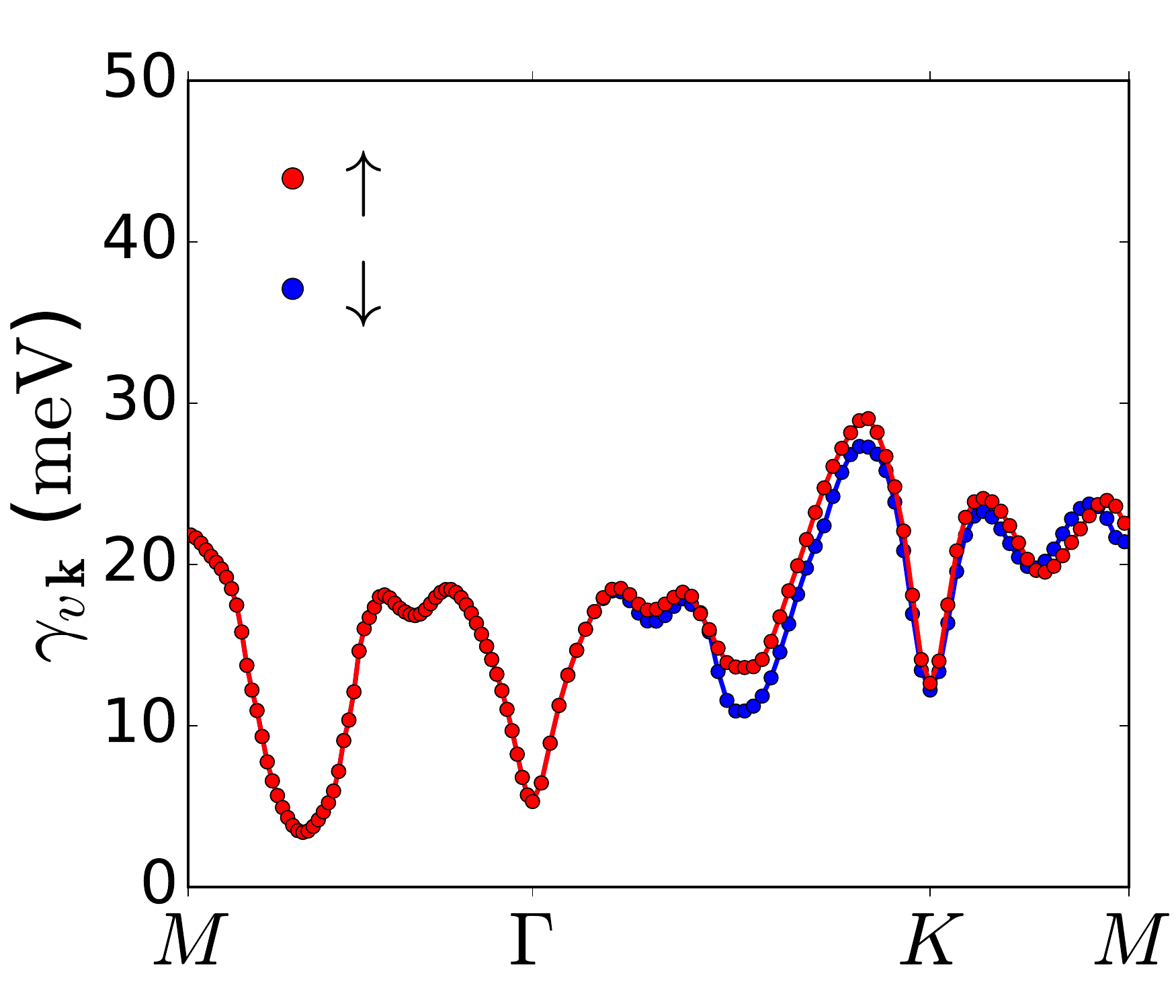}
  \includegraphics[width=0.49\linewidth]{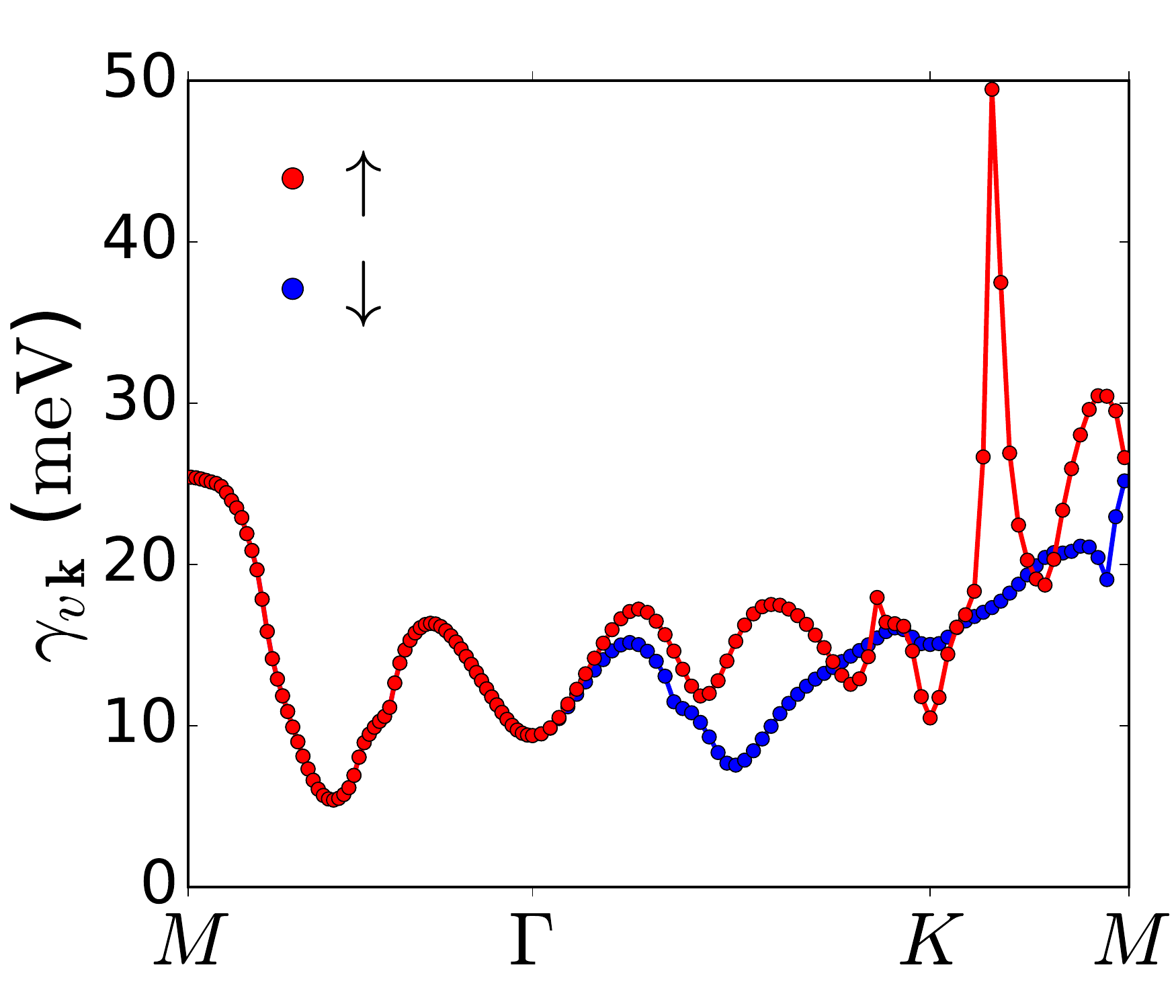}
  \caption{Disorder-induced linewidth broadening for the spectral functions
    in Fig.~\ref{fig:tmd_spectral}: (left) MoS$_2$, and (right) WSe$_2$;
    (top) conduction, and (bottom) valence band. The broadening has been
    obtained from the imaginary part of the on-shell self-energy via
    Eq.~\eqref{eq:gamma_nk}. Parameters: see caption of
    Fig.~\ref{fig:tmd_spectral}.}
\label{fig:tmd_gamma}
\end{figure}

The quasiparticle lifetime given by the broadening of the spectral function is
difficult to infer from Fig.~\ref{fig:tmd_spectral} due to the numerical
broadening $\eta$. In Fig.~\ref{fig:tmd_gamma} we therefore show the linewidth
broadenings obtained directly from the on-shell self-energy via
Eq.~\eqref{eq:gamma_nk}. Overall, the linewidths show a pronounced dependence on
$\bk$ with multiple peaks and dips along the considered path in the BZ. The
strong increase in the linewidth in the vicinity of the $\Gamma$ point in the
conduction bands (top plots) is due to overlap with higher-lying bands outside
the energy range shown in Fig.~\ref{fig:tmd_spectral}. The sharp peak in the
linewidth along the $K$-$M$ path in the lower right plot is due to resonant
scattering off the quasibound defect state introduced by the Se vacancy in the
valence band. For the defect concentration $c_i=1\,\%$ considered here, the
overall magnitude of the disorder-induced linewidth is comparable to the
phonon-induced linewidth at elevated temperatures~\cite{Wirtz:Temperature}.

\begin{figure}[!t]
  \centering
  \includegraphics[width=0.49\linewidth]{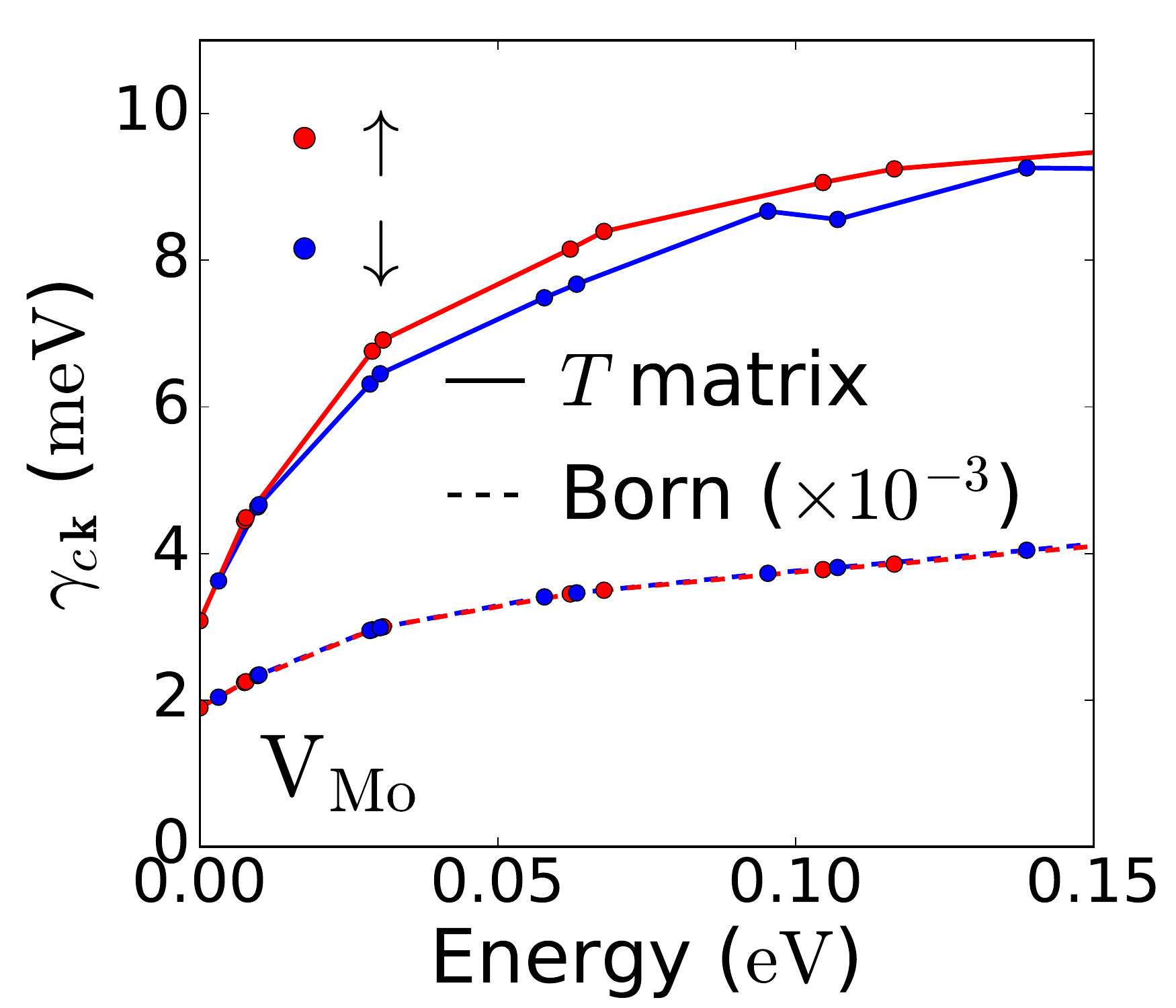}
  \includegraphics[width=0.49\linewidth]{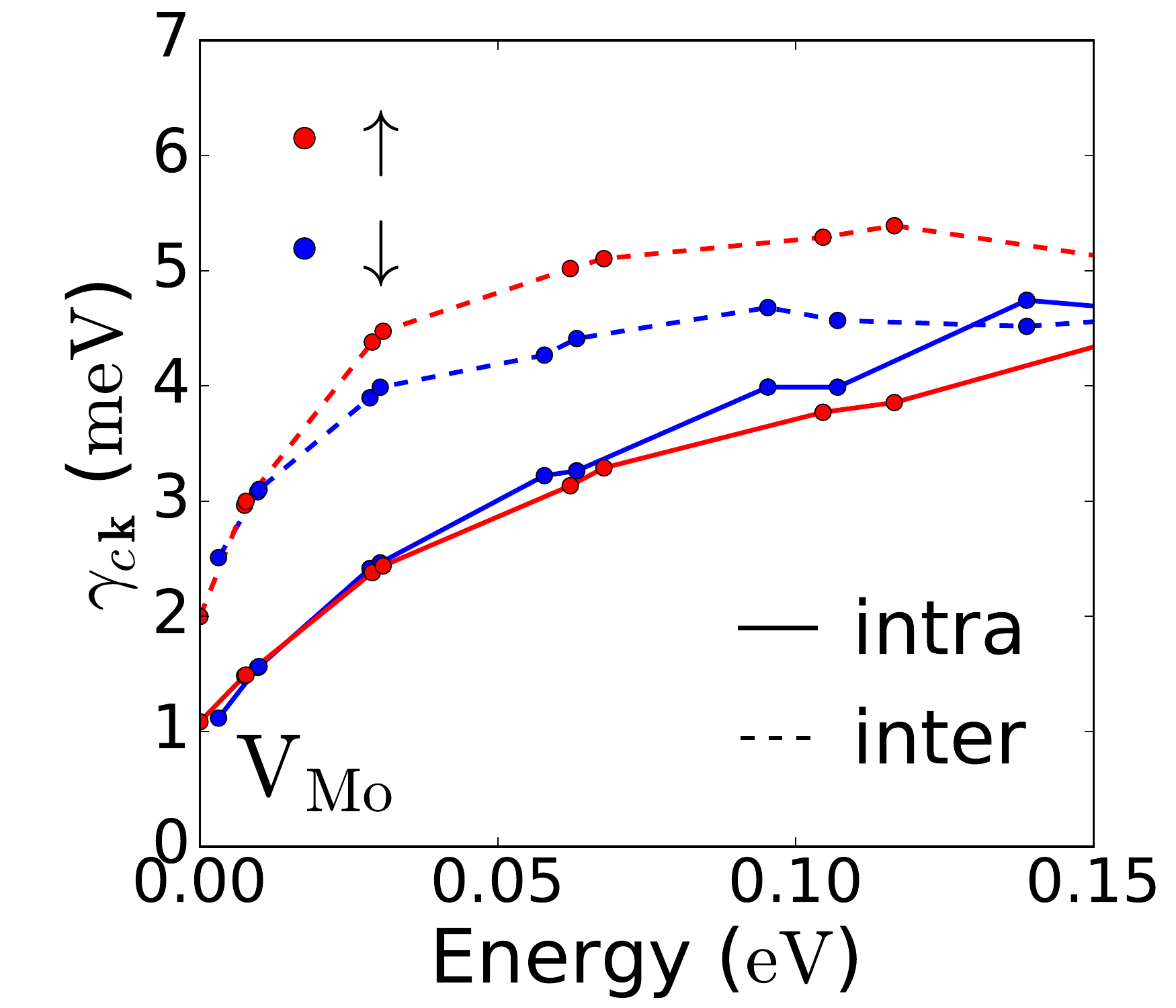}
  \includegraphics[width=0.49\linewidth]{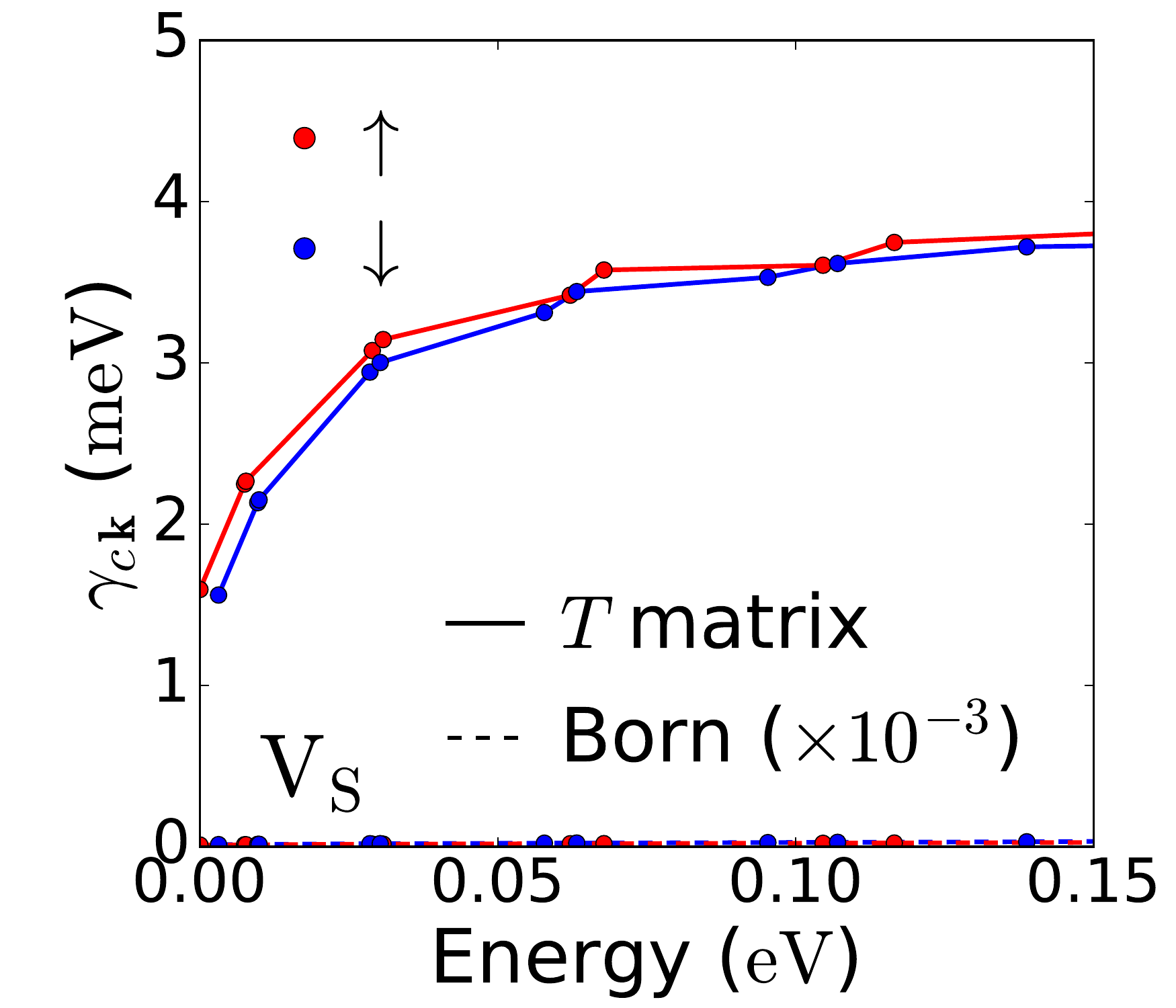}
  \includegraphics[width=0.49\linewidth]{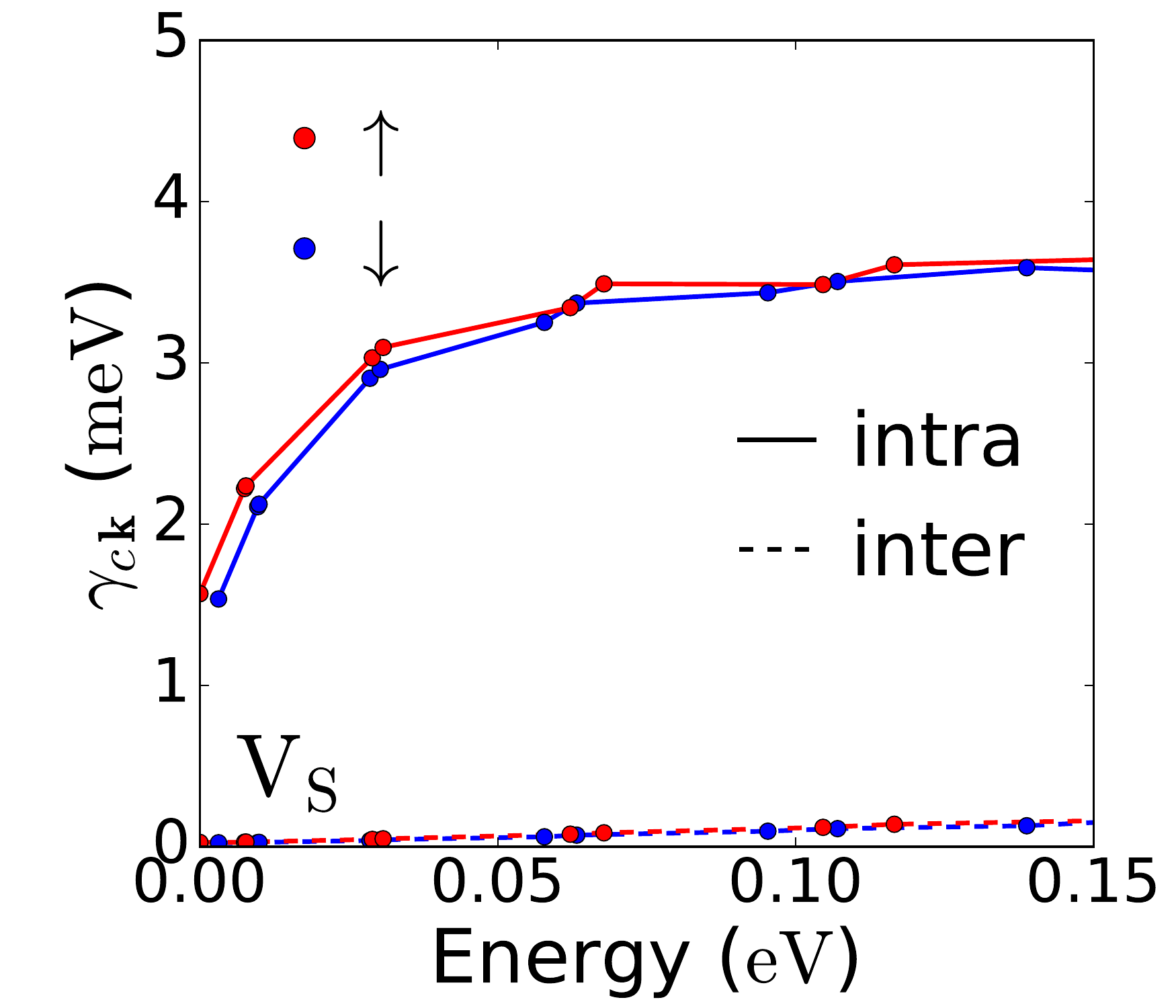}
  \caption{Energy dependence of the disorder-induced linewidth broadening in the
    $K$ valleys of the conduction band in MoS$_2$ due to (top) Mo, and (bottom)
    S vacancies. (left) Comparison between the Born and $T$-matrix
    approximations. (right) Intravalley vs intervalley scattering contributions to
    the $T$-matrix linewidth. The on-shell energy has been sampled along the
    $\Gamma$-$K$-$M$ path in the BZ, and is measured with respect to the
    conduction-band edge. Parameters: see caption of
    Fig.~\ref{fig:tmd_spectral}.}
\label{fig:mos2_gamma_vs_e}
\end{figure}

\begin{figure}[!t]
  \centering
  \includegraphics[width=0.49\linewidth]{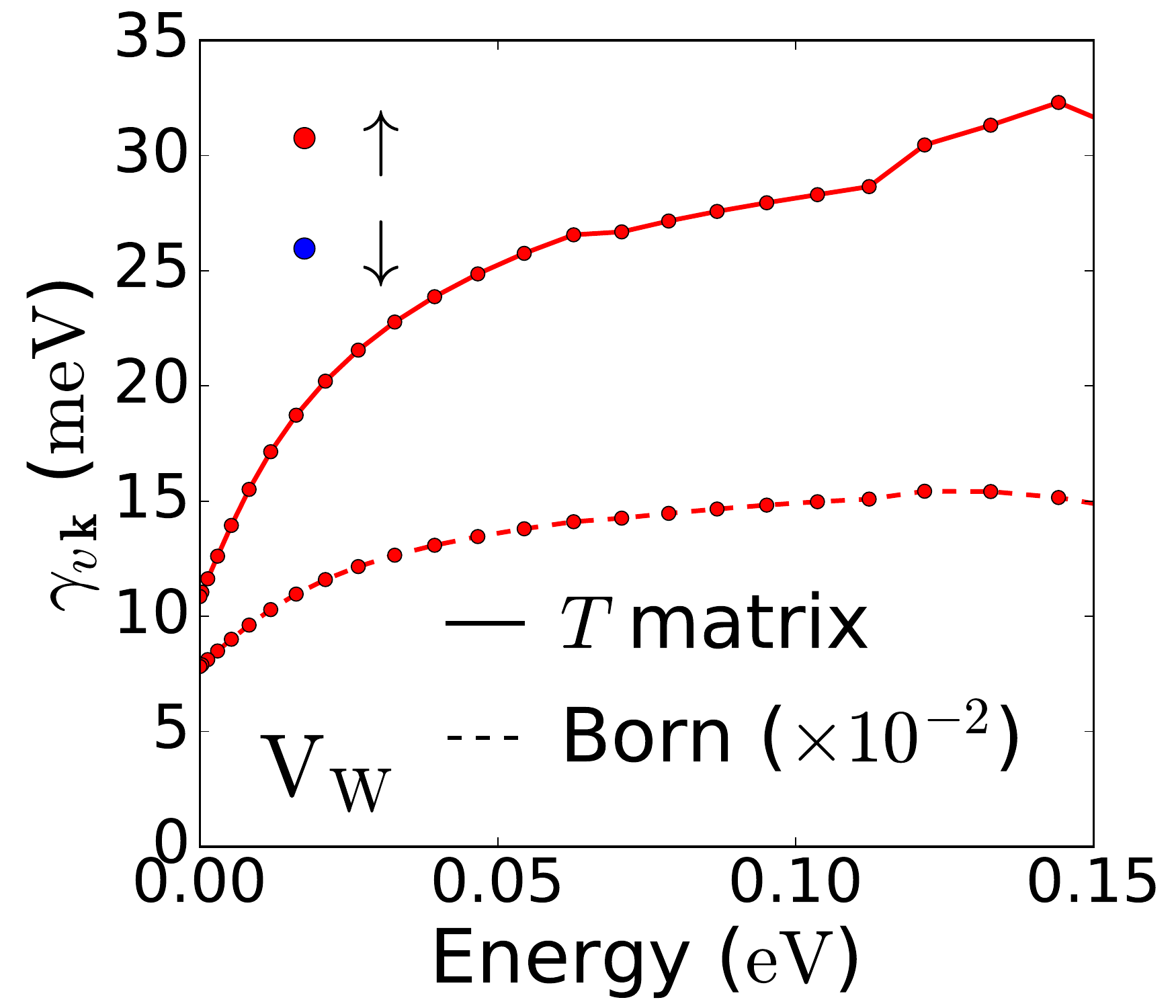}
  \includegraphics[width=0.49\linewidth]{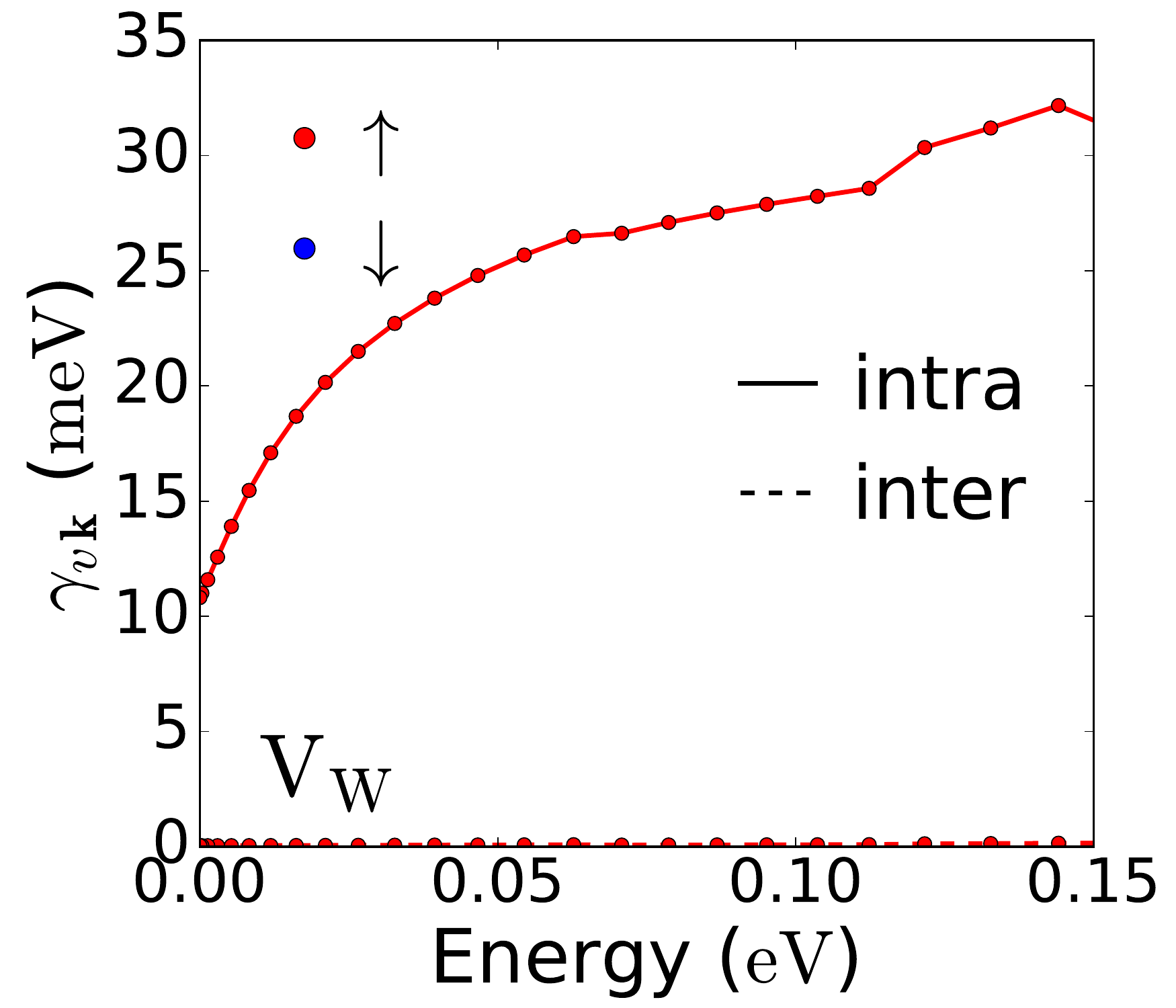}
  \includegraphics[width=0.49\linewidth]{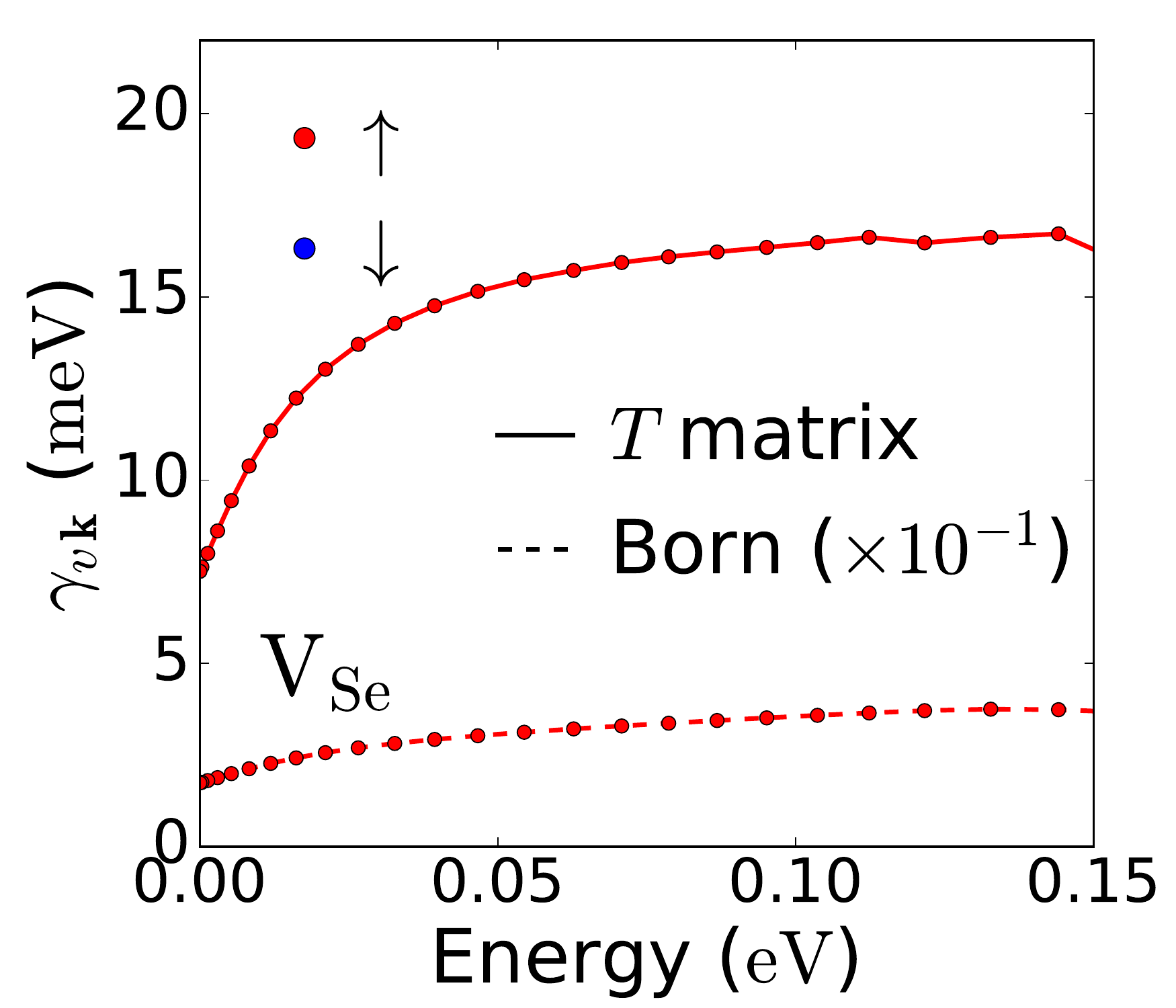}
  \includegraphics[width=0.49\linewidth]{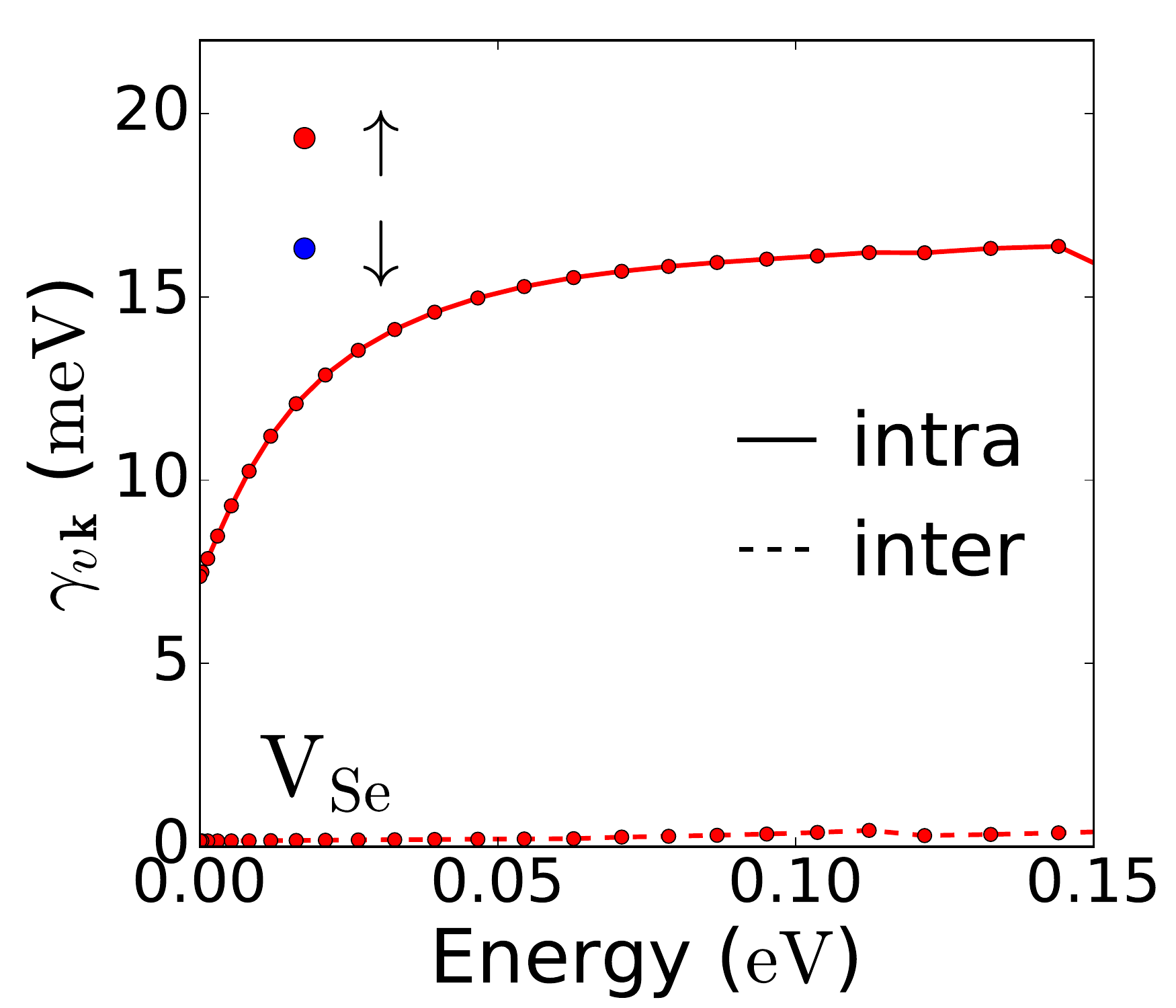}
  \caption{Energy dependence of the disorder-induced linewidth broadening in the
    $K$ valley of the valence band in WSe$_2$ due to (top) W, and (bottom) Se
    vacancies. (left) Comparison between the Born and $T$-matrix
    approximations. (right) Intravalley vs intervalley scattering contributions to
    the $T$-matrix linewidth. The on-shell energy has been sampled along the
    $\Gamma$-$K$-$M$ path in the BZ, and is measured positive with respect to
    the valence-band edge. Parameters: $99\times 99^*$ plus parameters in
    caption of Fig.~\ref{fig:tmd_spectral}.}
\label{fig:wse2_gamma_vs_e}
\end{figure}

In the $\Gamma$ and $K$ valleys close to the band edges, the linewidths show
characteristic dips with a particularly sharp shape. To analyze these features
in closer detail, we show in Figs.~\ref{fig:mos2_gamma_vs_e}
and~\ref{fig:wse2_gamma_vs_e} the linewidths for the spin up and down bands in
$K$ valley of, respectively, the conduction band of MoS$_2$ (Mo and S vacancies)
and the valence band of WSe$_2$ (W and Se vacancies) as a function of the band
energy $\varepsilon=\varepsilon_{n\bk}$ (measured with respect to the band
edges) instead of $\bk$. In the two figures, the left columns show a comparison
between the Born [Eq.~\eqref{eq:tau_born}] and $T$-matrix approximations,
whereas the right columns show the contributions to linewidth from intravalley and
intervalley scattering. Due to the large spin-orbit splitting in the valence
band of WSe$_2$, only the linewidth for the spin-up band appears in
Fig.~\ref{fig:wse2_gamma_vs_e}. Note that different $\bk$-point samplings have
been used in the two figures, hence the difference in energy resolution.

In Figs.~\ref{fig:mos2_gamma_vs_e} and~\ref{fig:wse2_gamma_vs_e}, the sharp dips
in the linewidths in Fig.~\ref{fig:tmd_gamma} mentioned above are manifested in
a strong energy dependence of the $T$-matrix linewidths close to the band
edges. For comparison, the linewidths in the Born approximation exhibit a weaker
energy dependence which can be traced back to the almost constant matrix
elements in Fig.~\ref{fig:M_BZ} inside the $K,K'$ valleys, implying that the
Born linewidth given by Eq.~\eqref{eq:tau_born} to a good approximation becomes
proportional to the DOS in the $K,K'$ valleys. The energy dependence of the Born
linewidths therefore reflects the gradual increase in the DOS in
Fig.~\ref{fig:tmd_dos} at the band edges. On the contrary, the sharp drop in the
$T$-matrix linewidths at the band edges is a consequence of higher-order
renormalization of the Born scattering amplitude by multiple scattering
processes. This makes the $T$-matrix amplitude strongly energy dependent, and
can modeled quantitatively with a simple analytic $T$-matrix model as we have
demonstrated in Ref.~\onlinecite{Kaasbjerg:Transport}.

Another consequence of the higher-order renormalization is a strong reduction of
the scattering amplitude between the Born and $T$-matrix approximations. This is
evident from Fig.~\ref{fig:mos2_gamma_vs_e} where the former overestimates the
scattering rates up to three orders of magnitude. The reduction is largest for
$M$ vacancies which are strong defects (cf. the matrix elements in
Fig.~\ref{fig:M_BZ}) for which the $T$ matrix leads to a giant renormalization
of the Born scattering amplitude. For the weaker $X$ centered defects (vacancies
and substitutional atoms), the matrix elements in Fig.~\ref{fig:M_BZ} are
smaller, but still large enough for the $T$ matrix to yield a nonnegligible
renormalization of the scattering amplitude. These observations point to a
concomitant breakdown of the Born approximation and stress the importance of a
$T$-matrix description of atomic defects in 2D TMDs.

In the plots in the right columns of Figs.~\ref{fig:mos2_gamma_vs_e}
and~\ref{fig:wse2_gamma_vs_e}, we have separated out the contributions to the
linewidth broadening from intravalley (solid lines) and intervalley (dashed lines)
scattering. Interestingly, the plots show that while the intravalley and intervalley
contributions are of the same order of magnitude for Mo vacancies in MoS$_2$,
the intervalley contribution is negligibly small for S vacancies as well as W
and Se vacancies in WSe$_2$. In the almost spin-degenerate conduction band of
MoS$_2$~\cite{Rossier:Large}, this is related to the symmetry-induced selection
rules for the intervalley matrix elements discussed in
Sec.~\ref{sec:examples_tmds} and Ref.~\onlinecite{Jauho:Symmetry}, which
strongly suppress $K\leftrightarrow K'$ intervalley scattering by $X$-centered
defects in 2D TMDs. In the valence band of WSe$_2$, this as well as the large
and opposite spin-orbit splitting in the $K,K'$ valleys~\cite{Schwing:GiantSO},
suppress intervalley scattering. Long valley lifetimes exceeding hundreds of ps
are therefore achievable even in highly disordered 2D TMDs if $M$-centered
defects can be eliminated.

Our finding for the suppression of intervalley scattering by chalcogen-centered
defects is also relevant for studies of weak localization/antilocalization in 2D
TMDs~\cite{Shen:Intervalley,Falko:SpinValley,Eda:Quantum,Houzet:Weak,Eda:Phase},
where, e.g., S vacancies in $n$-doped MoS$_2$ have often been mentioned as a
source pronounced intervalley
scattering~\cite{Eda:Quantum,Houzet:Weak,Eda:Phase}. As we have demonstrated
here, this is not the case and intervalley scattering must instead be attributed
to the existence of other point defects.

\subsection{Transport}

In studies of the disorder-limited transport properties of 2D TMDs, it is
important to consider the fact that defect-induced in-gap states can trap holes
or electrons as extrinsic carriers (i.e., gate induced) are introduced into the
bands. This holds, respectively, for occupied in-gap states in $p$-doped as well
as unoccupied in-gap states in $n$-doped samples, and results in charging of the
defect sites. A description of such charging effects within the framework of our
method in Sec.~\ref{sec:method} is beyond the scope of this work. Based on a
simple model, we recently demonstrated that the charge-impurity scattering
resulting from charging of defects has detrimental consequences for the carrier
mobility in 2D TMDs~\cite{Kaasbjerg:Transport}, and is therefore unfavorable in
order to realize high-mobility TMD samples.

In the following, we focus on cases where defect charging is not expected. As
witnessed by Fig.~\ref{fig:tmd_dos}, this situation is encountered in $p$-doped
WSe$_2$ with Se vacancies or O substitutionals which only introduce unoccupied
in-gap states above the intrinsic Fermi level. Upon $p$ doping the material,
i.e., moving the Fermi level into the valence band with a gate voltage, the
defects thus remain overall neutral as there are no occupied in-gap states to
deplete. The defect potentials in Sec.~\ref{sec:examples_tmds} obtained for the
charge-neutral defect supercells therefore give a realistic description of the
defects in the $p$-doped material.

\begin{figure}[!t]
  \centering
  \includegraphics[width=0.6\linewidth]{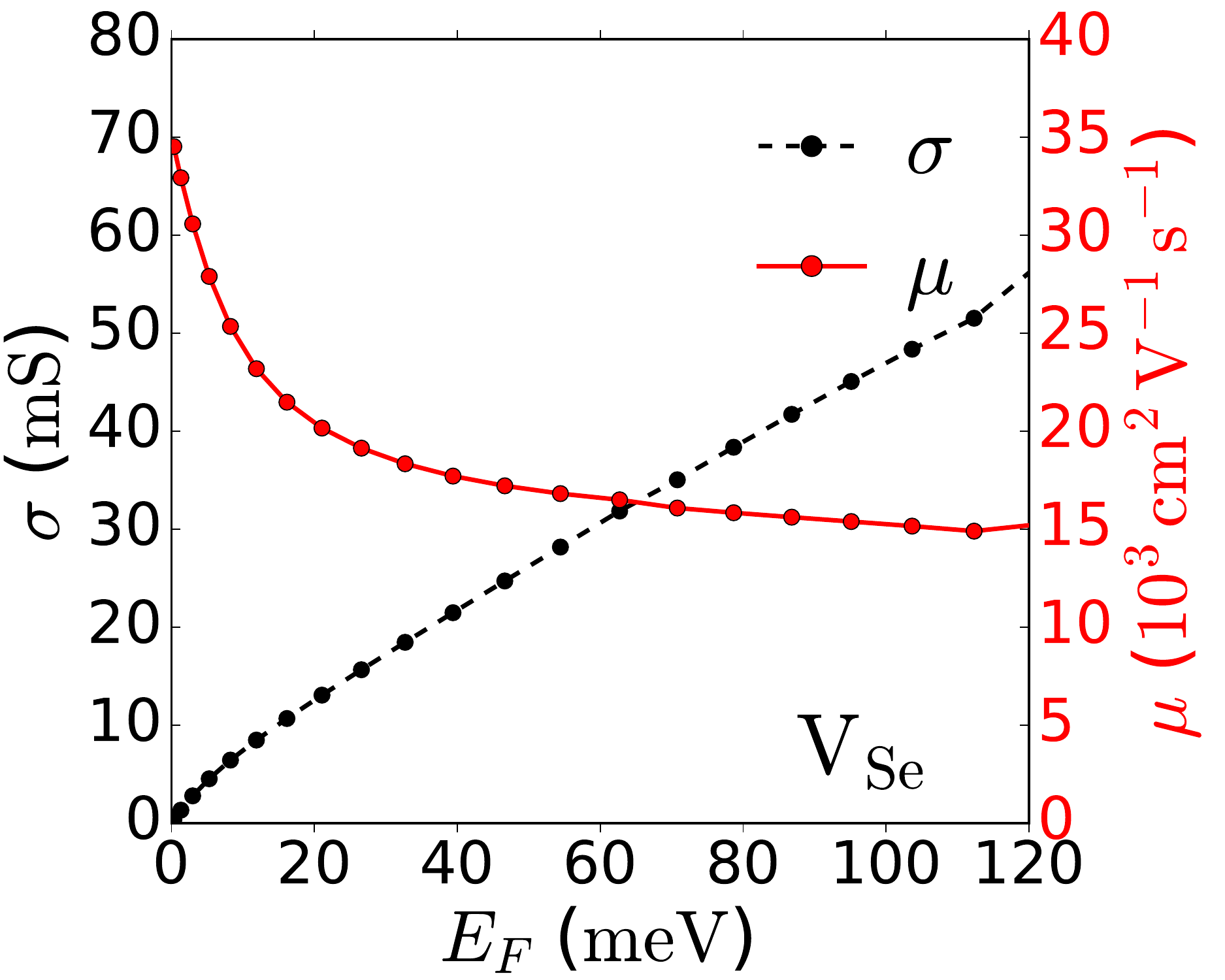}
  \includegraphics[width=0.6\linewidth]{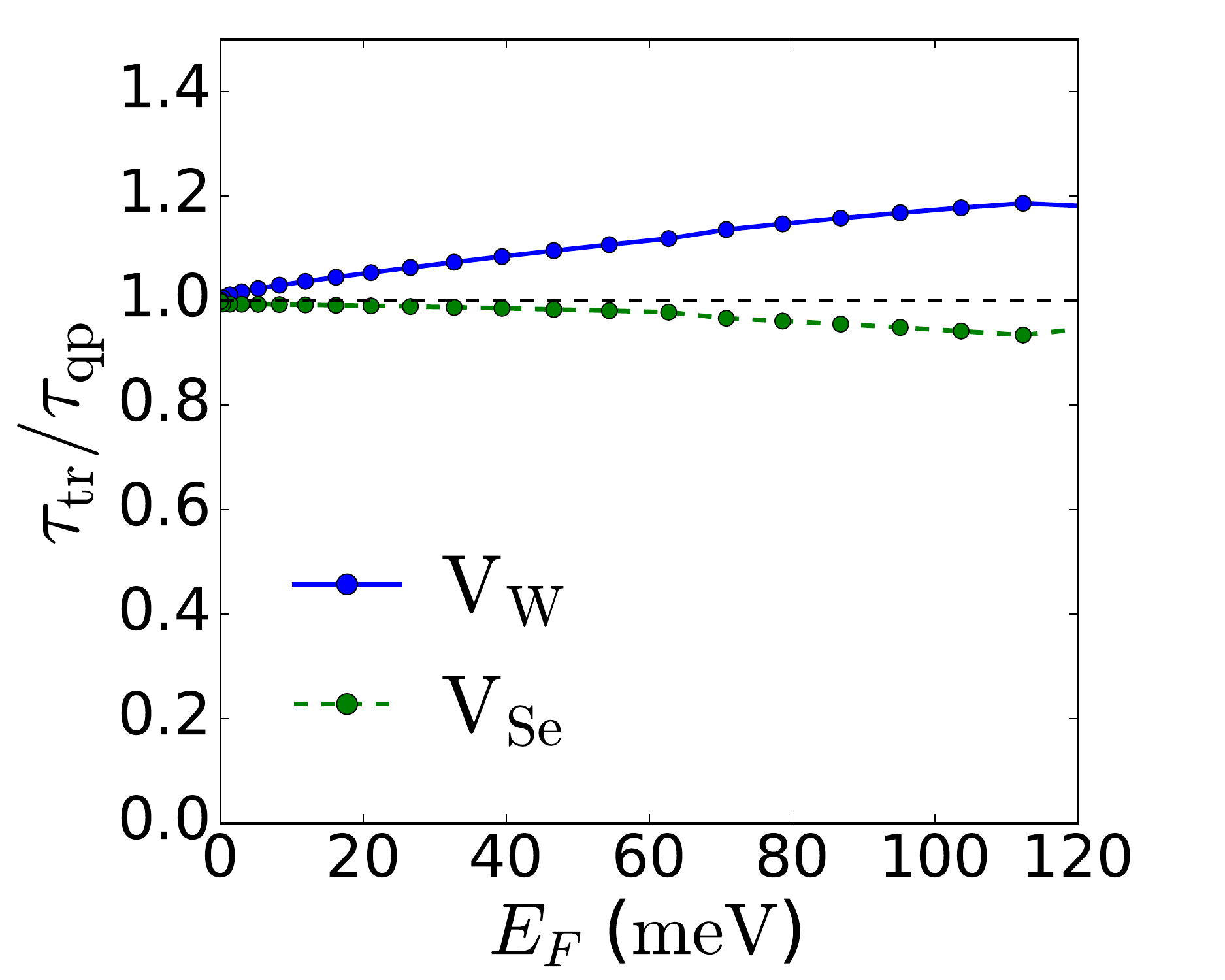}
  \caption{Low-temperature transport characteristics of disordered $p$-doped
    WSe$_2$ with a concentration of Se vacancies of $c_i = 0.01\,\%$
    ($n_i \approx 10^{11}\,\mathrm{cm}^{-2}$). (top) Conductivity and mobility vs
    Fermi level. (bottom) Ratio between the transport and quantum scattering
    times. Parameters: $m^*=0.46$ and caption of Fig.~\ref{fig:wse2_gamma_vs_e}.}
\label{fig:tmd_transport}
\end{figure}

In the top plot of Fig.~\ref{fig:tmd_transport} we show the low-temperature
transport characteristics of disordered $p$-doped WSe$_2$ with a $c_i=0.01\,\%$
concentration of Se vacancies corresponding to the defect density
($\sim 10^{11}$~cm$^{-2}$) in recently fabricated high-quality flux-grown
TMDs~\cite{Pasupathy:Approaching,Hone:Disorder}. The conductivity is obtained
from Eq.~\eqref{eq:drude} using our DFT-calculated hole mass ($m^*=0.46$), and
with the relaxation time calculated from Eq.~\eqref{eq:tau_tr_iso} using DFT
inputs for the band structure and $T$ matrix.

In $p$-doped WSe$_2$, the carrier density $n$ scales with the Fermi level as
$n \approx \tfrac{E_F}{5 \,\mathrm{meV}}\times 10^{12}\,\mathrm{cm}^{-2}$. The
energy range considered in Fig.~\ref{fig:tmd_transport} thus corresponds to
typical values of the carrier densities accessible in experiments. The
conductivity and mobility in Fig.~\ref{fig:tmd_transport} directly probe the
energy dependence of the scattering rate shown in
Fig.~\ref{fig:wse2_gamma_vs_e}. Since the scattering rate decreases with
energy, the conductivity exhibits an initial sublinear density dependence, which
translates into a mobility that decreases with carrier density. The
characteristic density scaling of the mobility in Fig.~\ref{fig:tmd_transport} is
therefore a direct fingerprint of the inherent energy dependence of the
$T$-matrix scattering amplitude for point defects in 2D TMDs.

Another important observation to make from Fig.~\ref{fig:tmd_transport} is the
overall large magnitude of the carrier mobility,
$\mu \sim 15,000$--$35,000$~$\mathrm{cm}^2\,\mathrm{V}^{-1}\,\mathrm{s}^{-1}$,
which exceeds all previously reported experimental values, so far not exceeding
$\mu \sim
5000$~$\mathrm{cm}^2\,\mathrm{V}^{-1}\,\mathrm{s}^{-1}$~\cite{Herrero:Intrinsic,Eda:Transport,Wang:Towards,Hone:Multi,Eda:Quantum,Tutuc:Shubnikov,Hone:Low,Ensslin:Gate,Dean:Ambi,Tutuc:Large,Ensslin:Interactions,Eda:Phase}.
The large theoretical mobility predicted here is a hallmark of (i) the low
defect density used in the calculation which corresponds to high-quality
TMDs~\cite{Hone:Disorder}, and (ii) the absence of the above-mentioned defect
charging which leads to a significant reduction of the mobility due to
charged-impurity scattering~\cite{Kaasbjerg:Transport}. Both factors are
essential for the realization of high-mobility monolayer TMD samples.

Finally, the bottom plot in Fig.~\ref{fig:tmd_transport} shows the ratio between
the transport scattering time and the quantum (quasiparticle) scattering time
which is accessible from Shubnikov--de Haas oscillations in
magnetotransport~\cite{Burkard:Landau}. The close-to-unity value of the ratio is
a direct manifestation of a weak $\bq=\bk-\bk'$ dependence of the $T$-matrix
scattering amplitude in the $K,K'$ valleys which is inherited from the Se
vacancy defect matrix element in Fig.~\ref{fig:M_BZ}. In this case, the
$\cos\theta_{\bk\bk'}$ term in the transport relaxation time in
Eq.~\eqref{eq:tau_tr_iso} vanishes, and the two scattering times become
identical.

Aside from the impact on the longitudinal conductivity considered here, other
theoretical works have studied the effect of disorder in 2D TMDs on various
other properties such as, e.g., the optical conductivity~\cite{Guinea:Effect},
excitons and optical absorption~\cite{Gunlycke:Optical},
localization~\cite{Shen:Intervalley,Falko:SpinValley,Houzet:Weak}, and spin and
valley Hall effects~\cite{Xiao:Spin,Sousa:Valley}. Extensions to studies based
on atomistic descriptions of the defect potential offer interesting perspectives
for future developments.

\section{Disordered graphene}
\label{sec:Graphene}

Graphene is known to host a wide variety of atomic-scale point defects which are
predicted to introduce resonant states on the Dirac cone associated with
quasibound defect
states~\cite{Neto:Disorder,Wiesendanger:Local,Neto:Electronic,Ducastelle:Electronic}.
The energy of such quasibound states depends on the interaction between the
defect and the graphene lattice which is highly sensitive to the position of the
defect~\cite{Ferreira:Impurity,Silva:Symmetry,Fabian:Resonant}. For vacancies
and substitutional atoms, quasibound states with energies in direct vicinity of
the Dirac point arise in a robust manner. In transport, such defects act as
resonant scatterers exhibiting a strong peak in the scattering cross section at
the resonance energy which suppresses the
conductivity~\cite{Mirlin:Electron,Basko:Resonant} and affects electron
cooling~\cite{Zeldov:Resonant,Mirlin:Resonant}.

In the following, we focus on monoatomic vacancies and nitrogen (N)
substitutionals on the $A$ and/or $B$ sublattice.

\subsection{DOS and spectral function}

The DOS of disordered graphene has been studied in numerous works (see,
e.g., Refs.~\onlinecite{Neto:Disorder,Neto:Electronic,Neto:Modeling,Sarma:Density})
addressing, e.g., defect-induced resonant states and band-gap openings. Below we
separate the discussion in two cases: (1) sublattice-asymmetric disorder,
i.e. defects located exclusively on one sublattice, and (2) sublattice-symmetric
disorder where the defects are distributed equally between the $A$ and $B$
sublattices.
\begin{figure}[!t]
  \centering
  \includegraphics[width=0.7\linewidth]{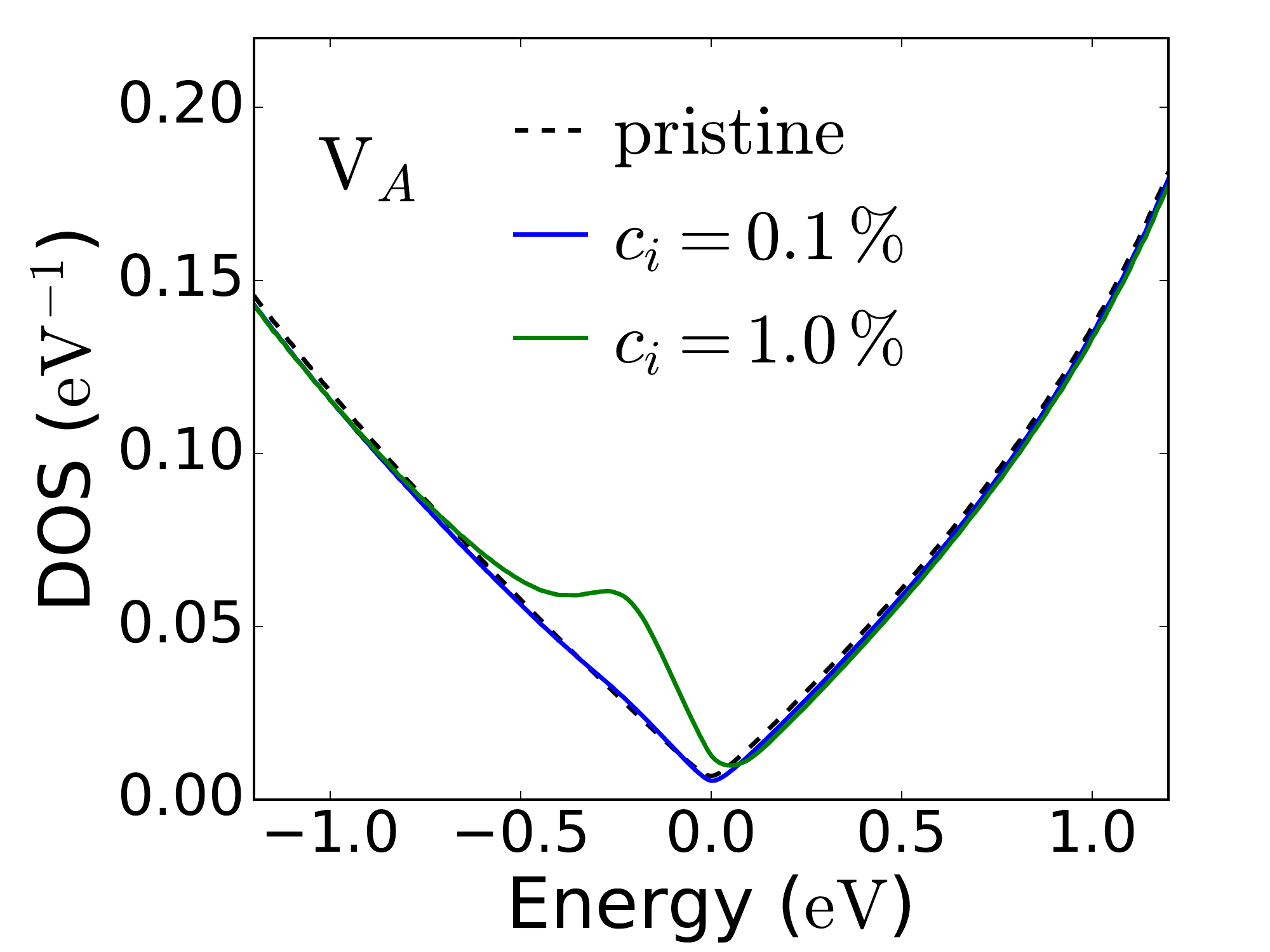}\\[4mm]
  \includegraphics[width=0.7\linewidth]{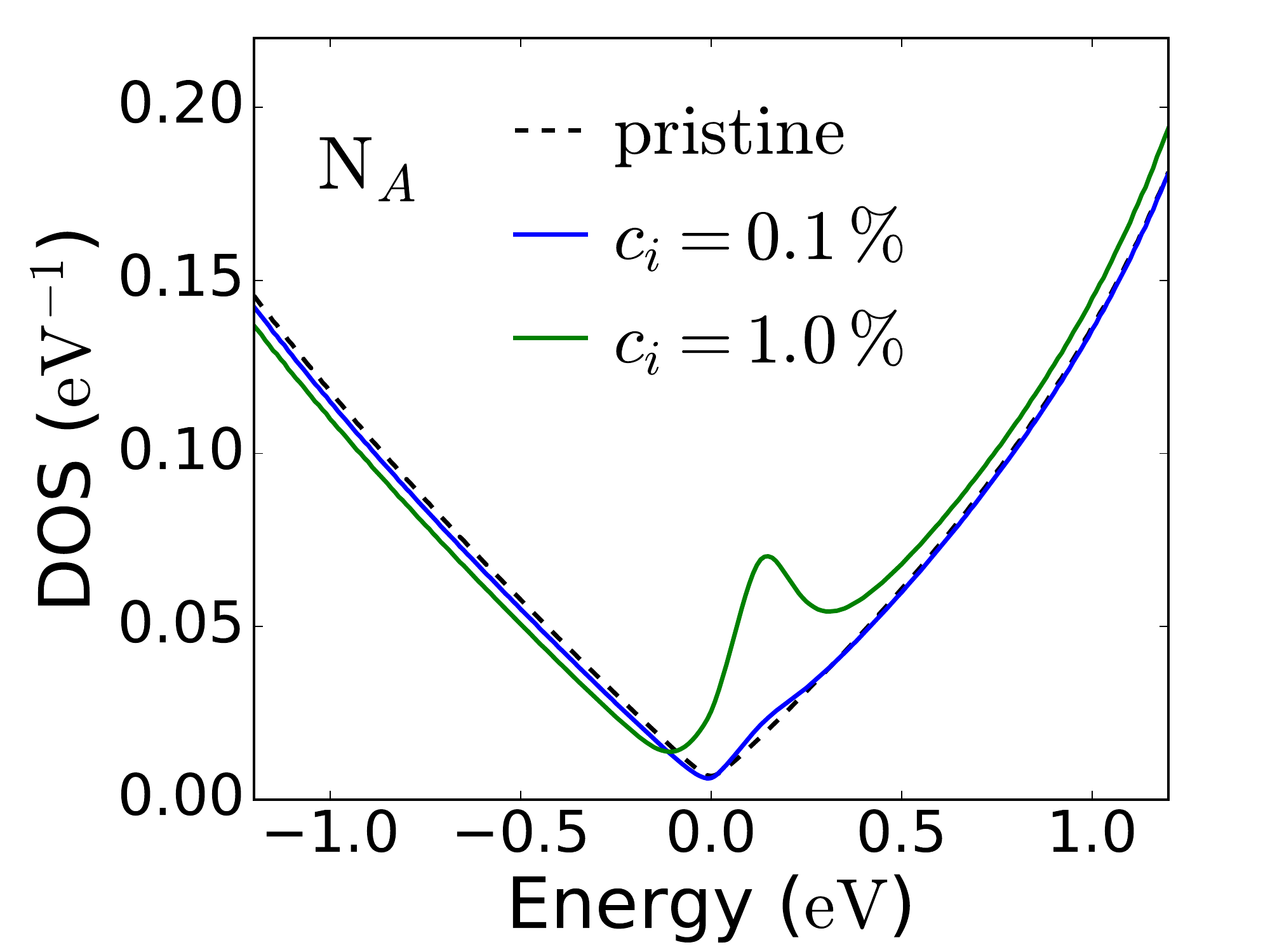}
  \caption{Density of states of disordered graphene with different defect
    concentrations of vacancies (top) and N substitutionals
    (bottom). Parameters: $99\times 99$$^*$ $\bk$ points ($300 \times 300$ for
    the pristine DOS), 2 bands, and $\eta=50$~meV ($15$~meV for pristine).}
\label{fig:graphene_dos}
\end{figure}

\subsubsection{Sublattice asymmetric disorder}
\label{sec:asymmetric}

In Fig.~\ref{fig:graphene_dos}, we show the DOS for disordered graphene with
vacancies (top) and N substitutionals (bottom) located on only the $A$
sublattice (V$_A$ and N$_A$). At low concentration, $c_i = 0.1\%$, the DOS is
hardly discernible from the DOS of pristine graphene. By contrast, at
$c_i = 1\%$ a clear peak in the DOS associated with a quasibound resonant state
emerges, respectively, below and above the Dirac point for V$_A$ and N$_A$
defects. This is in agreement with STM studies of the
LDOS~\cite{Pasupathy:Visualizing,Henrard:Localized,Andrei:Realization}. Our
finding for the position of the vacancy bound state contrasts previous
theoretical studies based on tight-binding
modeling~\cite{Neto:Electronic,Katsnelson:Resonant,Neto:Unified,Ducastelle:Electronic,Fabian:Resonant,Ast:Band},
which predict the resonant state to be at the Dirac point. In our analysis below
we comment on this discrepancy.

\begin{figure}[!t]
  \centering
  \includegraphics[width=0.49\linewidth]{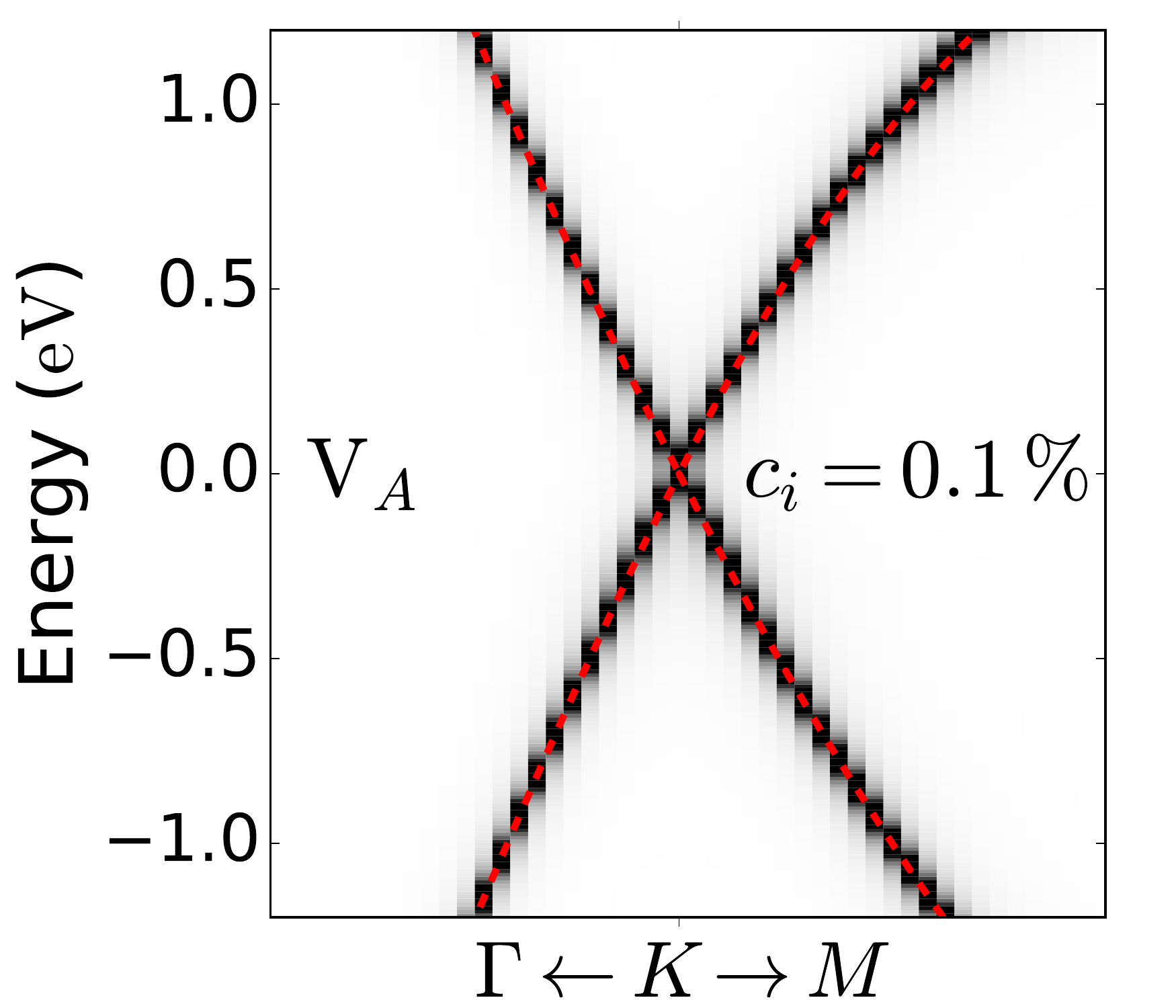}
  \includegraphics[width=0.49\linewidth]{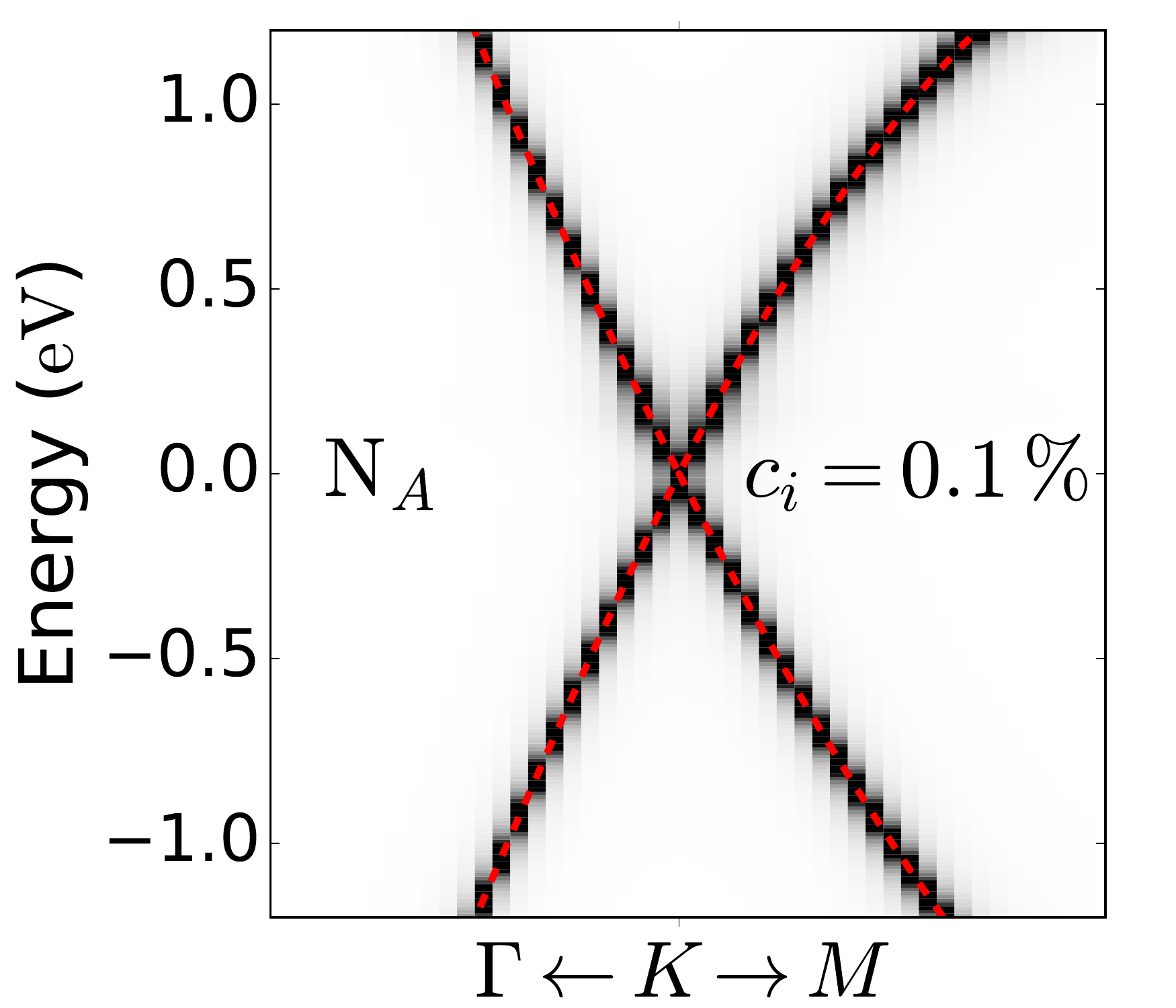}
  \\[4mm]
  \includegraphics[width=0.49\linewidth]{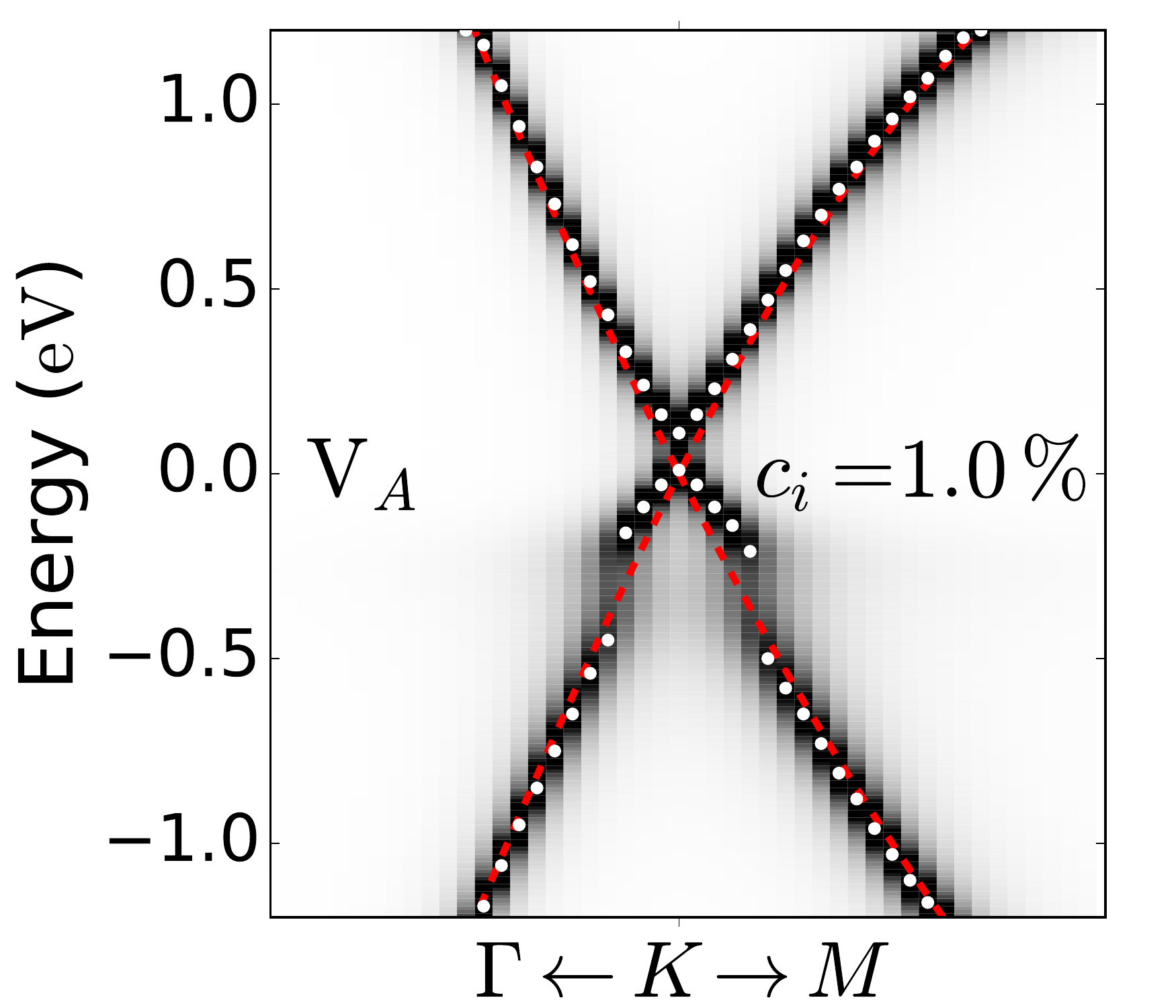}
  \includegraphics[width=0.49\linewidth]{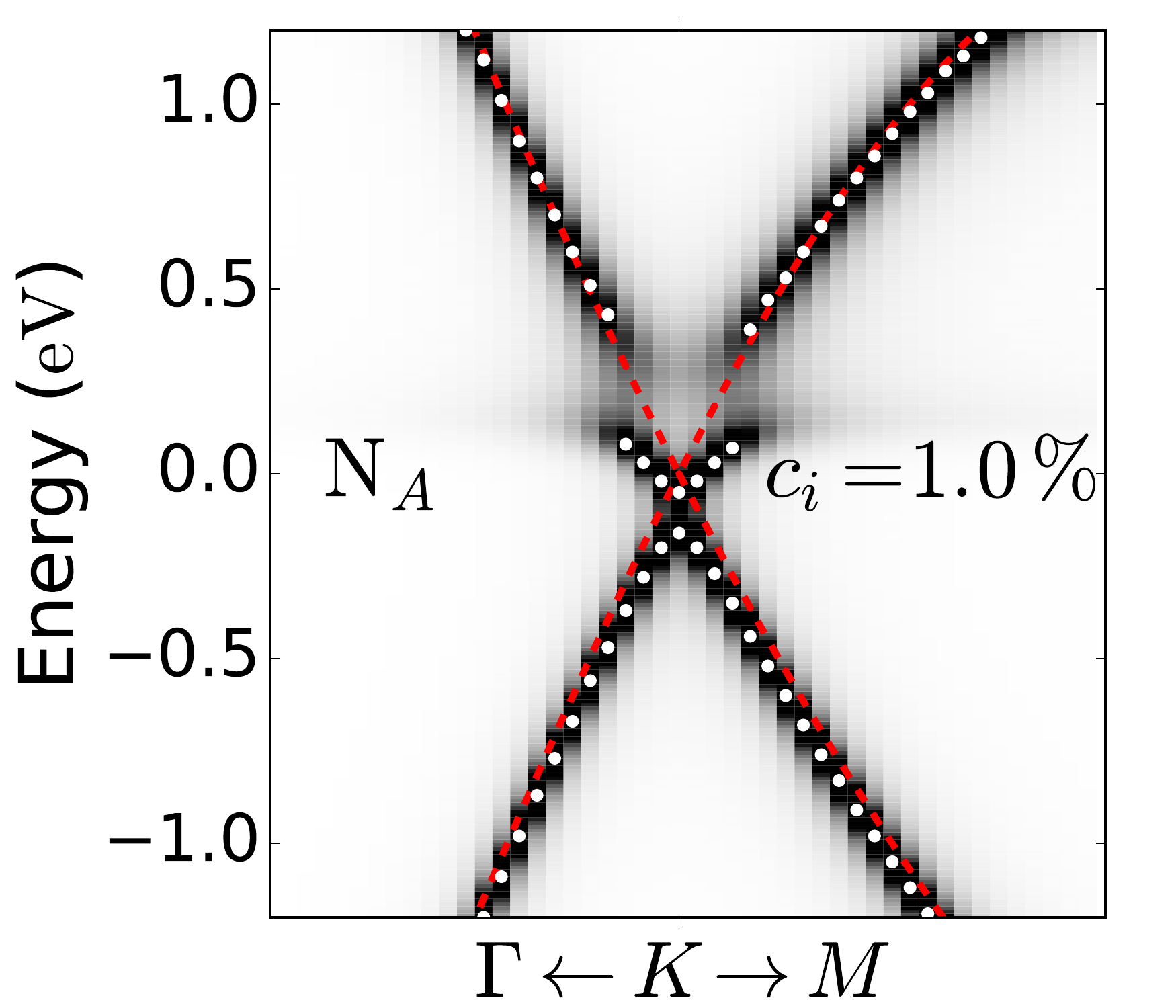}
  \caption{Spectral function for disordered graphene with different
    concentrations [(top) $c_i=0.1\%$, and (bottom) $c_i=1\%$] of $A$-sublattice
    (left) vacancy defects, and (right) N substitutionals. The red dashed lines
    show the dispersion of pristine graphene. The white dots mark the peak
    values of the spectral function. Parameters: $\eta=25$~meV, and caption of
    Fig.~\ref{fig:graphene_dos}.}
\label{fig:graphene_spectral1}
\end{figure}

Figure~\ref{fig:graphene_spectral1} shows plots of the spectral functions
corresponding to the different defects and concentrations in
Fig.~\ref{fig:graphene_dos}. The spectral function at $c_i=0.1\%$ reflects the
pristine bands (red dashed lines), although a finite broadening due to disorder
scattering, here masked by the numerical broadening $\eta$, is present (see
also Fig.~\ref{fig:graphene_gamma} below). For $c_i=1\%$, the Dirac cone is
strongly perturbed due to resonant scattering at the position of the quasibound
defect states in the DOS. In addition to a pronounced broadening of the states
which completely washes out the Dirac cone, this also produces a significant
renormalization of the bands below and above the position of the resonance.
Signatures of such effects in ARPES on nitrogen-doped graphene have so far not
been observed~\cite{Vyalikh:Nitrogen,Taut:Structural}, probably because the
concentration of N substitutionals is too low.

By closer inspection of the spectral functions in
Fig.~\ref{fig:graphene_spectral1}, a concentration-dependent band-gap opening
can be observed at the Dirac point which at $c_i=1\%$ is $\sim 100$~meV. This is
expected as defects located on a single sublattice break the sublattice
symmetry, effectively turning the disordered system into gapped
graphene~\cite{Cappelluti:Impurity} as also demonstrated in other theoretical
works considering sublattice-asymmetric
disorder~\cite{Neto:Modeling,Charlier:Electronic,Power:Electronic}. Band-gap
openings have also been reported experimentally in ARPES on nitrogen-doped
graphene~\cite{Vyalikh:Nitrogen} and graphene with hydrogen
adatoms~\cite{Hornekaer:Bandgap}, but the underlying mechanism is believed to be
of a different nature, i.e., not associated with sublattice asymmetry.

% Due to the finite numerical broadening $\eta$, it is difficult to resolve the
% band-gap opening in our DFT calculated spectral function.  
To shed additional light on the band-gap opening as well as the resonant
spectral features, the GF in the $2\times 2$ subspace spanned by the valence and
conduction bands. In this subspace, the diagonal elements obtained by matrix
inversion of the Dyson equation~\eqref{eq:dyson} take the form
\begin{equation}
  \label{eq:GF_NA}
  G_{\bk}^{nn}(\varepsilon) = 
      \frac{1}{\varepsilon - \varepsilon_{n\bk} 
        - \Sigma_{n\bk}^{\text{eff}}(\varepsilon)}
             , \quad A\;\text{or}\;B,
\end{equation}
where the \emph{effective} self-energy is given by
\begin{equation}
  \label{eq:Sigmaeff}
  \Sigma_{n\bk}^{\text{eff}}(\varepsilon) = \Sigma_\bk^{nn}(\varepsilon) +
      \frac{\Sigma_\bk^{n\bar{n}}(\varepsilon)\Sigma_\bk^{\bar{n}n}(\varepsilon)}
           {\varepsilon - \varepsilon_{\bar{n}\bk} -
            \Sigma_\bk^{\bar{n}\bar{n}}(\varepsilon)} ,
   \quad \bar{n}\neq n ,
\end{equation}
Here, the second term introduced by the matrix inversion describes a
defect-induced coupling between the valence and conduction bands, and is
responsible for the band-gap opening.

\begin{figure}[!t]
  \centering
  \includegraphics[width=0.99\linewidth]{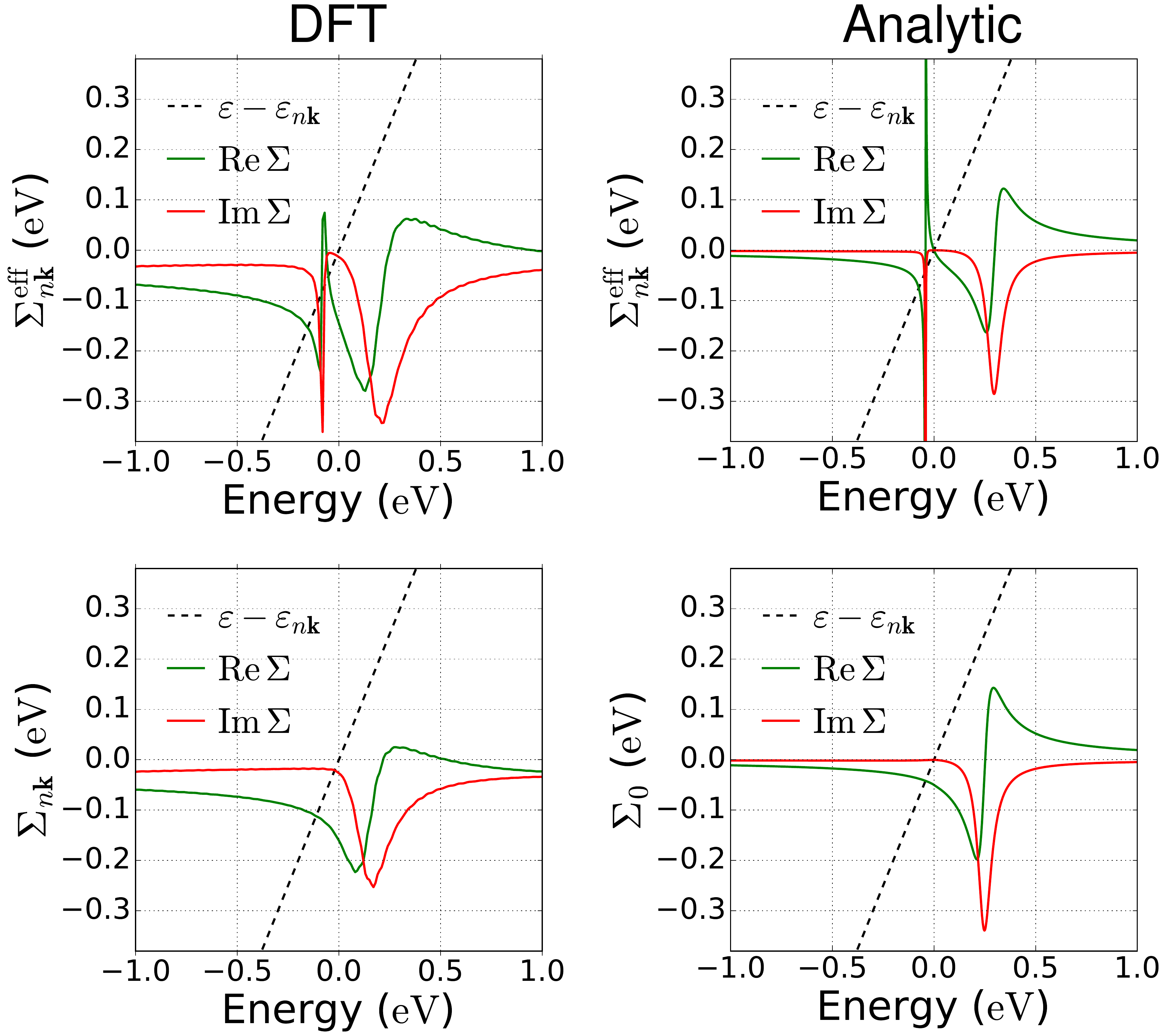}
  \caption{Disorder self-energy at $\bk=\mathbf{K}$ for (top) $A$-only, and
    (bottom) $A$+$B$ N substitutionals, respectively ($c_i=1\%$). (left) DFT
    results. (right) analytic results based on the self-energies in
    Eqs.~\eqref{eq:Sigmamn_N} and~\eqref{eq:Sigma_eff} with $V_0=-10$~eV,
    $A_\text{cell}=5.25$~{\AA}$^2$, $v_F=10^6$~m/s, $\Lambda=10^4$~eV. DFT
    parameters: $\eta=25$~meV, and caption of Fig.~\ref{fig:graphene_dos}.}
\label{fig:transcendental}
\end{figure}

The effective self-energy in Eq.~\eqref{eq:Sigmaeff} based on the DFT calculated
$T$ matrix for N substitutionals is shown in the left top plot of
Fig.~\ref{fig:transcendental} for $\bk=\mathbf{K}$. Here, the solution to the
transcendental equation for the QP equation in Eq.~\eqref{eq:qpeq} corresponds
to the intersection between the solid (green) and dashed lines. Clearly, the
effective self-energy shows a feature just below the Dirac point which gives
rise to two solutions of the QP equation, corresponding, respectively, to the
top of the valence band and the bottom of the conduction band, and hence mark a
band-gap opening.

The feature in the effective self-energy responsible for the band-gap opening
can be analyzed further based on the TB defect model introduced in
Sec.~\ref{sec:examples_graphene}. In this case, we can solve
Eq.~\eqref{eq:T_matrix} for the $T$ matrix analytically, yielding a
$\bk$-independent $T$ matrix which in the pseudospin basis inherits the matrix
structure of the defect potential in Eq.~\eqref{eq:Vab_i},
\begin{equation}
  \label{eq:Tab_i}
  \hat{T}_i(\varepsilon) = \frac{T_0(\varepsilon)}{2}
  \left( \hat{\sigma}_0 \pm \hat{\sigma}_z \right) ,
\end{equation}
where the prefactor is given by
\begin{equation}
  \label{eq:T_N}
  T_0(\varepsilon) = \frac{V_0}{1 - V_0 \bar{G}_0(\varepsilon)} ,
\end{equation}
and
$\bar{G}_0(\varepsilon)=\tfrac{1}{2}\mathrm{Tr}\,\hat{G}^0(\varepsilon) =
\tfrac{1}{2} \sum_{n\bk}G_{n\bk}^0(\varepsilon)$
is the $\bk$-summed GF for pristine graphene. In the Dirac model, it is given by
\begin{equation}
  \label{eq:Gbar_graphene}
  \bar{G}_0(\varepsilon) 
  = A_\text{cell} \frac{\bar{\rho}_0}{2}
  \left[ \varepsilon \ln \left\vert \frac{\varepsilon^2}{\varepsilon^2 - \Lambda^2} \right\vert 
         -i \pi \abs{\varepsilon}  \Theta(\Lambda - \abs{\varepsilon})
  \right] ,
\end{equation}
where $\bar{\rho}_0 = g_v /2\pi (\hbar v_F)^2$, $g_v=2$ is the valley
degeneracy, and $\Lambda$ is an ultraviolet cutoff.

Performing a unitary transformation to the eigenstate basis (as the results
below are independent on the valley index we omit it here), the $T$-matrix
self-energy in Eq.~\eqref{eq:sigma_T} becomes
\begin{equation}
  \label{eq:Sigmamn_N}
  \hat{\Sigma}_{\bk}^T(\varepsilon) = \Sigma_0(\varepsilon)
  \begin{pmatrix}
    1 & \pm 1 \\
    \pm 1 & 1
  \end{pmatrix}  , \quad
  \Sigma_0(\varepsilon) = \frac{c_i T_0(\varepsilon)}{2}  ,
\end{equation}
where the $\pm$ sign on the off-diagonal elements is for defects on the $A,B$
sublattice. In either case, the diagonal elements of the GF again take the form
in Eq.~\eqref{eq:GF_NA}, with the effective self-energy now given by,
\begin{equation}
  \label{eq:Sigma_eff}
  \Sigma_{n\bk}^{\text{eff}}(\varepsilon) =   
  \Sigma_0(\varepsilon) + 
    \frac{[\Sigma_0(\varepsilon)]^2}{\varepsilon - \varepsilon_{\bar{n}\bk} -
      \Sigma_0(\varepsilon)} , \quad \bar{n} \neq n.
\end{equation}
To see the role of the second term for the band-gap opening, we note that
$\bar{G}_0 \rightarrow 0$ in the vicinity of the Dirac point,
i.e. $\abs{\varepsilon} \rightarrow 0$, and hence the $T$ matrix in
Eq.~\eqref{eq:T_N} can be approximated as
$T_0(\varepsilon) \approx [ 1 + i \delta(\varepsilon)] V_0$, where
$\delta(\varepsilon)=V_0 \mathrm{Im}\, \bar{G}_0 (\varepsilon)$. In the
effective self-energy in Eq.~\eqref{eq:Sigma_eff} this leads to a pole in the
second term which for the Dirac-point self-energy, i.e. $\bk=\mathbf{K}$, is
located at $\varepsilon_0 = c_i V_0/2$.

The right top plot in Fig.~\ref{fig:transcendental} shows the analytic TB
self-energy in Eq.~\eqref{eq:Sigma_eff} for parameters corresponding to N
substitutionals (see caption). The parameters have been obtained by fitting to
the DFT self-energy as follows: We first fix $V_0$ to a value resembling the
average value of the intravalley and intervalley matrix elements in
Fig.~\ref{fig:M1_BZ}, and then treat $\Lambda$ as a fitting parameter in order
to match the DFT self-energies in Fig.~\ref{fig:transcendental}. This is in
contrast to the Debye-model inspired approach in
Ref.~\onlinecite{Neto:Electronic}, which is here found to be unable to yield a
satisfactory description of the DFT self-energy.
% Debye model
% \begin{itemize}
% \item Results highly dependent on $\Lambda$~\cite{Fabian:Resonant}.
% \item Ultra violet cutoff $\Lambda$ a fitting parameter. 
% \item Conserve number of states
%   $$ \frac{1}{N} \sum_\bk = 1 $$
%   implies that $k_c^2 = 4\pi^2 / A_\text{cell}$ and thus $\Lambda=\hbar v_F k_c$
%   becomes $\Lambda^2 = (\hbar v_F)^2 4 \pi / A_\text{cell} \rightarrow \Lambda \sim 7\,\mathrm{eV}$. 
% \item With this, the prefactor in $\bar{G}$ becomes $A_\text{cell} \bar{\rho}_0
%   / 2 = g_v / \Lambda^2$.
% \end{itemize}

Remarkably, the DFT- and TB-calculated self-energies are in almost perfect
agreement. However, due to (i) the nontrivial $\bq = \bk'-\bk$ dependence of the
matrix elements in Fig.~\ref{fig:M1_BZ}, and (ii) the finite numerical
broadening $\eta$ used in the DFT calculation of the $T$-matrix self-energy, some
quantitative differences arise.

Via the analytic self-energy in the top plot of Fig.~\ref{fig:transcendental},
the feature in the DFT self-energy responsible for the band-gap opening can now
be identified with the pole introduced by the second term in
Eq.~\eqref{eq:Sigma_eff}, which clearly emerges just below the Dirac point, and
is seen to give rise to the two solutions to the QP equation also found in the
DFT self-energy in Fig.~\ref{fig:transcendental}. Note that the QP equations for
the valence and conduction bands are identical at $\bk=\mathbf{K}$ (since
$\varepsilon_{n \mathbf{K}} = 0$ for $n=v,c$), implying that the states at the
band-gap opening are formed by a combination of the original valence and
conduction band states as in conventional gapped graphene.

In the top plots of Fig.~\ref{fig:transcendental}, the pole structure in the
self-energy at positive energy stems from the pole in the $T$ matrix associated
with the quasibound defect state in the DOS in Fig.~\ref{fig:graphene_dos}. For
the $T$ matrix in Eq.~\eqref{eq:T_N}, the pole is positioned at the energy where
$1/V_0 = \mathrm{Re}\,\bar{G}_0$, and is thus located above (below) the Dirac
point for $V_0<0$ ($V_0 > 0$) (see, e.g., Fig.~10 in
Ref.~\onlinecite{Henrard:Long}). This is in agreement with our discussion of the
$V_0$ parameter for the $\mathrm{V}_A$ and $\mathrm{N}_A$ defects in
Sec.~\ref{sec:examples_graphene}. Note that $\abs{V_0} \rightarrow \infty$
produces a bound state at the Dirac point as discussed in several TB studies of
vacancy
defects~\cite{Neto:Electronic,Katsnelson:Resonant,Ducastelle:Electronic,Fabian:Resonant,Ast:Band}. However,
as our DFT calculations here demonstrate, this limiting value of $V_0$ provides
an unrealistic description of the vacancy potential and consequently also the
position of the quasibound defect state.

In addition to introducing the defect state itself, poles in the $T$ matrix also
account for resonant scattering off the defect state which strongly perturbs the
bands. As evident from the bottom plots in Fig.~\ref{fig:graphene_spectral1},
this introduces a splitting of the conduction band at the resonance energy which
resembles a gap opening, but the broadening of the states due to resonant
scattering prevents the opening of a spectral gap.

\subsubsection{Sublattice symmetric disorder}

\begin{figure}[!t]
  \centering
  \includegraphics[width=0.49\linewidth]{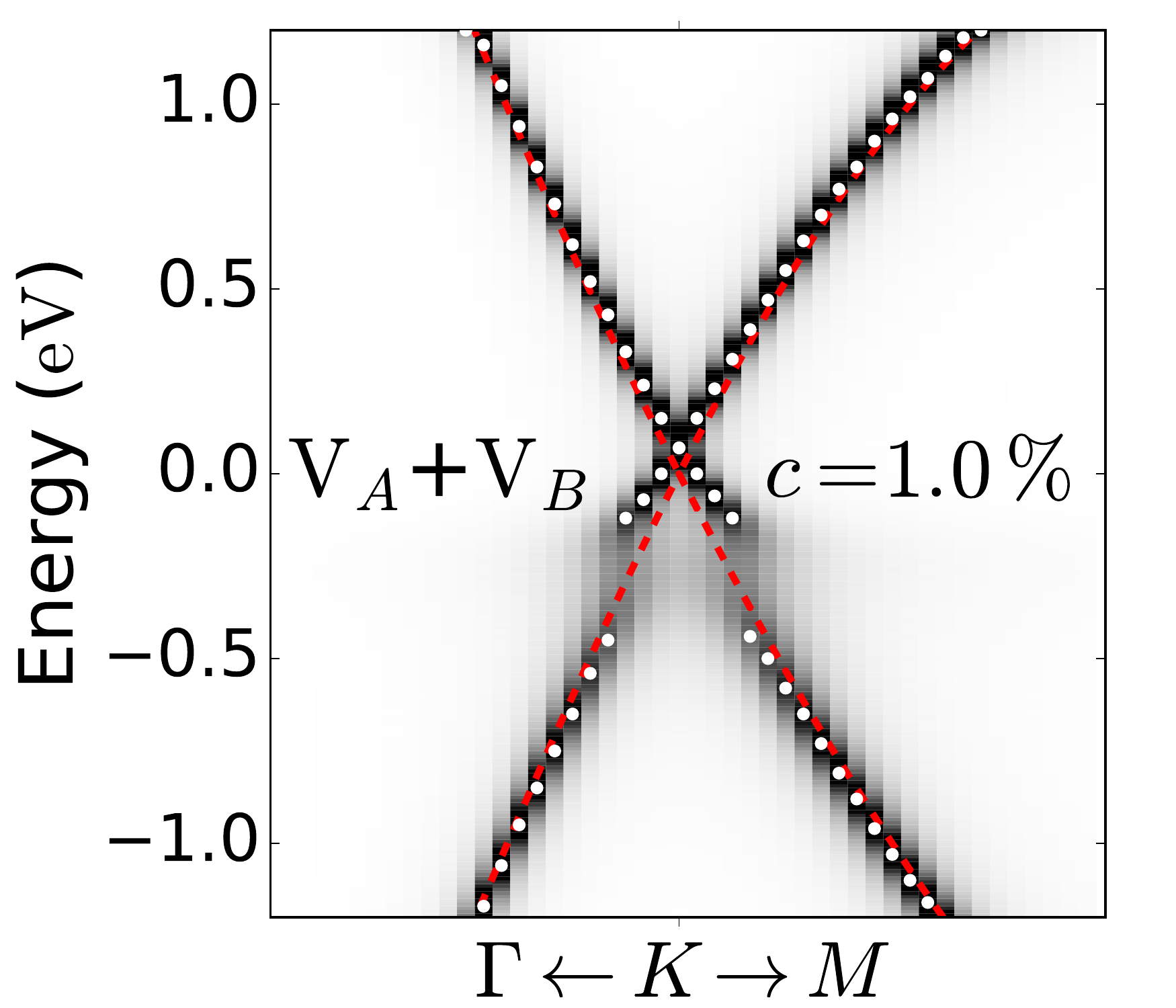}
  \includegraphics[width=0.49\linewidth]{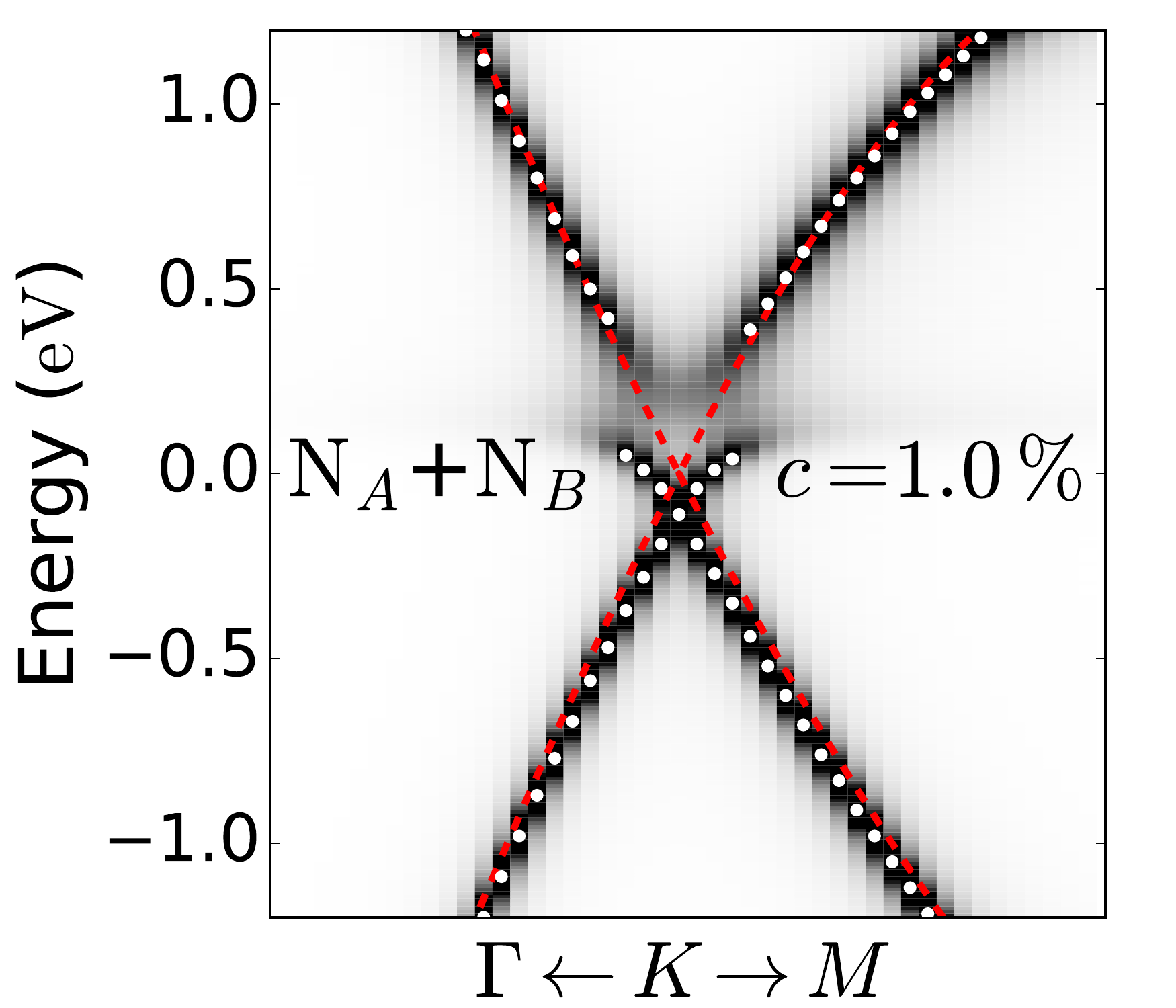}
  \\[4mm]
  \includegraphics[width=0.49\linewidth]{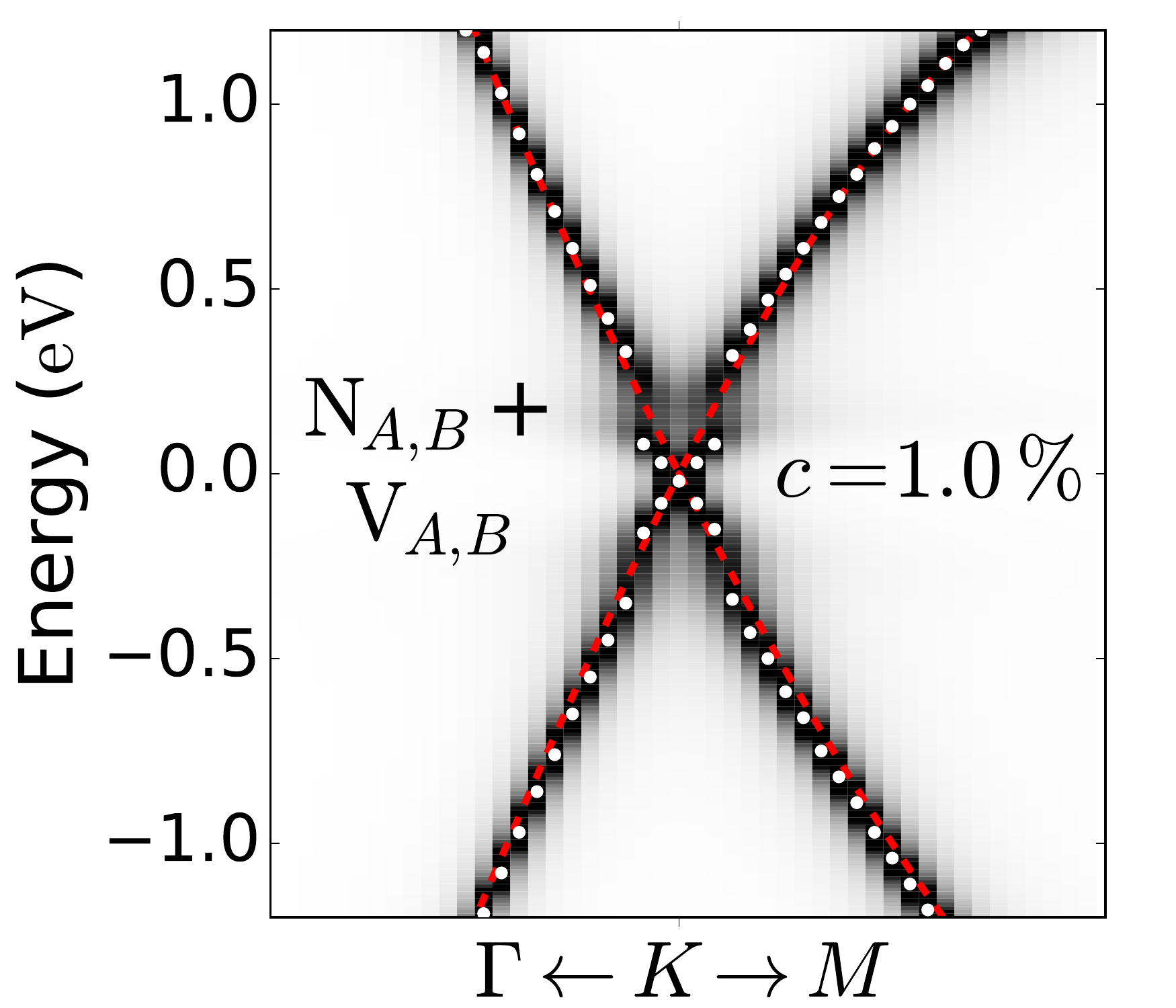}
  \caption{Spectral function for disordered graphene with equal amounts of $A$
    and $B$ sublattice defects ($c=1\%$). The plots show the spectral function
    along the $\Gamma$-$K$-$M$ BZ path for (top left) vacancies, (top right) N
    substitutionals, and (bottom) a combination of vacancies and N
    substitutionals. The red dashed lines show the dispersion of pristine
    graphene and the white dots mark the peak values of the spectral
    function. Parameters: $\eta=25$~meV, and caption of
    Fig.~\ref{fig:graphene_dos}.}
\label{fig:graphene_spectral2}
\end{figure}

In Fig.~\ref{fig:graphene_spectral2} we show the spectral function of disordered
graphene ($c=1\%$) with defects distributed equally on the $A$ and $B$
sublattices for vacancies (top left), N substitutionals (top right), and both
vacancies and N substitutionals (bottom). The two top plots to a large extent
resemble the bottom plots for $c_i=1\%$ of $A$-only defects in
Fig.~\ref{fig:graphene_spectral1}, however, with the important difference that
the spectral functions in Fig.~\ref{fig:graphene_spectral2} do not feature a
band-gap opening. In the presence of both $\mathrm{V}_{A,B}$ and
$\mathrm{N}_{A,B}$ defects, the renormalization of the bands is strongly reduced
at energies between the two resonant states in the DOS in
Fig.~\ref{fig:graphene_dos} as if the two types of defects cancel the effect of
each other.

We can again rationalize these findings by considering the GF in the valence and
conduction band subspace. As otherwise identical defects on the $A$ and the $B$
sublattices must be considered as different types of defects, the total
self-energy is given by the average over the self-energies for the individual
$A$,$B$ defects as in Eq.~\eqref{eq:sigma_avg}, which here amounts to averaging
the matrix structure of the self-energies for $A$ and $B$ sublattice
defects~\cite{Mirlin:Electron,Basko:Resonant}.

For the DFT self-energy, we find that the sublattice-averaged self-energy is
almost perfectly diagonal (not shown), such that the diagonal elements of the
GF to a good approximation become
\begin{equation}
  \label{eq:GF_NAB}
  G_{\bk}^{nn}(\varepsilon) = 
      \frac{1}{\varepsilon -\varepsilon_{n\bk} - \Sigma^{nn}_\bk(\varepsilon)} ,
      \quad A\;\text{and}\;B.
\end{equation}
This could also have been anticipated from the TB self-energy where the
sublattice average obviously eliminates the off-diagonal elements in
Eq.~\eqref{eq:Sigmamn_N} and
$\Sigma^{nn}_\bk(\varepsilon) = \Sigma_0(\varepsilon)$. Thus, for identical
defects distributed equally on the $A$ and $B$ sublattices, overall sublattice
symmetry and, hence the chirality of the graphene states, is conserved.

The two bottom plots in Fig.~\ref{fig:transcendental} show, respectively, the
DFT and TB self-energies for N substitutionals on both sublattices. Again, there
is an excellent qualitative agreement between the TB and DFT self-energies, and
quantitative differences can be attributed to the factors mentioned in
Sec.~\ref{sec:asymmetric} above. While the structure in the self-energy due to
the pole in the $T$ matrix is retained, the form of $\mathrm{Re}\Sigma$ in the
vicinity of the Dirac point does evidently not give rise to a band-gap opening,
but only a small downshift of the bands also visible in the right plot of
Fig.~\ref{fig:graphene_spectral2}.

In the case of both $\mathrm{V}_{A,B}$ and $\mathrm{N}_{A,B}$ defects, the
reduction of the band renormalization at energies immediately above and below
the Dirac point can be attributed to a partial cancellation between the real
parts of the two self-energies in this energy range. This follows
straight forwardly from the fact that the self-energy for
$\mathrm{V}_A+\mathrm{V}_B$ defects resembles a shifted version of the
self-energy for $\mathrm{N}_A+\mathrm{N}_B$ defects in the bottom plot of
Fig.~\ref{fig:graphene_spectral2} with the pole structure centered around the
position of the bound-state in Fig.~\ref{fig:graphene_dos}.

\subsection{Quasiparticle scattering and transport}

In this section, we study in further detail the disorder-induced quasiparticle
scattering responsible for the spectral linewidth broadening in
Figs.~\ref{fig:graphene_spectral1} and~\ref{fig:graphene_spectral2} as well as
its impact on the transport properties of graphene. In order to avoid
complicating the discussion with potential band-gap openings, we here focus on
sublattice-symmetric disorder.

\begin{figure}[!t]
  \centering
  \includegraphics[width=0.49\linewidth]{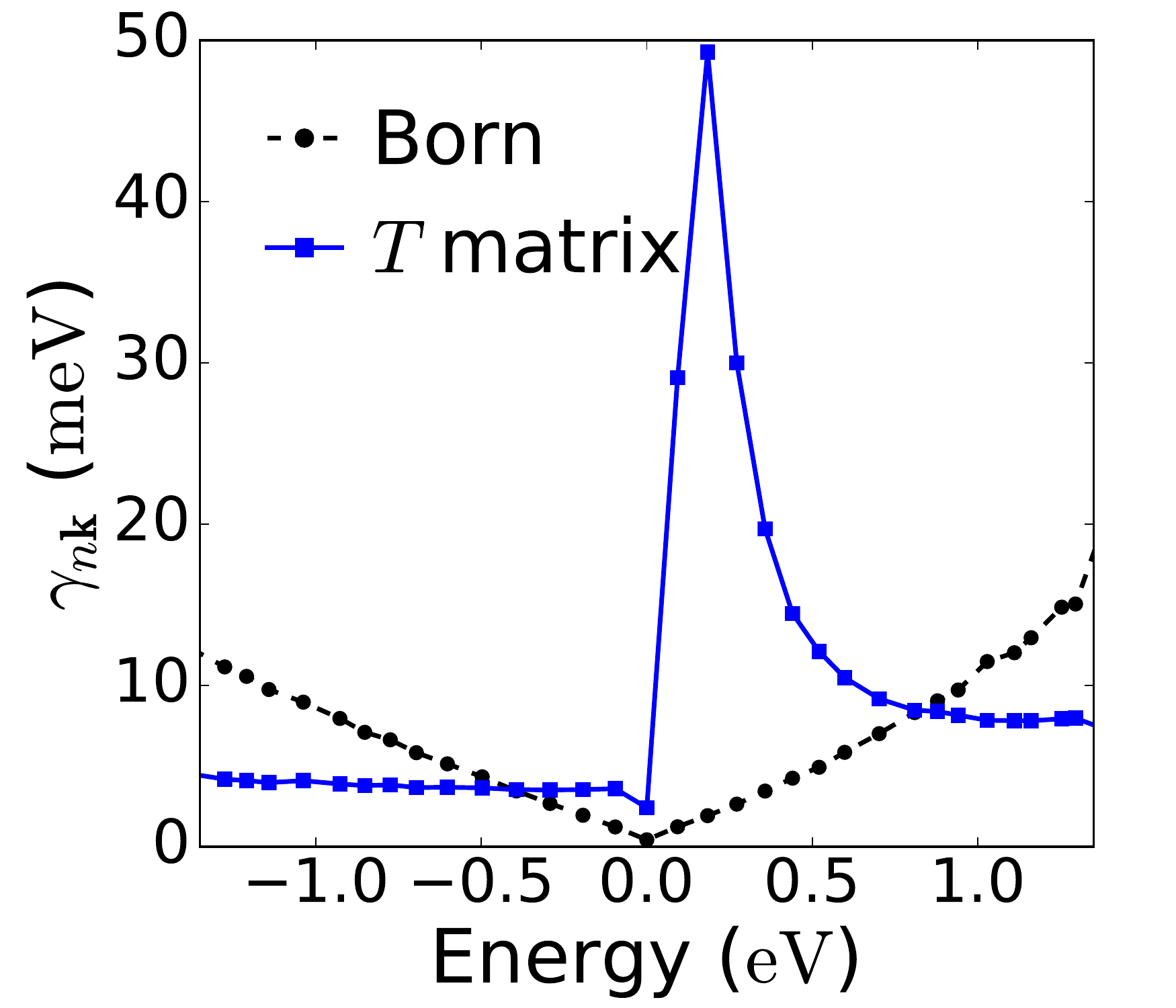}
  \includegraphics[width=0.49\linewidth]{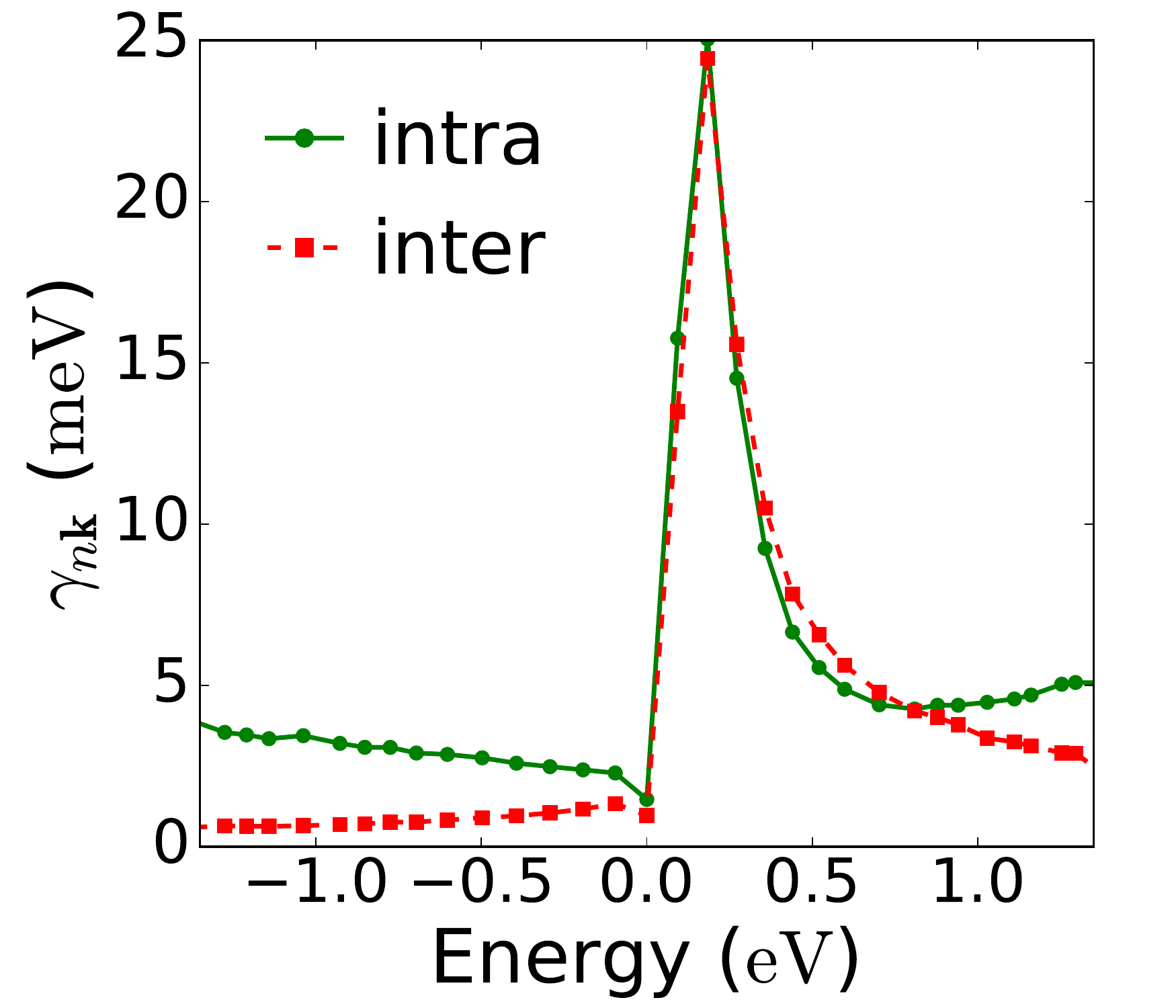}
  \caption{Energy dependence of the linewidth broadening due to sublattice
    symmetric N substitutionals in graphene. (left) Comparison between the Born
    and $T$-matrix approximations. (right) Intravalley and intervalley
    contributions to the total $T$-matrix linewidth. Points acquired along the
    $\Gamma$-$K$-$M$ path. Parameters: $c_\text{dis}=0.1\%$, $\eta=25$~meV, and
    caption of Fig.~\ref{fig:graphene_dos}.}
\label{fig:graphene_gamma}
\end{figure}

In Fig.~\ref{fig:graphene_gamma} we show the linewidth broadening in graphene
with $A$+$B$ nitrogen substituationals ($c=0.1\%$) as a function of the on-shell
energy on the Dirac cone. The left plot shows a comparison between the Born and
$T$-matrix approximations, whereas the right plot shows the individual intravalley
and intervalley contributions to the $T$-matrix linewidth in the left plot.
Note that the energy dependence of the linewidth has been obtained from $\bk$
points along the $\Gamma$-$K$-$M$ path in the BZ, and the fact that it forms a
single continuous curve along the two line segments shows that it is highly
isotropic.

While the energy dependence of the linewidth broadening in the Born
approximation reflects the energy dependence of the density of states of
pristine graphene in Fig.~\ref{fig:graphene_dos}, the $T$-matrix linewidth is
strongly electron-hole asymmetric with a pronounced peak on the electron side
due to resonant scattering. Also, on the hole side where the DOS of pristine and
nitrogen substituted graphene are almost identical
(cf. Fig.~\ref{fig:graphene_dos}), does the $T$ matrix yield a strong
renormalization of the Born approximation, with an almost energy independent
linewidth broadening.

\begin{figure*}[!t]
  \centering
  \includegraphics[width=0.3\linewidth]{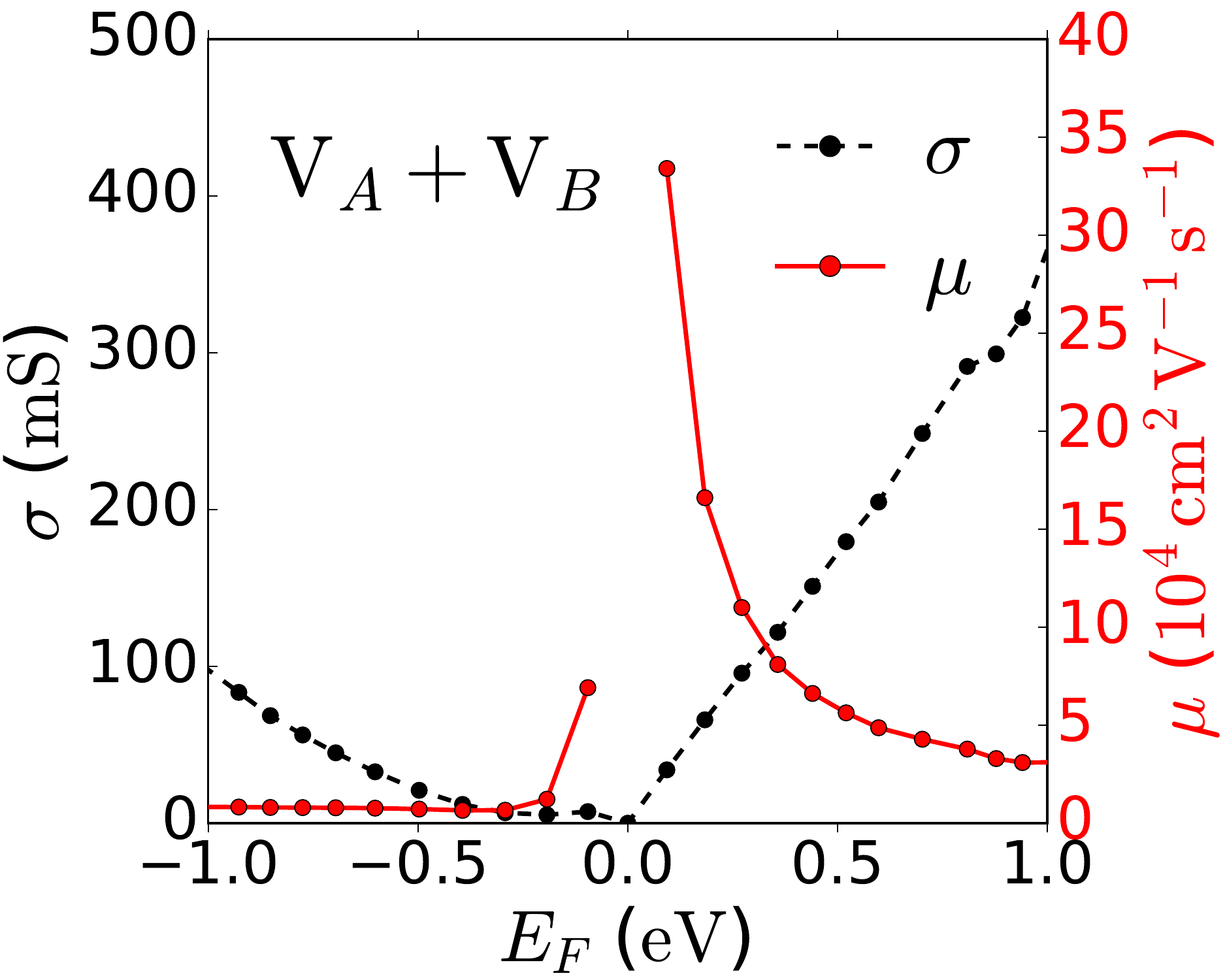}
  \includegraphics[width=0.3\linewidth]{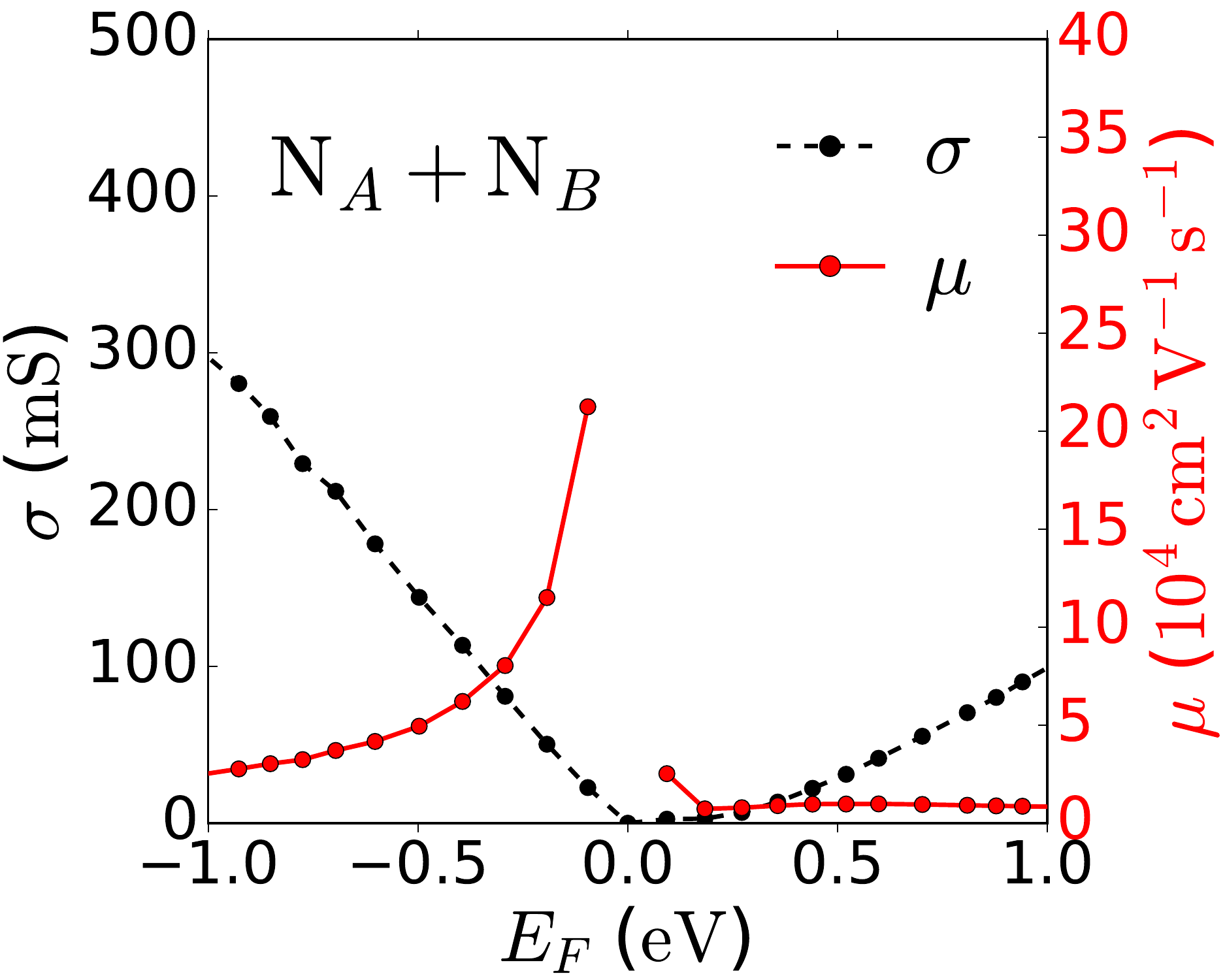}
  \includegraphics[width=0.3\linewidth]{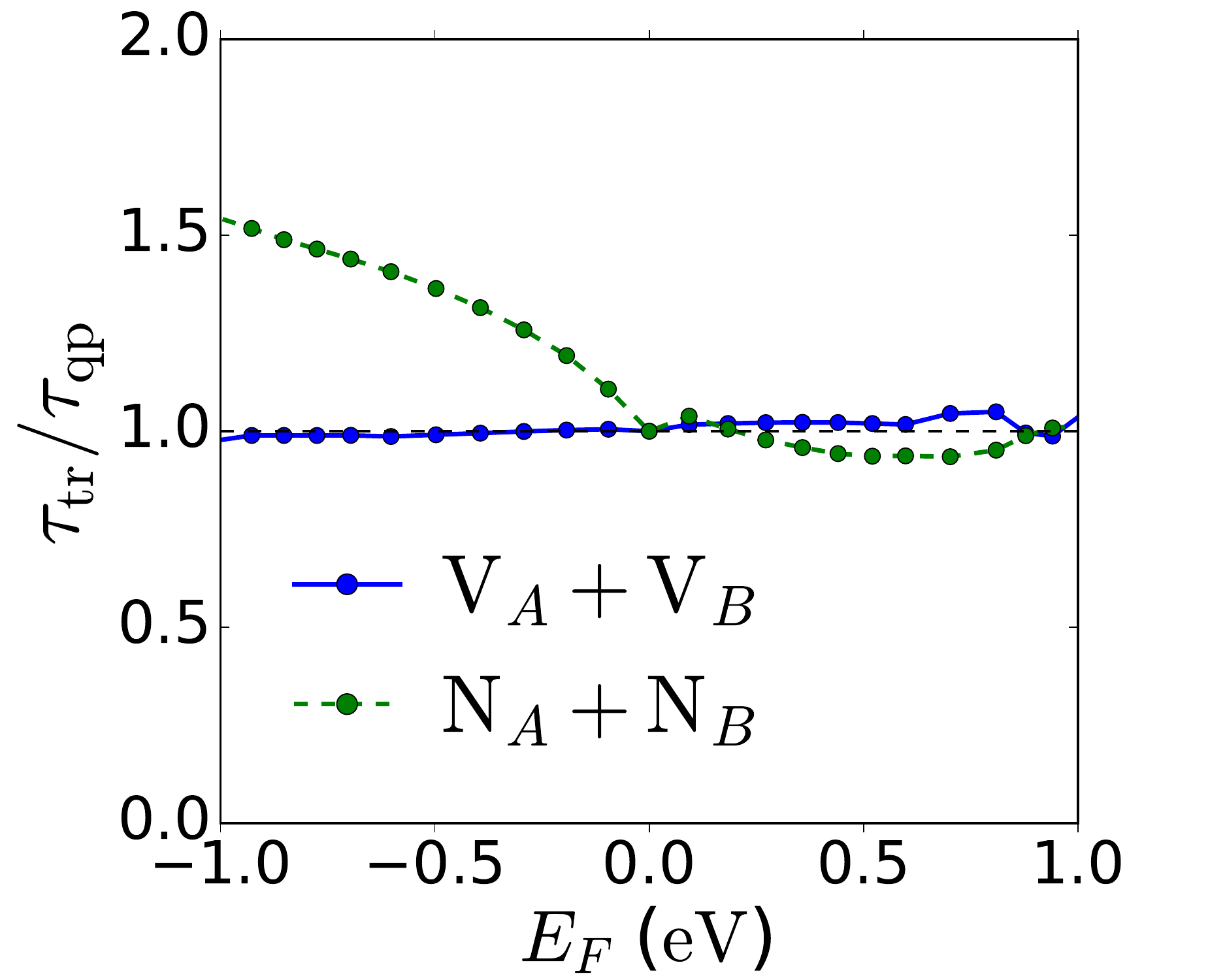}
  \caption{Low-temperature transport characteristics of disordered graphene with
    a $c = 0.01\,\%$ concentration
    ($n \approx 2\times 10^{11}\,\mathrm{cm}^{-2}$) of sublattice symmetric
    defects. (left) vacancy defects ($\mathrm{V}_A + \mathrm{V}_B$), and (center)
    nitrogen substitutionals ($\mathrm{N}_A + \mathrm{N}_B$). The plots show the
    conductivity (left $y$ axis) and mobility (right $y$ axis) as a function of
    the Fermi level. (right) Ratio between the quantum and transport scattering
    times. Parameters: $\eta=25$~meV, $v_F=10^6$~m/s, and caption of
    Fig.~\ref{fig:graphene_dos}.} 
\label{fig:graphene_transport}
\end{figure*}
The separation into intravalley and intervalley scattering contributions in the
right plot of Fig.~\ref{fig:graphene_gamma} reveals that the two types of
scattering processes contribute equally to the linewidth broadening at positive
energies where resonant scattering dominates. This is in agreement with the TB
model in Eqs.~\eqref{eq:GF_NAB} and~\eqref{eq:Sigmamn_N}. On the contrary, this
is not the case on the hole side where intravalley scattering is stronger than
intervalley scattering, which can be attributed to the different intravalley and
intervalley matrix elements in Fig.~\ref{fig:M1_BZ}. Our finding for the strong
electron-hole asymmetry in the intervalley scattering rate is in excellent
qualitative agreement with recent magnetotransport measurements where the
intervalley rate was extracted from the weak localization (WL) correction to the
conductivity in nitrogen-doped graphene~\cite{Xu:Electron} and graphene with
point defect created by ion bombardment~\cite{Wu:Electrical}.

In Fig.~\ref{fig:graphene_transport}, the left and center plots show the
low-temperature transport characteristics of disordered graphene with a
$c=0.01\%$ concentration of sublattice symmetric vacancies
($\mathrm{V}_A + \mathrm{V}_B$) and N substitutionals
($\mathrm{N}_A + \mathrm{N}_B$), respectively. The plots show the conductivity
(left $y$ axis) and mobility (right $y$ axis) as a function of the Fermi level,
with the corresponding carrier density scaling as
$n \approx \left(\tfrac{E_F}{120 \,\mathrm{meV}}\right)^2 \times
10^{12}\,\mathrm{cm}^{-2}$.
For both types of defects, the transport exhibits a strong electron-hole
asymmetry in the conductivity/mobility which is inherited from the
resonant-scattering-induced asymmetry in the underlying scattering rates
(cf. Fig.~\ref{fig:graphene_gamma}). Similar electron-hole asymmetries in the
transport characteristics of disordered graphene have been addressed in other
theoretical works~\cite{Basko:Resonant,Roche:Charge,Charlier:Electronic}, and
demonstrated experimentally in defected and nitrogen-doped
graphene~\cite{Williams:Defect,Xu:Electron}.

Due to the fact that the quasibound states for the two defects considered in
Fig.~\ref{fig:graphene_transport} are, respectively, on the hole and electron
sides of the Dirac point (cf. Fig.~\ref{fig:graphene_dos}), the two sets of
conductivities/mobilities are almost mirror symmetric versions of each
other. Considering the large difference in the value of the matrix elements for
$\mathrm{V}_{A}$ and $\mathrm{N}_{A}$ defects in Fig.~\ref{fig:M1_BZ}, it is
perhaps surprising that the magnitude of the conductivities/mobilities are
almost identical when comparing the electron (hole) side for
$\mathrm{V}_A + \mathrm{V}_B$ with the hole (electron) side for
$\mathrm{N}_A + \mathrm{N}_B$. However, as witnessed by the Born vs $T$-matrix
comparison in the left plot of Fig.~\ref{fig:graphene_gamma}, the bare matrix
element which together with the DOS determines the overall magnitude of the Born
scattering rate, simply does not reflect the magnitude of the true scattering
rate given by the $T$ matrix. This holds, in particular, for strong defects
where the renormalization of the Born scattering amplitude is most significant.

In the right plot of Fig.~\ref{fig:graphene_transport} we show the ratio between
the transport and quantum scattering times as a function of the Fermi energy. In
spite of the fact that the transport characteristics for
$\mathrm{V}_A + \mathrm{V}_B$ and $\mathrm{N}_A + \mathrm{N}_B$ defects are
similar, the ratios between the scattering times for the two types of defects
show qualitative differences, in particular, on the hole side. On the basis of
the $\bk,\bq=\bk'-\bk$ dependence of the matrix elements in
Fig.~\ref{fig:M1_BZ}, vacancies are expected to behave as short-range disorder
(constant matrix element) for which $\tau_\mathrm{tr}/\tau_\mathrm{qp}\sim 1$ in
agreement with Fig.~\ref{fig:graphene_transport}. On the other hand, the strong
anisotropy and $\bq$ dependence of the matrix element for nitrogen
substitutionals reflect a dual short-range and charged-impurity character as
also discussed in Sec.~\ref{sec:examples_graphene} above. Since the transport
scattering time is less sensitive to small-angle scattering, this results in a
ratio larger than unity $\tau_\mathrm{tr}/\tau_\mathrm{qp} > 1$ on the hole
side~\cite{Sarma:Single}. On the electron side of the Dirac point, the ratio is
close to unity as resonant scattering stemming from the short-range nature of
the defect potential [cf. the TB model in Eqs.~\eqref{eq:Vab_i}
and~\eqref{eq:T_N}] dominates.

In order for a complete characterization of the nature of the defects via
transport studies, it is thus advantageous to combine measurements of the
longitudinal conductivity with measurements of the Shubnikov--de Haas
oscillations in the magnetoconductivity from which the quantum scattering time
can be inferred.

\subsection{Adatoms and adsorbates}

In addition to the in-plane defects in graphene considered above, adatoms and
molecular adsorbates sitting on top of graphene are also of high relevance as
they can be used to
functionalize~\cite{Wu:Engineering,Mauri:Phonon,Ferreira:Impact} and
dope~\cite{Cohen:First} graphene, but may at the same time dominate the
transport properties due to resonant scattering off the adatom and adsorbate
levels~\cite{Falko:Adsorbate,Lichtenstein:Impurities,Katsnelson:Resonant,Fabian:Resonant}.

In a recent work on disordered Li-decorated graphene~\cite{Kaasbjerg:Spectral},
we demonstrated that in $T$-matrix descriptions of adatoms (we expect that the
same holds for other types of adatoms and adsorbates), it is essential to
express the $T$ matrix and the Dyson equation in Eqs.~\eqref{eq:Tmatrix}
and~\eqref{eq:dyson} in a ``complete'' Bloch-state basis; i.e., the basis must
include bands which describe the electronic structure of both graphene and the
surface region where the adatoms are located.

From the point of view of first-principles calculations, the importance of using
a complete basis is not a surprising observation. On the other hand, $T$-matrix
studies of graphene with adatoms or adsorbates have consistently been based on
simple TB models considering a single impurity level coupled to the $\pi$ bands
of graphene~\cite{Falko:Adsorbate,Katsnelson:Resonant,Fabian:Resonant}. While
this can be expected to capture, e.g., resonant scattering at a qualitative
level, it does not account for the fact that the impurity level itself may
depend on the impurity concentration via their coupling to socalled surface
states~\cite{Kaasbjerg:Spectral}. The two, i.e. resonant scattering and the
position of the impurity level, are obviously interconnected and must hence be
treated in a self-contained framework.

In the $T$-matrix approach outlined here in Secs.~\ref{sec:method}
and~\ref{sec:Tmatrix}, electronic levels of adatoms and adsorbates enter through
the second term in the defect potential in Eq.~\eqref{eq:V_paw} and emerge as
poles in the $T$ matrix. In this respect, the method presented here must be
expected to give a more complete description of the spectral properties and
defect scattering in disordered 2D materials with adatoms and
adsorbates~\cite{Kaasbjerg:Spectral}.

\section{Discussion and outlook}
\label{sec:discussion}

The first-principles $T$-matrix methodology for modeling the electronic
properties of disordered materials presented here, is a natural step beyond
first-principles methods based on the Born approximation, see, e.g.,
Refs.~\onlinecite{Pantelides:First,Aberg:Charge,Mertig:Impact,Windl:A,Bernardi:Efficient}.
As witnessed by the examples included here, this is a critical step in 2D
materials where the Born approximation often breaks down and fails to capture
even the qualitative picture.

Our work has identified some of the main technical challenges associated with
first-principles $T$-matrix calculations as described in
Sec.~\ref{sec:numerics}. For example, the large matrix dimensions and memory
requirements encountered in the solution of the matrix equation in
Eq.~\eqref{eq:Tmatrix_matrix} call for careful parallelization considerations,
beyond the simple multithreading/shared memory approach adopted
here. Alternatively, the $T$-matrix equation can be solved in real-space as
described in Appendix~\ref{sec:Tmatrix_lcao} by using the supercell LCAO basis
$\{\ket{\phi_{\mu l}}\}$ in Eq.~\eqref{eq:bloch_sum} and with a subsequent
transformation to the Bloch-state basis as in Eq.~\eqref{eq:V_lcao}. This has
the immediate advantage that the dimensions of the matrices in
Eq.~\eqref{eq:Tmatrix_matrix} will be fixed to the number of LCAO basis
functions in the supercell, which is rather low ($\sim 1000$--$10000$ for the
supercell sizes considered here). However, this comes at the cost of having to
perform the transformation in Eq.~\eqref{eq:V_lcao} to the desired $\bk$-point
grid for each energy in the $T$ matrix, but this is manageable and can be more
efficient when only the $\bk$-diagonal elements of the $T$ matrix (self-energy)
are needed.

Irrespective of the strategy chosen for the solution of the $T$-matrix equation,
it is essential to use nonuniform $\bk$-point samplings of the BZ in order to
achieve a satisfactory energy resolution in subsequent calculations of, e.g.,
spectral properties, scattering rates, or transport properties; see, e.g.,
Refs.~\onlinecite{Gibertini:Mobility,Giustino:Towards,Bernardi:Efficient,Stokbro:ATK}
for other recent developments in this direction. As this allows for energy
resolutions of the order of meV, our $T$-matrix method is advantageous in
comparison to Kubo based
approaches~\cite{Roche:Charge,Charlier:Electronic,Guinea:Effect,Roche:Linear}
for the calculation of the low-temperature longitudinal conductivity and its
dependence on the Fermi energy (carrier density) in dilute, disordered
materials.

In addition to the technical aspects of the implementation and the electronic
properties discussed in this work, there are several interesting extensions to
be considered in future works. For example, a generalization to spin-dependent
defect potentials,
$\hat{V}_i = \sum_s \hat{V}_{i,s}(\hat{\br}) \otimes \hat{\sigma}_s$, accounting
for the local change in the spin-orbit interaction around the defect is
straight forward, and would allow to address spin-orbit and spin-flip
scattering, and hence defect-mediated spin relaxation~\cite{Mertig:Impact}.

A generalization of our method to the treatment of charged defects due to
filling of bound defect states by extrinsic carriers presents another highly
relevant extension of this work. This requires a self-consistent framework
as well as a proper treatment of the resulting long-range Coulomb contribution
to the defect potential like in calculations of long-range electron-phonon
interactions~\cite{Giustino:Frohlich,Mauri:Two}. Preliminary steps for modeling
charged defects have recently been reported~\cite{Lischner:Resonant}.

Finally, extensions to other 2D materials and vdW multilayer
structures~\cite{Grigorieva:vdW} as well as to solutions of the Boltzmann
equation based on first-principles inputs for the band structure, band
velocities, and $T$-matrix scattering amplitude (cf. Appendix~\ref{sec:BE}) will be
important for the future characterization of the electronic and transport
properties of new 2D materials. Also, our method for calculating the defect
matrix elements paves the way for new diagrammatic first-principles treatments
of, e.g., excitons and optical properties~\cite{Gunlycke:Optical}, as well as
transport phenomena such as, e.g., localization~\cite{Lee,Lee:rmp} and anomalous
Hall~\cite{Sinova:Anomalous,Neto:Extrinsic,Ferreira:Quantum,Ferreira:Anomalous,Roche:Linear}
effects in disordered 2D materials.
% \begin{itemize}
% \item Higher densities require other methods such as,
%   e.g., the coherent potential approximation (CPA)~\cite{RevModPhys.46.465}.
% \item Noneq. GF calculations (Guo)
% \item This vs CPA ??? See Jiawei Yan and Youqi Ke, arXiv:1511.09182.
% \end{itemize}

\section{Conclusions}

In conclusion, we have presented a DFT based first-principles method for the
calculation of defect matrix elements for realistic descriptions of impurities,
defects, substitutionals, adatoms, adsorbates, etc., in 2D materials. In
combination with a full first-principles based evaluation of the $T$-matrix
approximation for the disorder self-energy, we have developed a powerful
parameter-free first-principles framework for the description of bound defect
states, spectral properties, quasiparticle and carrier scattering, and transport
in disordered 2D materials. In spite of the fact that the focus here has been on
2D materials, the method is completely general and can be applied also to 1D and
3D materials. The method is implemented in the GPAW electronic structure
code~\cite{GPAW,GPAW1,GPAW2}.

We first applied the method to defects in the two monolayer TMDs MoS$_2$ and
WSe$_2$. We demonstrated that both vacancies and substitutional oxygen give rise
to a series of in-gap bound states with some of the states exhibiting a large
spin-orbit-induced splitting. As we have discussed in a recent
work~\cite{Kaasbjerg:Transport}, the presence of in-gap states leads to charging
of the defect sites in the extrinsic (i.e., gated) materials, and the resulting
charged-impurity scattering has detrimental consequences for the achievable
mobility. However, interestingly we find that Se vacancies and oxygen
substitutionals in WSe$_2$ only introduce empty in-gap states above the
intrinsic Fermi level, implying that these defects will remain charge neutral in
extrinsic $p$-type WSe$_2$. In the transport characteristics of high-quality vdW
WSe$_2$ devices ($n_\text{dis}\sim 10^{10}$--$10^{11}$~cm$^{-2}$) free from
charged impurities in the substrate~\cite{Hone:Disorder}, this manifests itself
in a record-high low-temperature mobility,
$\mu \sim 15000$--$35000$~$\mathrm{cm}^2\,\mathrm{V}^{-1}\,\mathrm{s}^{-1}$,
which surprisingly decreases with the carrier density (Fermi energy). The
unconventional density dependence of the mobility can be traced back to a strong
renormalization of the Born scattering amplitude by multiple-scattering
processes accounted for by the $T$ matrix. As a consequence, the quantum and
transport scattering times become strongly energy dependent and increase away
from the band edge. In conjunction with the severe overestimation of the
scattering rate by the Born approximation, this underlines the importance of a
$T$-matrix treatment of point defects in disordered 2D semiconductors.

We also discussed our previously reported symmetry-induced protection against
intervalley scattering by defects in 2D TMDs~\cite{Jauho:Symmetry}, and showed
that it completely suppresses intervalley scattering by, e.g., S vacancies in
the conduction band of MoS$_2$, which has often been suggested as the origin of
the intervalley scattering extracted from WL/WAL in 2D
MoS$_2$~\cite{Eda:Quantum,Eda:Phase}. This finding furthermore points to the
possibility of achieving extremely long valley lifetime even in disordered 2D
TMDs.

In the last part, we studied the effect of carbon vacancies and nitrogen
substitutionals on the electronic properties of graphene. Here, we found that
the two types of defects give rise to quasibound resonant states, respectively,
below and above the Dirac point. While the latter is in agreement with
experimental studies of nitrogen substitutionals in
graphene~\cite{Pasupathy:Visualizing,Henrard:Localized}, our finding for the
position of the vacancy-induced resonant state below the Dirac point is in
contrast to numerous tight-bindings studies where it appears at the Dirac
point~\cite{Neto:Electronic,Katsnelson:Resonant,Ducastelle:Electronic,Fabian:Resonant,Ast:Band}. For
now, we can only speculate that this is due to an oversimplified treatment of
the vacancy defect potential in the tight-binding models.

Studying the spectral properties of disordered graphene with, respectively,
sublattice-asymmetric and sublattice-symmetric distributions of vacancies and
nitrogen substitutionals, we demonstrated defect concentrations of the order of
$c\sim1\,\%$ ($n_\text{dis}\sim 10^{13}$~cm$^{-2}$) are required in order to see
fingerprints of the resonant states in the spectral function measured in
ARPES. We furthermore showed that sublattice-asymmetric disorder with the
defects located exclusively on one of the sublattices, opens a
concentration-dependent band gap in graphene, which for the above-mentioned
concentration is of the order of $\sim 100$~meV. For sublattice-symmetric
disorder, the spectrum again becomes gapless, but retains its characteristic
form at the position of the quasibound states caused by strong resonant
scattering. In the presence of equal concentrations of sublattice-symmetric
vacancies and nitrogen substitutionals, the band renormalization due to the two
types of defects cancel each other, which results in a less dramatic deformation
of the Dirac cone in the vicinity of the quasibound states.

Finally, we demonstrated that the transport characteristics of disordered
graphene become strongly electron-hole asymmetric in the presence of quasibound
resonant states.  

Altogether, our first-principles-based $T$-matrix method is an important step
toward accurate modeling of realistic defects and their impact on the
electronic properties of disordered materials.

\begin{acknowledgments}
  The author would like to thank J.~H.~J.~Martiny, J.~J.~Mortensen, N.~Papior,
  M.~Brandbyge, T. Olsen and A.-P. Jauho for stimulating discussions and
  comments on the manuscript. K.K.~acknowledges support from the European
  Union's Horizon 2020 research and innovation programme under the Marie
  Sklodowska-Curie Grant Agreement No.~713683 (COFUNDfellowsDTU). The Center for
  Nanostructured Graphene (CNG) is sponsored by the Danish National Research
  Foundation, Project No. DNRF103.
\end{acknowledgments}

\appendix

\section{Numerical solution of the Boltzmann equation}
\label{sec:BE}

The Boltzmann equation in Eq.~\eqref{eq:BE} can be recast as a matrix equation
in the composite band and wave-vector index $(n,\bk)$,
\begin{equation}
  \label{eq:BE_matrix}
  \mathbf{C}\, \mathbf{\tilde{f}} = \mathbf{b}, \quad  
  \tilde{f}_{n\bk} = \frac{\delta f_{n\bk}}
     {q \abs{\mathbf{E}} 
        \frac{\partial f^0}{\partial \varepsilon} 
        \vert_{\varepsilon=\varepsilon_{n\bk}}}  ,
\end{equation}
which is solved for $\tilde{f}_{n\bk}$. Here, the matrix elements of the
collision matrix and the vector on the right-hand side are given, respectively,
by
\begin{align}
  C_{n\bk,n'\bk'} & = - \delta_{n\bk,n'\bk'} \sum_{n''\bk''}
  P_{n\bk,n''\bk''} + P_{n\bk,n'\bk'}  ,
  \\ 
  % \quad \text{and} \quad
  b_{n\bk} & = \mathbf{v}_{n\bk} \cdot \mathbf{\hat{E}} ,
\end{align}
where $P_{n\bk,n'\bk'}$ is the $T$-matrix transition rate in
Eq.~\eqref{eq:Pnknk}, $\mathbf{v}_{n\bk}$ is the band velocity, and
$\mathbf{\hat{E}} = \mathbf{E}/\abs{\mathbf{E}}$ is a unit vector in the
direction of the applied electric field. It should be noted that no
approximations have been invoked in the solution of the Boltzmann equation
outlined above which applies in the case of \emph{elastic} scattering.

The matrix form of the Boltzmann equation in Eq.~\eqref{eq:BE_matrix} appended
with the additional particle-conserving constraint
$\sum_{n\bk} \delta f_{n\bk}=0$ on the distribution function, can be solved with
a standard least-squares method based on a singular-value decomposition of the
collision matrix.

\section{$T$-matrix equation in LCAO basis}
\label{sec:Tmatrix_lcao}

As an alternative to the Bloch-state formulation of the $T$-matrix equation in
Eq.~\eqref{eq:Tmatrix} of the main text, it may for practical reason be
advantageous to use the LCAO supercell basis as discussed in
Sec.~\ref{sec:discussion}.

In the nonorthogonal LCAO basis $\{\ket{\phi_{\mu k}}\}$, the completeness
relation takes the form
$\sum_{k\mu,l\nu} \ket{\phi_{\mu k}}(S^{-1})_{kl}^{\mu\nu}\bra{\phi_{\nu l}} =
\hat{1}$,
where $S^{-1}$ is the inverse of the overlap matrix defined by
$S_{kl}^{\mu\nu} = \braket{\phi_{\mu k}}{\phi_{\nu l}}$. Considering the matrix
elements
$T_{i,kl}^{\mu\nu}(\varepsilon) = \bra{\phi_{\mu k}} \hat{T}_i(\varepsilon)
\ket{\phi_{\nu l}}$
and inserting completeness relations in the $T$-matrix equation
$\hat{T}_i(\varepsilon) = \hat{V}_i + \hat{V}_i \hat{G}^0(\varepsilon)
\hat{T}_i(\varepsilon)$, the latter becomes
\begin{equation}
  \label{eq:Tmatrix_lcao}
  \hat{T}_{i,kl}(\varepsilon) = \hat{V}_{i,kl}
  + \sum_{k'l'} \hat{V}_{i,kk'} \hat{\tilde{G}}_{k'l'}^0(\varepsilon)
  \hat{T}_{i,l'l}(\varepsilon) ,
\end{equation}
where the orbital index has been omitted for brevity and
$\hat{\tilde{G}}^0 = \hat{S}^{-1} \hat{G}^0 \hat{S}^{-1}$.

Like in Sec.~\ref{sec:Tmatrix}, this can be recast as a matrix equation in the
cell and orbital indices,
\begin{equation}
  \label{eq:Tmatrix_lcao_matrix}
  \left[\mathbf{1} - \mathbf{V} \mathbf{\tilde{G}}^0(\varepsilon) \right]
  \mathbf{T}(\varepsilon) =  \mathbf{V},
\end{equation}
with the GF given by
\begin{equation}
  \label{eq:GF_lcao}
  \mathbf{\tilde{G}}^0(\varepsilon) = 
  \left[\varepsilon \mathbf{S} - \mathbf{H}_0  \right]^{-1} .
\end{equation}
The dimension of the matrices is here given by the number of LCAO orbitals in
the supercell.

% \section{BZ sums}
% 
% % \begin{figure}[!t]
% %   \centering
% %   \includegraphics[width=0.49\linewidth]{bzgrid_uniform}
% %   \includegraphics[width=0.49\linewidth]{bzgrid_nonuniform}
% %   \caption{(a) Uniform BZ grid. (b) Nonuniform BZ grid with a denser sampling in
% %     a small region around the high-symmetry points (red squares) of particular
% %     interest.} 
% % \label{fig:bzgrid}
% % \end{figure}
% 
% Sum to integral conversion
% \begin{equation}
%   \label{eq:sum2inte}
%   \frac{1}{N} \sum_\bk \rightarrow
%   A_\text{cell} \int_\text{BZ} \! \frac{d\bk}{(2\pi)^2} ,
% \end{equation}
% where $N$ is the number of unit cells in the system.
% 
% Discretization of the BZ integral:
% \begin{equation}
%   \label{eq:weight}
%   \Delta\bk = \Omega_\text{BZ} w_\bk = \frac{(2\pi)^2}{A_\text{cell}} w_\bk
% \end{equation}
% 
% BZ samplings:
% \begin{itemize}
% \item Uniform: $w_\bk = 1 / N_\bk$.
% \item Nonuniform: $w_{i, \bk} = Z_i / N_{i,\bk}$ where $Z_i$ is the fraction of
%   the BZ sampled with $N_{i,\bk}$ $\bk$ points.
% \end{itemize}
% 
% \newpage

\bibliography{journalabbreviations,references}
% \bibliography{paper}

\end{document}